\documentclass[a4paper,11pt]{article}
 \pdfoutput=1
 \usepackage{jheppub} 
 \usepackage{graphicx}
 \usepackage{amssymb}
\usepackage{dcolumn}% Align table columns on decimal point
\usepackage{bm}% bold math 
\usepackage{color}
\usepackage{multirow}
 
\newcommand{\met}{{/\!\!\! E_T}} 
\newcommand{\mpt}{{\;/\!\!\!\! \vec{P}_T}} 
\usepackage{relsize}
\usepackage{cancel}
\usepackage[normalem]{ulem}

\newcommand{\mPT}{\vec{\cancel{P}}_\ms{T}}

\newcommand{\A}[1]{A_{#1}}
\newcommand{\B}[1]{B_{#1}}
\newcommand{\C}[1]{C_{#1}}
\renewcommand{\a}[1]{a_{#1}}
\renewcommand{\b}[1]{b_{#1}}
\newcommand{\ms}[1]{{\!\mathsmaller{#1}}}
\newcommand{\mA}{m_\ms{A}}
\newcommand{\mB}{m_\ms{B}}
\newcommand{\mC}{m_\ms{C}}
\newcommand{\pa}[1]{p_{a_{#1}}}
\newcommand{\pb}[1]{p_{b_{#1}}}
\newcommand{\qA}[1]{q_{\ms{A}_{#1}}}
\newcommand{\qB}[1]{q_{\ms{B}_{#1}}}
\newcommand{\qC}[1]{q_{#1}}
\newcommand{\tmA}{\tilde{m}_\ms{A}}
\newcommand{\tmB}{\tilde{m}_\ms{B}}
\newcommand{\tmC}{\tilde{m}_\ms{C}}

\newcommand{\Ea}[1]{E_{a_{#1}}}
\newcommand{\Eb}[1]{E_{b_{#1}}}
\newcommand{\EC}[1]{E_{\ms{C}_{#1}}}

\newcommand{\GeV}{\;\mathrm{GeV}}
\newcommand{\TeV}{\;\mathrm{TeV}}

 \newcommand{\lsim}{{\;\raise0.3ex\hbox{$<$\kern-0.75em\raise-1.1ex\hbox{$\sim$}}\;}}
\newcommand{\gsim}{{\;\raise0.3ex\hbox{$>$\kern-0.75em\raise-1.1ex\hbox{$\sim$}}\;}}
\newcommand{\beq}{\begin{equation}}
\newcommand{\eeq}{\end{equation}}
\newcommand{\bea}{\begin{eqnarray}}
\newcommand{\eea}{\end{eqnarray}}

\def\baa{\begin{array}}
\def\eaa{\end{array}}

\mathchardef\minus="002D

\def\met{E_T\hspace{-0.45cm}/\hspace{0.25cm}}

\title{\boldmath Kinematic Focus Point Method for Particle Mass Measurements in Missing Energy Events}
 
\author[a]{Doojin Kim,}
\author[b]{Konstantin T.~Matchev,} 
\author[b]{Prasanth Shyamsundar}

\affiliation[a]{Department of Physics, University of Arizona, Tucson, AZ 85721, USA}
\affiliation[b]{Institute for Fundamental Theory, Physics Department, University of Florida, Gainesville, FL 32611, USA}

\abstract{We investigate the solvability of the event kinematics in missing energy events at hadron colliders, as a function of the particle mass ansatz. 
To be specific, we reconstruct the neutrino momenta in dilepton $t\bar{t}$-like events, without assuming any prior knowledge of the mass spectrum.
We identify a class of events, which we call {\em extreme events}, 
with the property that the kinematic boundary of their allowed region in mass parameter space passes through the true mass point. 
We develop techniques for recognizing extreme events in the data and demonstrate that they are abundant in a realistic data sample, 
due to expected singularities in phase space. We propose a new method for mass measurement
whereby we obtain the true values of the mass parameters as the focus point of the kinematic 
boundaries for {\em all} events in the data sample. Since the masses are determined from a relatively sharp peak structure 
(the density of kinematic boundary curves), the method avoids some of the systematic errors associated with other techniques.
We show that this new approach is complementary to previously considered methods 
in the literature where one studies the solvability of the kinematic constraints throughout the mass parameter space. 
In particular, we identify a problematic direction in mass space of nearly 100\% solvability, 
and then show that the focus point method is effective in lifting the degeneracy. }
 
\date{June 6, 2019}

\begin{document} 
\maketitle
\flushbottom

\section{Introduction}
\label{sec:introduction}

At the Large Hadron Collider (LHC), information about the underlying physics is extracted by 
studying the kinematic properties of the final state objects (jets, leptons, photons, etc.) which are
reconstructed in each collision. Therefore, understanding the unique kinematic features arising from different event topologies 
is an important step in any LHC data analysis, be it for a new physics search, or a parameter measurement.

For example, the traditional method to discover a new heavy resonance (as well as to measure its mass) is to 
look for a bump in the invariant mass distribution of its decay products (daughter particles). This procedure is straightforward if
the daughter particles are all visible in the detector and their energies and momenta are measured. However,
things become more involved if some of the daughter particles are invisible, e.g., they can be neutrinos or new 
weakly interacting massive particles, as predicted in many models of new physics with dark matter candidates and/or dark sectors \cite{Feng:2010gw}.
A prime example of this situation is provided by semi-leptonic $t\bar{t}$ events, in which one of the top quarks, say $\bar{t}$, decays hadronically, 
while the other decays as 
\beq
t \to bW^+\to b \ell^+\nu,
\label{topdecay}
\eeq
where a neutrino $\nu$ goes missing. At this point a natural approach would be to attempt to {\em compute} the 
momentum of the neutrino from the available kinematic information in the event, perhaps supplemented with some theoretical assumptions.
For example, assuming that there are no other invisible particles in the event, we can identify the measured missing transverse momentum $\mpt$
in the event with the transverse component $\vec{p}_{\nu T}$ of the neutrino momentum
\beq
\vec{p}_{\nu T} = \mpt.
\label{pnuTeqMET}
\eeq
On the other hand, the neutrino arises from the decay of a $W$-boson, which is typically produced on-shell, in turn implying that 
\beq
(p_{\ell^+}+p_\nu)^2 = m_W^2,
\label{Wconstraint}
\eeq
where $p_{\ell^+}$ and $p_\nu$ are the 4-momenta of the lepton and neutrino, respectively, and $m_W$ is the physical mass of the $W$-boson.
Finally, the energy $E_\nu$ and 3-momentum $\vec{p}_\nu$ of the neutrino are related by the mass shell condition
\beq
p^2_\nu=m_\nu^2,
\label{nuconstraint}
\eeq
where $p_\nu = (E_\nu,\vec{p}_\nu)$. Equations (\ref{pnuTeqMET}-\ref{nuconstraint}) provide a total of $2+1+1=4$ constraints on the four 
components of $p_\nu$, allowing to solve for the neutrino 4-momentum, up to discrete ambiguities\footnote{The mass constraints 
(\ref{Wconstraint}) and (\ref{nuconstraint}) lead to a quadratic equation, which in general has two solutions. Additional discrete combinatorial ambiguities arise 
if there are more than one lepton and/or more than one $b$-jet present in the event.}.
In practice, this is typically done through an eventwise kinematic fit which accounts for the detector resolution and the combinatorial uncertainties
\cite{Abbott:1998dc,Gripaios:2011jm,Sirunyan:2018gqx}.

Now and for the rest of the paper we shall turn our attention to the much more challenging case of
{\em dilepton} $t\bar{t}$ events, in which {\em both} top quarks decay leptonically as in (\ref{topdecay}).
This event topology is relevant not only for top physics, but also for new physics searches where 
the decay chains of the newly produced particles terminate in a generic dark matter candidate, i.e.,
a stable, neutral, weakly interacting particle which is invisible in the detector just like the neutrino.
In many such models, e.g., supersymmetry (SUSY) with $R$-parity \cite{Martin:1997ns}, 
universal extra dimensions with KK-parity \cite{Appelquist:2000nn,Rizzo:2001sd,Cheng:2002ab},
Little Higgs with $T$-parity \cite{ArkaniHamed:2002qy,Cheng:2004yc,Schmaltz:2005ky,Perelstein:2005ka}, etc., 
the lifetime of the dark matter particle is protected by an exact $Z_2$
symmetry, which ensures that new particles are pair-produced at the LHC, leading to symmetric events with two decay chains.
From that point of view, dilepton $t\bar{t}$ events are an excellent testbed which allows us to test 
ideas originally developed for the study of new physics signatures, see, e.g., \cite{Chatrchyan:2013boa}.

Once we consider dilepton $t\bar{t}$ events, we have to face the complication that now there are twice as many 
unknown momentum components, i.e., the four momenta, $p_\nu$ and $p_{\bar\nu}$, of the two missing neutrinos.
The number of on-shell constraints is also doubled --- (\ref{Wconstraint}) is replaced by
\begin{subequations}
\bea
(p_{\ell^+}+p_\nu)^2 &=& m_W^2, \\ [2mm]
(p_{\ell^-}+p_{\bar\nu})^2 &=& m_W^2,
\eea
\label{Wconstraint2}
\end{subequations}
while (\ref{nuconstraint}) is replaced by
\begin{subequations}
\bea
p^2_\nu&=&m_\nu^2, \\ [2mm]
p^2_{\bar\nu}&=&m_\nu^2.
\eea
\label{nuconstraint2}
\end{subequations}
We also gain a new constraint, that the two parent particles (top quarks) have equal mass:
\beq
(p_b+p_{\ell^+}+p_\nu)^2=(p_{\bar b}+p_{\ell^-}+p_{\bar\nu})^2,
\label{topequal}
\eeq
but the missing transverse momentum constraint (\ref{pnuTeqMET}) remains a single 2D vector equation:
\beq
\vec{p}_{\nu T} + \vec{p}_{\bar\nu T} = \mpt.
\label{pnuTeqMET2}
\eeq
As a result, the seven constraints (\ref{Wconstraint2}-\ref{pnuTeqMET2}) are not sufficient to determine the eight unknown
components $p_\nu$ and $p_{\bar\nu}$ uniquely,
and therefore, one cannot reconstruct the top as a mass bump.
The problem is exacerbated in new physics applications, where {\em a priori} one does not know the masses of the 
new particles playing the roles of the $W$-boson and the neutrino, so that the best one can do is to apply the equal mass constraints
\bea
p^2_\nu&=&p^2_{\bar\nu} \label{nuequal}\\ [2mm]
(p_{\ell^+}+p_\nu)^2 &=& (p_{\ell^-}+p_{\bar\nu})^2 \label{Wequal}
\eea
in place of (\ref{Wconstraint2}-\ref{nuconstraint2}), leaving us with only 5 constraints (\ref{topequal}-\ref{Wequal}) for 8 unknowns.

There are several possible approaches to tackle this conundrum \cite{Barr:2010zj}:
\begin{itemize}
\item {\em Gain additional constraints.} This can be done in several ways. First, one may consider longer decay chains which may occur in
new physics scenarios. The classic examples are the three-step squark decay chain and the four-step gluino decay chain in supersymmetry,
where the trick of analyzing several events simultaneously could provide the required number of constraints to solve for all unknown momenta, up to
discrete ambiguities \cite{Nojiri:2003tu,Kawagoe:2004rz,Cheng:2007xv,Cheng:2008mg,Cheng:2009fw,Webber:2009vm}. 
One may also impose constraints resulting from existing preliminary measurements of the kinematic endpoints of suitable variables
\cite{Nojiri:2007pq,Djilkibaev:2008pj,Casadei:2010nf}. Finally, one could preselect a sub-sample of events with special properties,
e.g., events located at the kinematic endpoint of a suitable variable, in which case the kinematics of the decays is further 
constrained \cite{Kersting:2009ne,Kang:2009sk}.
\item {\em Adopt an ansatz for the undetermined components of the invisible momenta.} The idea here is to assign values for the 
remaining invisible momentum components through a suitable ansatz which preserves some useful property, 
e.g., that a parent mass bound is not exceeded   \cite{Barr:2011xt,Kim:2017awi}. 
Operationally the values are assigned by extremizing some relevant function of the invisible momenta. 
Often the minimum of the function itself becomes a useful kinematic variable --- some well-known examples include the 
Cambridge $M_{T2}$ variable \cite{Lester:1999tx,Barr:2003rg} and its variants \cite{Burns:2008va,Barr:2009jv,Konar:2009wn,Konar:2009qr}, 
the $\sqrt{s}_{min}$ variable \cite{Konar:2008ei,Konar:2010ma,Swain:2014dha},
a variety of constrained transverse mass variables \cite{Barr:2009mx,Gross:2009wia,Barr:2011he,Barr:2011ux},
the $M_{CT2}$ variable \cite{Cho:2009ve,Cho:2010vz}, 
the $M_{2C}$ variable \cite{Ross:2007rm,Barr:2008ba}, 
the MAOS method \cite{Cho:2008tj,Choi:2009hn,Park:2011uz,Guadagnoli:2013xia},
the $M_2$ class of variables \cite{Mahbubani:2012kx,Cho:2014naa,Cho:2014yma,Kim:2014ana,Cho:2015laa,Konar:2015hea,Konar:2016wbh,Goncalves:2018agy}, etc. 
While this approach has useful practical applications, 
it still only represents an approximate treatment and does not lead to a mass reconstruction through a bump.
\item {\em Use kinematic endpoints of observable invariant mass distributions.} Another logical possibility is to ignore the invisible momenta altogether and
design the analysis entirely in terms of the measured momenta of the reconstructed visible objects in the event. Historically this was the 
original approach which led to the classic analyses of SUSY mass measurements from kinematic endpoints in invariant mass distributions
\cite{Hinchliffe:1996iu,Bachacou:1999zb,Allanach:2000kt,Gjelsten:2004ki,Gjelsten:2005aw,Matchev:2009iw,Matchev:2019sqa}. However, 
as in the previous case, the method relies on the measurement of kinematic endpoints instead of mass bumps. Endpoint measurements
are generally challenging --- they may be difficult to extract in the presence of background, 
their location is rather sensitive to the effects of particle widths and detector resolution,
and they require fitting to a suitable profile \cite{Agashe:2010tu,Curtin:2011ng}, which introduces additional systematics.
\item {\em Check the solvability of an enlarged set of kinematic constraints.} Here the idea is first to enlarge the existing set of kinematic constraints 
by assuming test values for the unknown masses, and then instead of focusing on the actual solutions for the invisible momenta, 
simply ask the question whether such solutions exist or not. If solutions do not exist, the corresponding test ansatz for the particle masses is
inconsistent with the kinematics of the particular event and those masses are disfavored \cite{Kawagoe:2004rz,Cheng:2007xv,Cheng:2008hk}.
By repeating the procedure over the full event sample, one can gradually restrict the allowed mass parameter space, 
hopefully until it shrinks to a point.\footnote{A variant of this technique assigns weights derived from the 
parton distribution functions to different points in the allowed region of mass parameter space 
in order to arrive at the most likely value of the particle masses \cite{Anagnostou:2011aa}.} 
Indeed, the true masses should be compatible with every signal event, while
for any wrong choice of particle masses, one would hope that, given sufficient statistics, there would be at least one signal event in the data 
which would have incompatible kinematics. We will illustrate these ideas more explicitly later on in Sections~\ref{sec:solvability} and \ref{sec:fd}.
For now we just mention that this method appears to have the desired property that the true particle masses are revealed by a bump
in mass parameter space (for example, when plotting the fraction of compatible events). However, as we shall show in Section~\ref{sec:fd},
this expectation is misleading, as there is generally a whole ``flat direction" of near 100\% solvability in mass parameter space.
\end{itemize}

In this paper we present a new approach to measuring masses in events with $\mpt$, 
by building up on the solvability method discussed in the last bullet above. Our method makes crucial use of 
the singularities in the phase space density which arise from projecting the full phase space 
of final state momenta into the visible phase space 
\cite{Kim:2009si}. This method, which we shall call the ``kinematic focus points method"\footnote{Not to be confused with the 
focus point phenomenon in the RGE running of certain SUSY mass parameters \cite{Feng:1999hg,Feng:1999zg}.}, 
has similarities to a traditional mass-bump search in the sense that
the density of an appropriately defined quantity will peak at the true values of the unknown masses in the event topology. 
Previous attempts to design an analysis for measuring the unknown masses in SUSY-like events with missing energy from such bump-like features 
were limited to the case of a single decay chain \cite{Lester:2013aaa,Agrawal:2013uka,Debnath:2016gwz,Altunkaynak:2016bqe,Debnath:2018azt}.
Instead, here we will be interested in the more typical situation of two symmetric decay chains. 
While our method is applicable more generally, we shall introduce it
in the context of the dilepton $t\bar{t}$ event topology discussed above.

The paper is organized as follows. In Section~\ref{sec:notation} we specify the process under study, introduce our conventions and terminology, and
provide some details on our simulations. In Section~\ref{sec:solvability} we list the kinematic constraints for our event topology, 
and briefly review the idea of using their solvability in order to perform mass measurements (or at the very least, to restrict the allowed 
region of mass parameter space). In Section~\ref{sec:fd} we discuss the expected outcome from this approach and highlight
its advantages as well as potential pitfalls. In particular, we identify a flat direction in mass parameter space of nearly 100\% solvability, 
along which the masses are consistent with the kinematics of nearly all events in the data. We then show that by performing 
complementary measurements of kinematic endpoints one is able to lift the flat direction and identify the true mass point.\footnote{To the
careful reader, this should not come as a surprise, since kinematic endpoint measurements by themselves are 
already able to fully determine the mass spectrum \cite{Burns:2008va,Chatrchyan:2013boa}.}

In the remaining sections, we define the kinematic focus point method 
and illustrate its performance. First in Section~\ref{sec:extremeevents} we discuss the event-wise kinematic boundaries in 
mass parameter space, and introduce the notion of ``extreme" events, i.e., events for which the kinematic boundary passes through the 
true mass point. We argue that extreme or near-extreme events are quite abundant in a realistic data sample. More importantly, we show that 
the density of extremeness boundaries in the mass parameter space is singular at the true mass point\footnote{This is a purely 
mathematical statement which is valid in the zero width approximation and in the absence of detector smearing. 
Both of those effects will lead to some smearing of the singularity, the extent of which is studied later on in Sections~\ref{sec:resolution} and \ref{sec:ttbar}.}. 
Using the results from Section~\ref{sec:extremeevents}, in Section~\ref{sec:focus} we motivate the 
kinematic focus points method and proceed to investigate its performance for two cases --- a SUSY-like mass spectrum (where 
top events are part of the SM background) as well as for top pair-production itself, thus allowing for a measurement 
of the top, $W$-boson and neutrino masses independently\footnote{An analogous procedure based on kinematic endpoints
was advertised in~\cite{Burns:2008va} and tested in \cite{Chatrchyan:2013boa}.}.  
In Section~\ref{sec:conclusions}, we present our conclusions and outlook for future studies.
In Appendix~\ref{app:flatdirection} we derive the parametric equation for the flat direction of low mass sensitivity encountered in Sections~\ref{sec:fd} and \ref{sec:focus}.

\section{Notations and Setup}
\label{sec:notation}

\subsection{Conventions}

\begin{figure}[t]
 \centering
 \includegraphics[width=.5\textwidth]{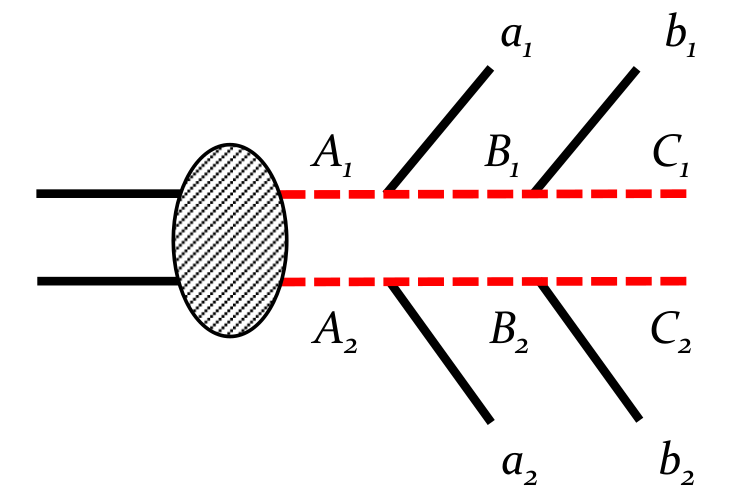}
 \caption{\label{fig:feynmandiag} A schematic depiction of the dilepton $t\bar{t}$ event topology. 
 Two heavy particles, $\A{1}$ and $\A{2}$, are pair-produced, and each $\A{i}$ decays via two successive 
 two-body decays as $\A{i}\rightarrow\a{i}\B{i}\rightarrow\a{i}\b{i}\C{i}$. 
 In general, $\A{i}$, $\B{i}$ and $\C{i}$ are new particles of {\em a priori} unknown masses. 
 The two branches are assumed to be identical and the intermediate resonances $\A{i}$ and $\B{i}$ are assumed to be on-shell. 
 The final state particles $\a{i}$ and $\b{i}$ are visible Standard Model particles which for simplicity will be assumed massless throughout the analysis.
  The final state particles $\C{i}$ are invisible in the detector and their momenta are not measured.}
\end{figure}

In this paper, we consider the $t\bar{t}$-like event topology depicted in Fig~\ref{fig:feynmandiag}. 
A pair of heavy particles, $\A{1}$ and $\A{2}$, is initially produced, and each $\A{i}$ decays via two successive two-body decays:
\begin{subequations}
\bea
pp &\rightarrow& \A{1}\A{2}, \\ [2mm]
\A{i}&\rightarrow& \a{i}\B{i}\rightarrow\a{i}\b{i}\C{i} \quad (i=1,2),
\eea
\end{subequations}
to two visible Standard Model particles $\a{i}$ and $\b{i}$ and an invisible particle $\C{i}$.
The two branches in Fig.~\ref{fig:feynmandiag} are assumed to be identical  
and the intermediate resonances $\A{i}$ and $\B{i}$ will be taken to be on-shell. 
The true masses of particles $\A{i}$, $\B{i}$ and $\C{i}$ are {\em a priori} unknown, and will be denoted by $\mA$, $\mB$ and $\mC$, respectively. 
The goal is to measure $\mA$, $\mB$ and $\mC$ independently from a sample of events with the topology of Fig.~\ref{fig:feynmandiag}.

Since the final state particles $\a{i}$ and $\b{i}$ are visible in the detector,
their 4-momenta, $\pa{i}$ and $\pb{i}$, are measured known quantities for each event. 
In many new physics models with this event topology, $\a{i}$ and $\b{i}$ are bottom quarks and leptons, respectively,
just like the case of dilepton $t\bar{t}$ events. Then, the $\a{}$'s can be distinguished from the $\b{}$'s, 
but there remains a twofold ambiguity in partitioning the two $\a{}$'s and the two $\b{}$'s into the two branches. 
As there are already several suggestions in the literature on how to address this combinatorial 
issue \cite{Baringer:2011nh,Choi:2011ys,Debnath:2017ktz}, 
we shall not dwell on the combinatorial problem any further.
Note that our proposed method below will be robust against combinatorial issues, since
it will rely on detecting a peak structure which is absent from the combinatorial background.
Therefore, for clarity of the presentation, in what follows 
we shall assume that the visible particles have been properly assigned to the two branches. 

The final state particles $\C{i}$ are invisible in the detector ($C$ is typically a dark matter candidate). 
Consequently, their 4-momenta, $\qC{i}$, are {\em a priori} unknown. 
The 4-momenta of $\B{i}$ and $\A{i}$ are denoted by $\qB{i}(\equiv \pb{i}+\qC{i})$ and $\qA{i}(\equiv \pa{i}+\pb{i}+\qC{i})$,
respectively. Note that the notation here follows that of \cite{Barr:2011xt}, where
the letter $p$ is used to denote known momenta, while $q$ is used for {\em a priori} unknown momenta. 
An overhead arrow will be used to denote the spatial component $\vec{p}$ of a 4-momentum $p$. 
The beam axis (the longitudinal direction) is chosen to be the $z$ axis, while
the transverse components of a 4-momentum $p$ will be denoted 
by $\vec{p}_T$. The product of two 4-momenta will refer to their Lorentz inner product.

We shall often consider hypothesized values for the {\em a priori} unknown masses, i.e., {\em test} or {\em trial} masses, which will 
be denoted by a tilde: $\tmA$, $\tmB$, and $\tmC$. For convenience, the singular form \emph{true mass} and \emph{test mass} 
will sometimes be used to refer to a set of true masses $(\mA, \mB, \mC)$ and a set of test masses $(\tmA, \tmB, \tmC)$. 
For notational convenience, $\qC{i}$, $\qB{i}$ and $\qA{i}$ will be used interchangeably as variables or 
particular values for these variables, and the meaning will be clear from the context.

\subsection{Simulation details}

The analyses in this study were performed on events generated as follows.

\begin{itemize}
 \item Obviously, the event topology of Fig.~\ref{fig:feynmandiag} can be applied directly to the study of top quark pair production, 
in which case the particles $A$, $B$ and $C$ are the top quark, the $W$-boson and the neutrino, respectively. 
While their masses $m_t$, $m_W$ and $m_\nu$, are already known, the techniques discussed below can still be 
used to perform more sensitive measurements of those SM parameters. Our primary target, however, is a
new physics event topology of the type shown in Fig.~\ref{fig:feynmandiag}, where the three masses 
$\mA$, $\mB$ and $\mC$ are {\em a priori} unknown.
For example, in SUSY $A$ can be a top squark, $B$ a chargino and $C$ a sneutrino. Thus 
for our main simulation we choose a study point with a spectrum of $\mA=1000\GeV$, $\mB=800\GeV$ and $\mC=700\GeV$.
Note that the mass differences are relatively modest, so that the $p_T$ and energy distributions of the jets and leptons in the signal are 
similar to those of the main background (dilepton $t\bar{t}$ events), so that a template-matching method is unlikely to be very sensitive.
 \item For clarity of the presentation, the resonances $\A{i}$ and $\B{i}$ were kept on-shell. 
 Including small widths for these particles will lead to a slight smearing of the peaks and 
 endpoints discussed in the study \cite{Grossman:2011nh,Kim:2017qdi}.
 \item The visible final state particles $\a{i}$ and $\b{i}$ were taken to be massless. This is a very good approximation 
 in the typical scenario where $\a{i}$ and $\b{i}$ are bottom quarks and leptons, respectively. 
 Regardless, the mass measurement methods considered here do not rely on this choice, and this assumption can be easily relaxed if needed.
 \item For definiteness the decay vertices were taken to be non-chiral, i.e., there are no non-trivial spin effects. 
 This assumption also does not impact our results, 
 since the methods studied below are based on purely kinematics arguments and are therefore model independent. 
 \item The parent particles $\A{i}$ were pair produced at LHC center of mass energy of $13\TeV$. 
 For simplicity, initial state radiation (ISR) was not turned on. Again, this assumption has little bearing on our results ---
 for example the ISR jets are unlikely to be $b$ jets in which case one might worry that they can be confused with particles $\a{i}$ 
 in Fig.~\ref{fig:feynmandiag}, complicating the combinatorial issue. 
\item The events were generated at the parton level with {\sc MadGraph} \cite{Alwall:2011uj} and resonances were decayed by phase space. 
Initially, we will show results with no detector simulation applied,
before studying the effects of the detector resolution in Sections~\ref{sec:resolution} and \ref{sec:ttbar}. 
\end{itemize}

\section{Solvability of Kinematic Constraints} 
\label{sec:solvability}

The general problem with missing energy events is that there exist final state momenta ($\qC{i}$ in our case) which are not measured.
However, if we could somehow {\em calculate} the invisible momenta $\qC{i}$ (and from there $\qB{i}$ and $\qA{i}$), 
the unknown masses can be simply obtained by $\mA^2 = \qA{i}^2$, $\mB^2 = \qB{i}^2$ and $\mC^2 = \qC{i}^2$. 
This motivates us to study the kinematic constraints obeyed by the invisible momenta.

With the notation from Section~\ref{sec:notation}, the five kinematic constraints (\ref{topequal}-\ref{Wequal}) 
for the event topology in Fig.~\ref{fig:feynmandiag} become
\begin{subequations}
\label{eqn:on-shell-orig}
\bea
\qC{1}^2 &=& \qC{2}^2\\ [1mm]
(\pb{1}+\qC{1})^2 &=& (\pb{2}+\qC{2})^2\\ [1mm]
(\pa{1}+\pb{1}+\qC{1})^2 &=& (\pa{2}+\pb{2}+\qC{2})^2
\eea
\end{subequations}
\beq
\vec{q}_{1T} + \vec{q}_{2T} = \mPT~~~~~~~~
\label{eqn:met}
\eeq
These five constraints are clearly insufficient to determine all 8 unknown components of $\qC{i}$. 
To this end, we introduce test masses $(\tmA, \tmB, \tmC)$ with $\tmA > \tmB > \tmC$. 
The constraints in eq.~\eqref{eqn:on-shell-orig} are thus enlarged to
\begin{subequations}
\label{eqn:on-shell}
\bea
\qC{1}^2 &=& \tmC^2,\\ [1mm]
\qC{2}^2 &=& \tmC^2\\ [1mm]
(\pb{1}+\qC{1})^2 &=& \tmB^2\\ [1mm]
(\pb{2}+\qC{2})^2 &=& \tmB^2\\ [1mm]
(\pa{1}+\pb{1}+\qC{1})^2 &=& \tmA^2\\ [1mm]
(\pa{2}+\pb{2}+\qC{2})^2 &=& \tmA^2
\eea
\end{subequations}
Together with eq.~(\ref{eqn:met}), these six constraints make up a total of 8 kinematic constraints 
which can be solved by standard means \cite{Sonnenschein:2005ed,Sonnenschein:2006ud,Betchart:2013nba} 
in order to yield test values for the unknown momenta $\qC{i}$ 
corresponding to the given choice of test masses $(\tmA, \tmB, \tmC)$.

At this point it might seem that we have not made much progress, since we have traded one set of unknowns for another.
The key idea now is to focus not on the actual solutions for the invisible momenta, 
but on the existence (or lack thereof) of viable solutions in the first place \cite{Kawagoe:2004rz,Cheng:2007xv,Cheng:2008hk}.
Indeed, not all choices of test masses will lead to real solutions for $\qC{i}$.
In general, the kinematic constraints (\ref{eqn:met}) and (\ref{eqn:on-shell}) allow four complex solutions. 
Non-real solutions come in complex conjugate pairs \cite{Gripaios:2011jm}, and  
therefore, there can be either zero, two or four real (possibly degenerate) solutions for a given test mass. 
As a result, each event separates the three-dimensional mass parameter space
$(\tmA, \tmB, \tmC)$ into \emph{allowed} and \emph{disallowed} regions, 
depending on whether there exists a real solution for $\qC{i}$ or not. 
The event will be said to be \emph{solvable} by the test mass if there exists a real solution for $\qC{i}$. 
Note that for signal events the true mass point will always solve the event\footnote{Barring the effects from the detector resolution and finite particle widths.}, 
and thus can never be ruled out. 

\begin{figure}
 \centering
 \includegraphics[width=.45\textwidth]{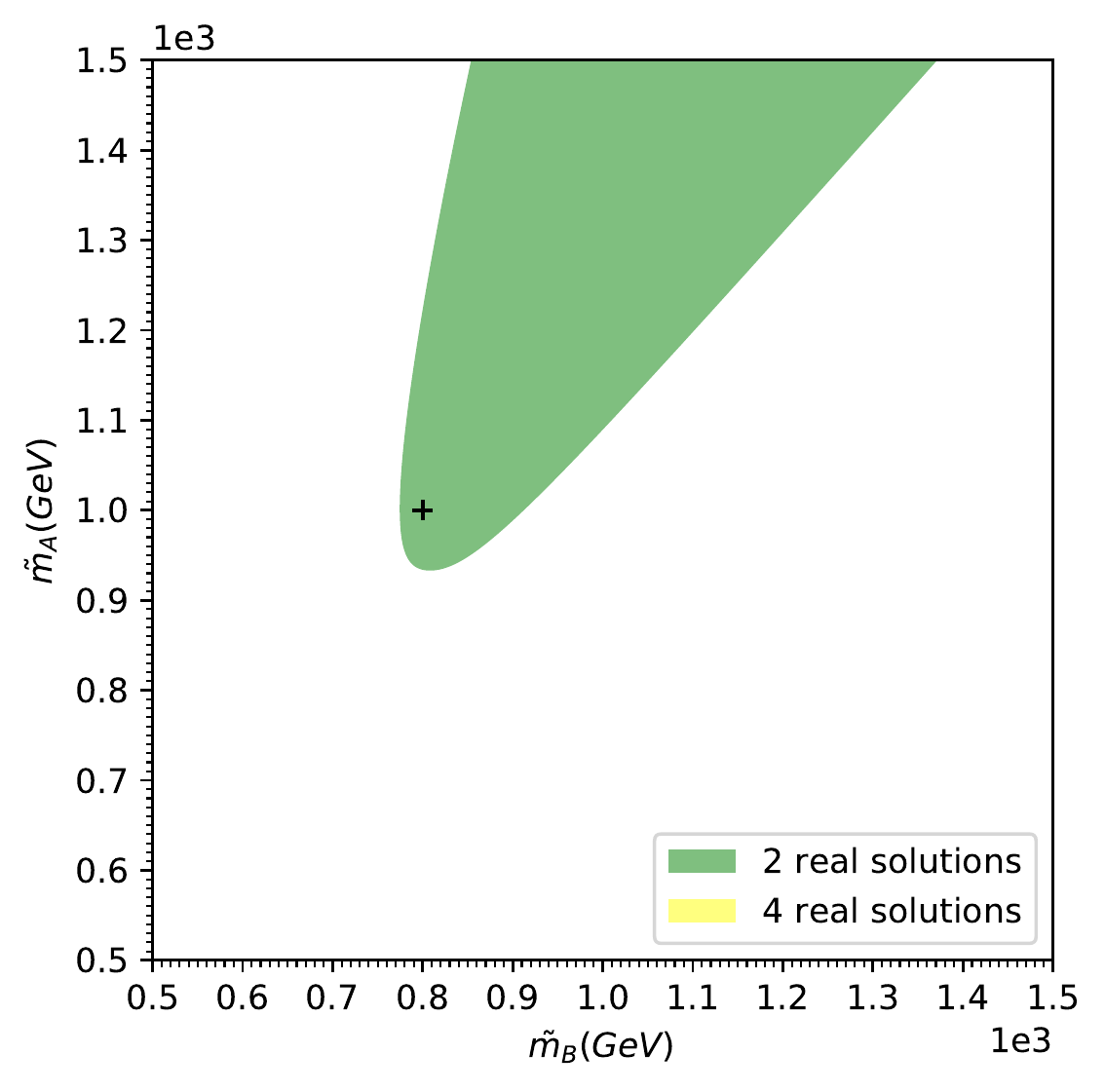}
 \hskip 5mm
 \includegraphics[width=.45\textwidth]{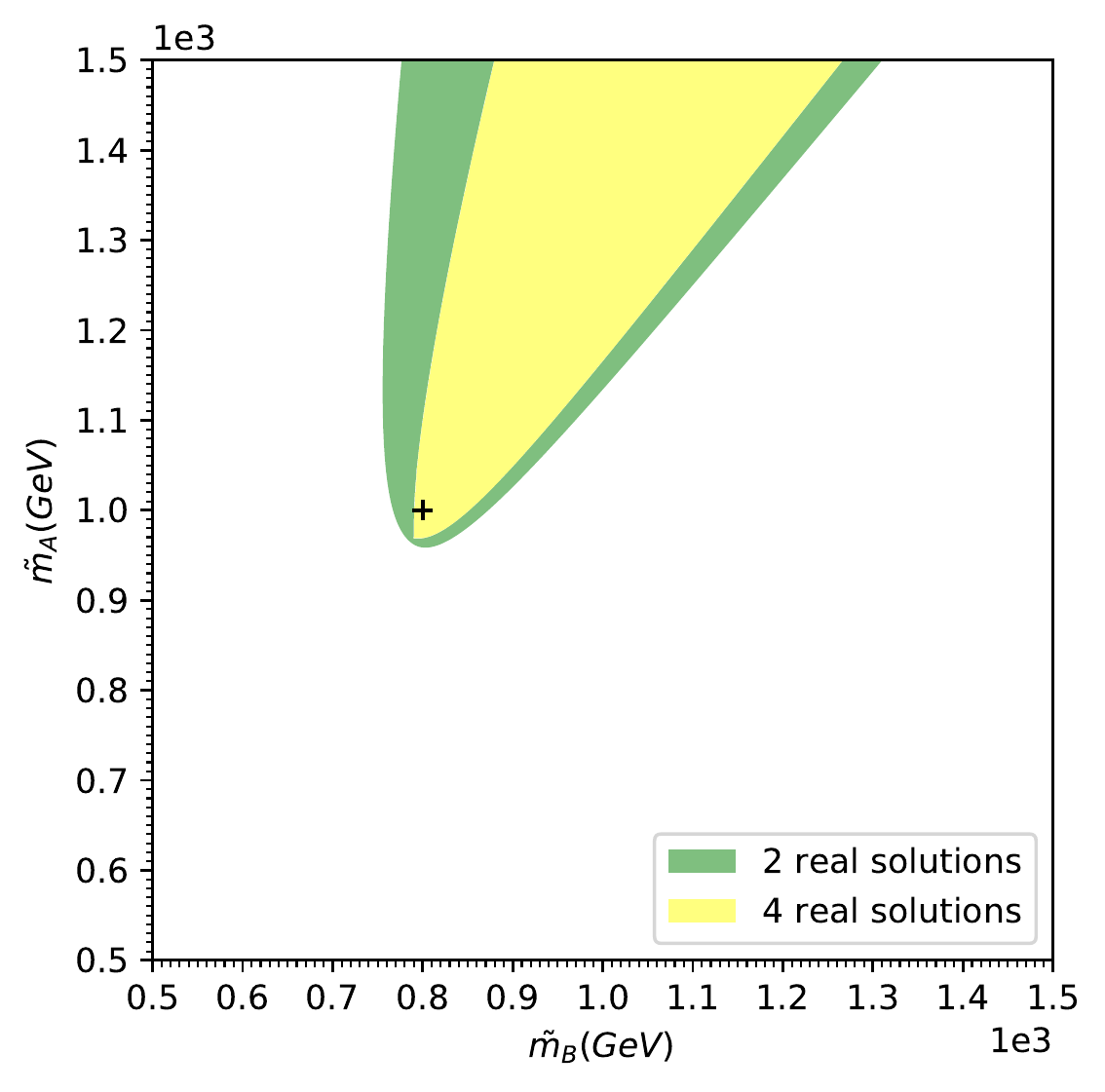}
 \caption{\label{fig:700tongues}Solvability plots for two different events, 
 showing the number of solutions allowed by different values of $(\tmB, \tmA)$. 
 The test mass $\tmC$ has been fixed to be the true mass, $\tmC=\mC=700\GeV$. 
 Uncolored areas in the plots correspond to zero real solutions, while green (yellow) regions correspond to two (four) solutions. 
 The cross marks the true mass point $(\mB,\mA)$. 
 For the event on the left, the plot only shows an allowed region with 2 real solutions 
 (the allowed region with four solutions is outside the plot range). 
 For the event on the right, both a region with 2 solutions and a region with 4 solutions can be seen.}
\end{figure}

\begin{figure}[t]
 \centering
 \includegraphics[width=.45\textwidth]{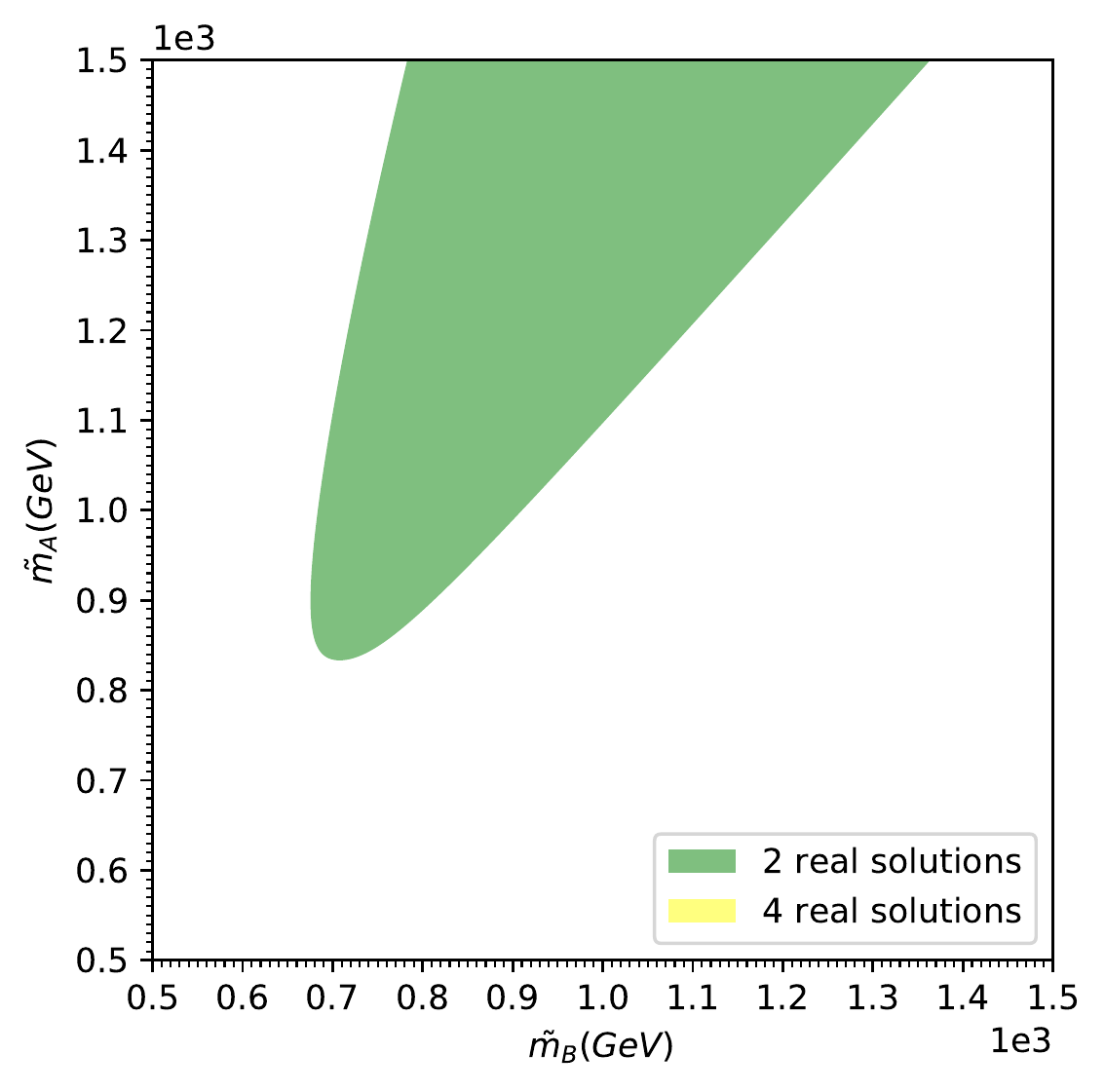}
 \hskip 5mm
 \includegraphics[width=.45\textwidth]{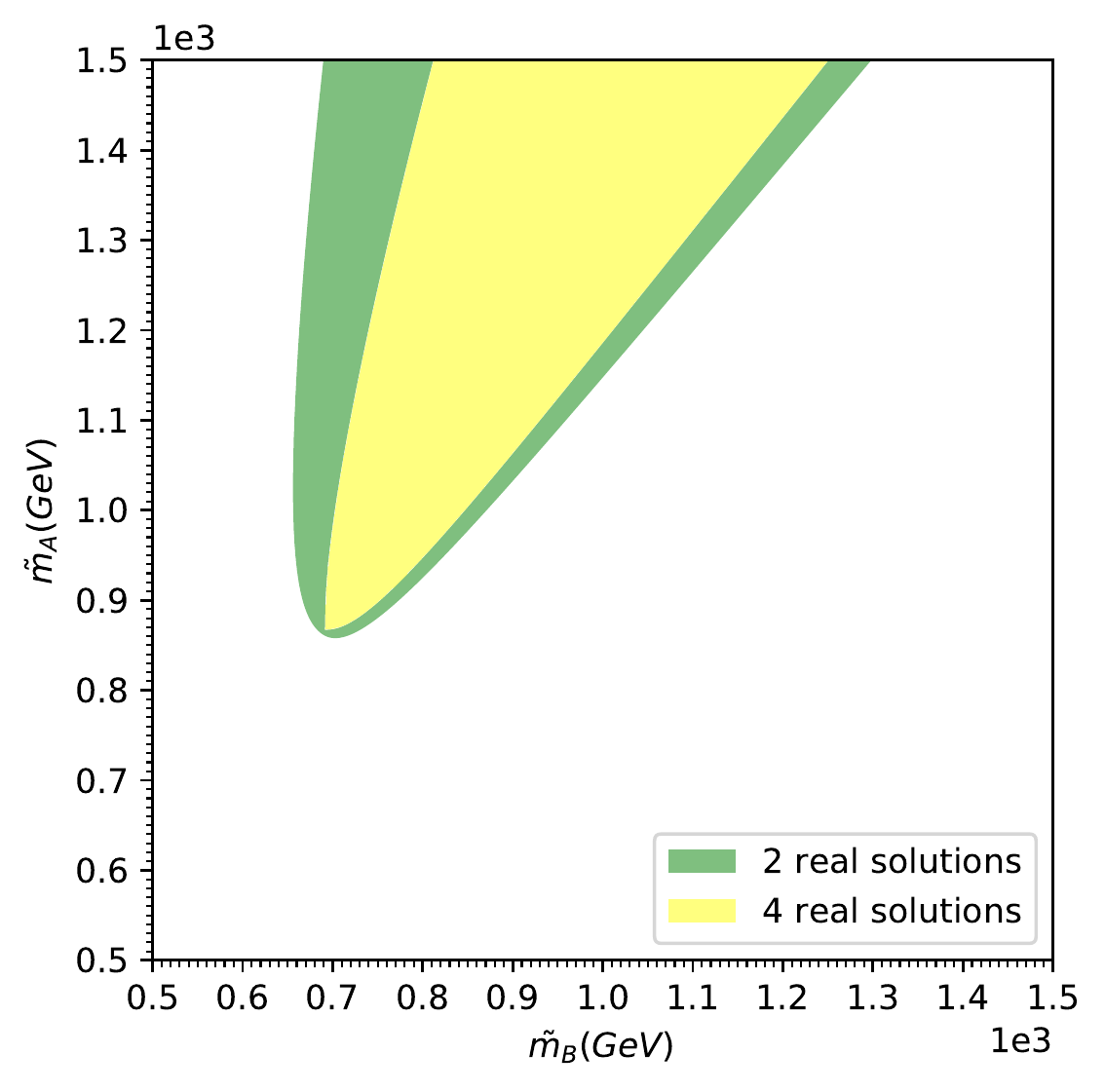}
 \caption{\label{fig:600tongues} The same as Figure~\ref{fig:700tongues}, except the trial mass $\tmC$ is now chosen to be smaller than the true mass:
 $\tmC=600\GeV<\mC$. Since $\tmC$ is different from $\mC$, the true mass point is not seen in these plots.}
\end{figure}

\begin{figure}[t]
 \centering
 \includegraphics[width=.45\textwidth]{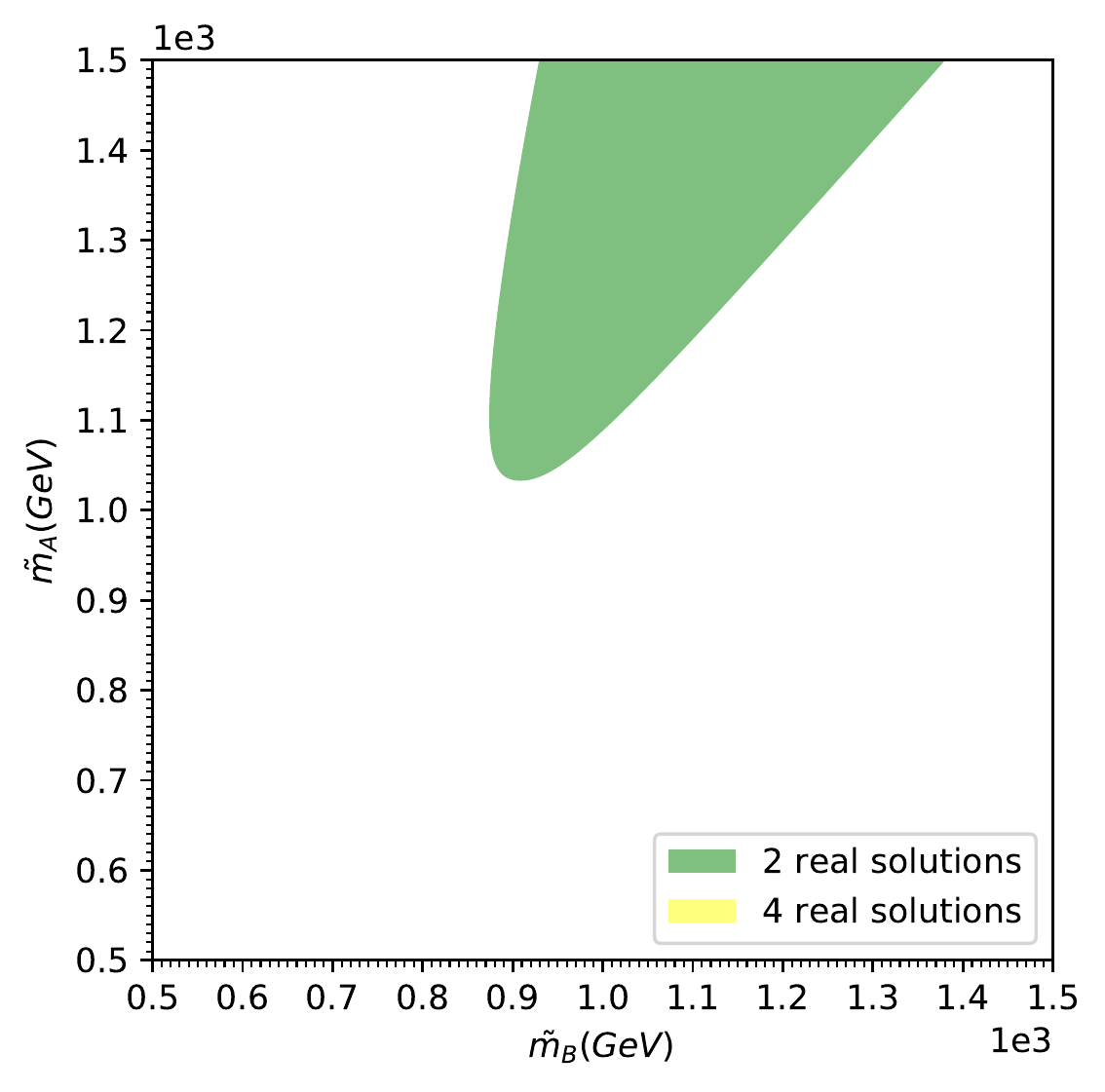}
 \hskip 5mm
 \includegraphics[width=.45\textwidth]{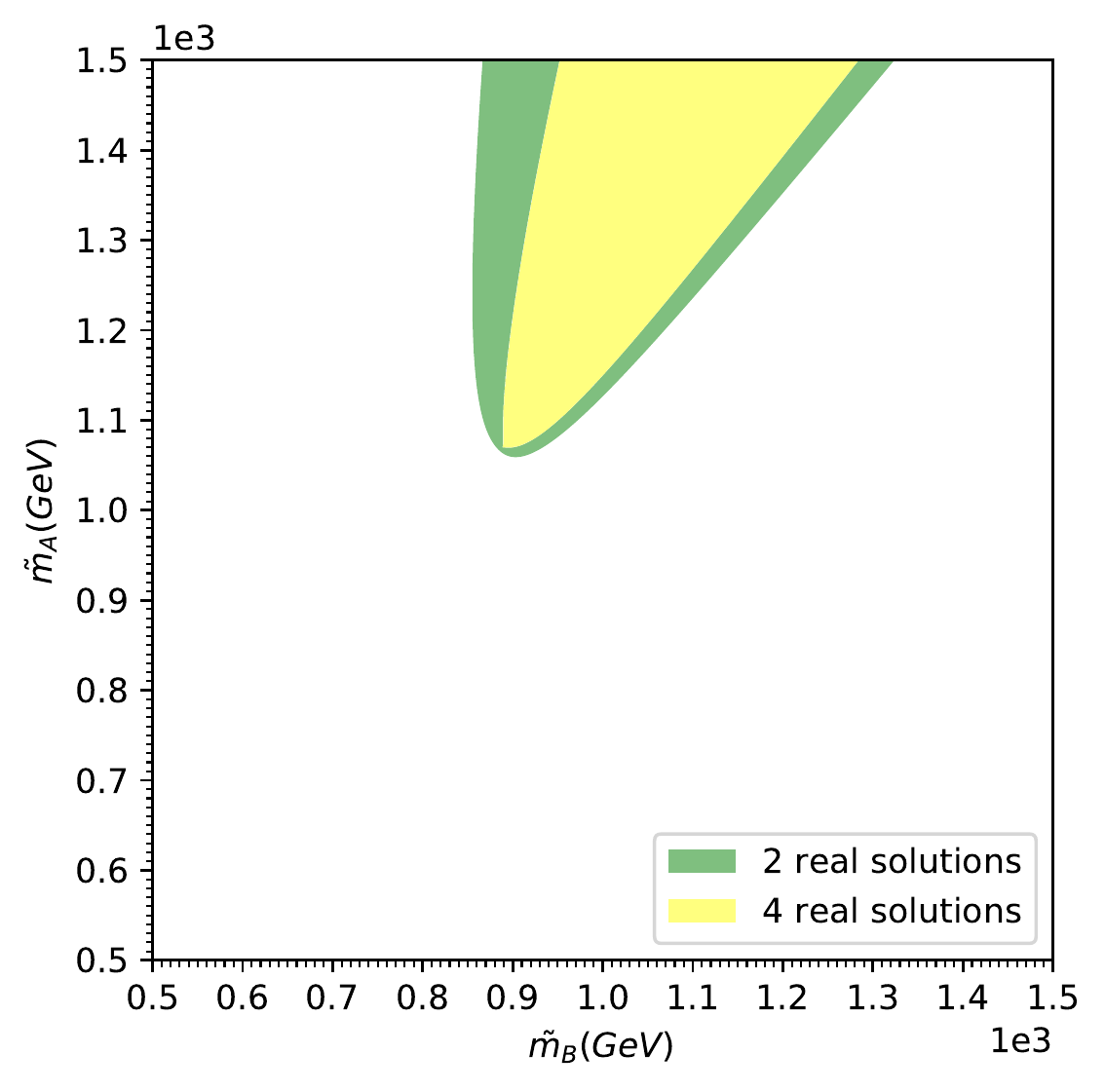}
 \caption{\label{fig:800tongues} The same as Figure~\ref{fig:600tongues}, but for a choice of test mass $\tmC$ 
larger than the true mass: $\tmC=800\GeV>\mC$. }
\end{figure}

Figs.~\ref{fig:700tongues}-\ref{fig:800tongues} illustrate the above discussion for two representative signal events.
In the figures, we show ```solvability plots" constructed as follows. Given that it is difficult to visualize the full
three-dimensional mass parameter space, we choose to present slices at fixed values of $\tmC$: in Fig.~\ref{fig:700tongues}
$\tmC$ is set to the true value of $\mC=700\GeV$, while in Fig.~\ref{fig:600tongues} we take
$\tmC=600\GeV$ and in Fig.~\ref{fig:800tongues} we have $\tmC=800\GeV$.
In each panel, we color-code the $(\tmB, \tmA)$ plane according to the 
number of real solutions to the kinematic constraints (\ref{eqn:met}) and (\ref{eqn:on-shell}):
white (uncolored) areas correspond to no real solutions, green areas have two real solutions and 
yellow areas allow four real solutions. The cross in Fig.~\ref{fig:700tongues} marks the true mass point 
$\mB=800$ GeV, $\mA=1000$ GeV (the cross is absent in Figs.~\ref{fig:600tongues} and \ref{fig:800tongues}
since the test mass $\tmC$ in those figures is different from the true value $\mC$). 

There are several important lessons which can be learned from these figures. First, it is rather remarkable that the large majority 
of the plotted areas are actually ruled out by this single event --- notice how the white color is the dominant one on the plots.
We also observe that as the third (fixed) test mass $\tmC$ is decreased (increased), the allowed region in the
$(\tmB, \tmA)$ plane correspondingly shifts towards lower (higher) values of $\tmB$ and $\tmA$. 
In turn, the allowed region itself has two sub-regions, with 2 and 4 real solutions, correspondingly. The 
(yellow-colored) region with 4 real solutions sometimes appears at relatively large masses, and 
this just happens to be the case with the event used for the left panels in Figs.~\ref{fig:700tongues}-\ref{fig:800tongues}.
Finally, it is worth noting that for both of these randomly chosen events, 
the true mass point marked with the cross in Fig.~\ref{fig:700tongues} 
is located rather close to the boundary of the allowed region. 
At this point, this may seem to be just a coincidence, but a more extensive search throughout the signal event sample
reveals that this property is actually quite generic. This observation then brings up the question whether
it is possible for the true mass point to be located {\em exactly} on the boundary of the allowed region.
This is one of the central issues in this paper, which will be addressed below in Section~\ref{sec:extremeevents}.
But first we shall present and discuss the mass measurement method based on solvability.

\section{Solvability as a Mass Measurement Method}
\label{sec:fd}

\subsection{Superposition of individual events}
\label{sec:superposition}

In the previous section, we saw that by requiring solvability, a single event can already rule out a sizable chunk of the 
three-dimensional mass parameter space $(\tmA, \tmB, \tmC)$. Repeating the same analysis with a different event,
we would expect a slightly different portion of the mass space to be disfavored. Then, by taking the intersection of the
two regions allowed by the individual events, we will further constrain the mass 
parameter space \cite{Kawagoe:2004rz,Cheng:2007xv,Cheng:2008hk,Lester:2013aaa}.

\begin{figure}[t]
 \centering
 \includegraphics[width=.45\textwidth]{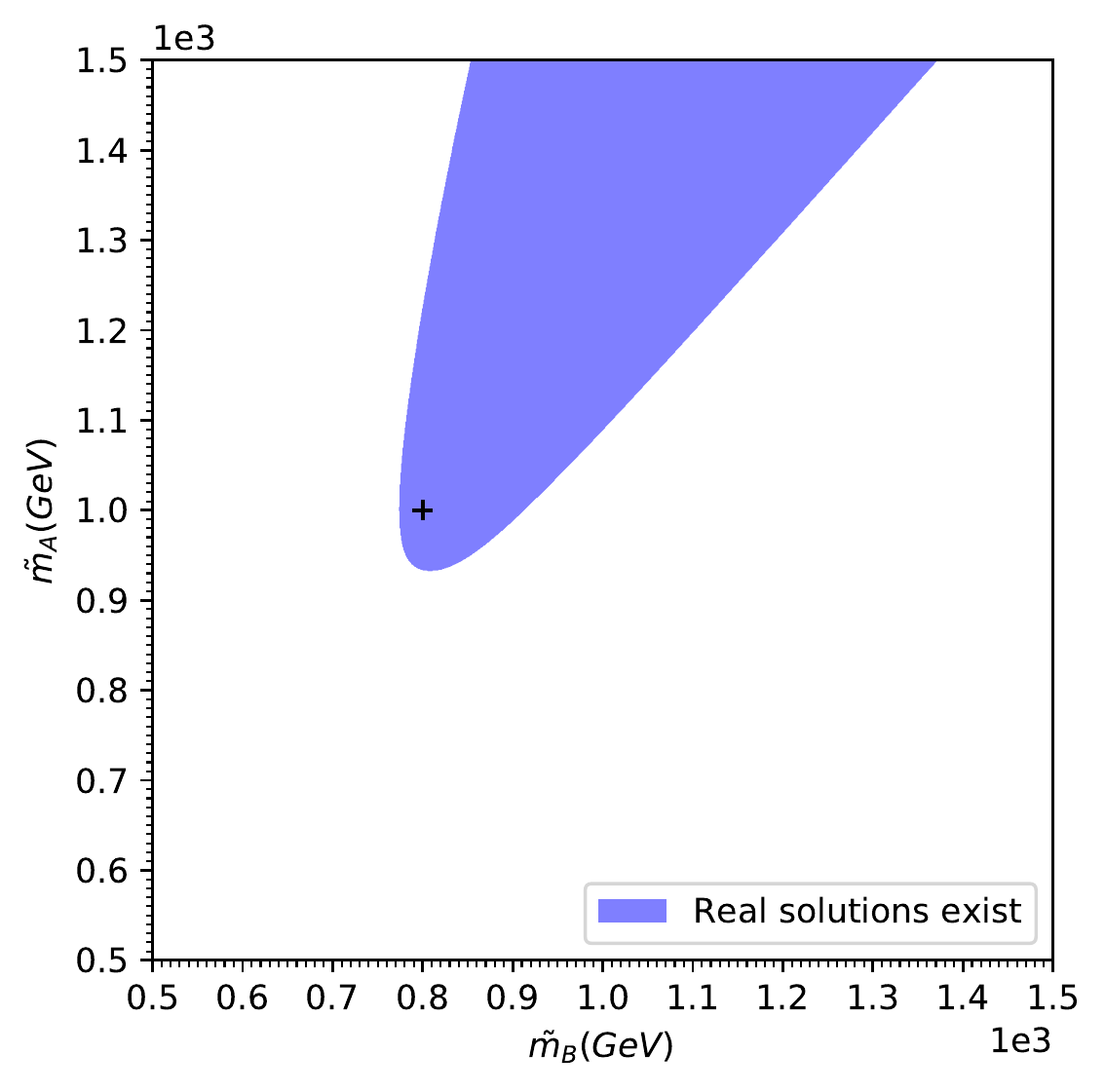}
 \hskip 5mm
 \includegraphics[width=.45\textwidth]{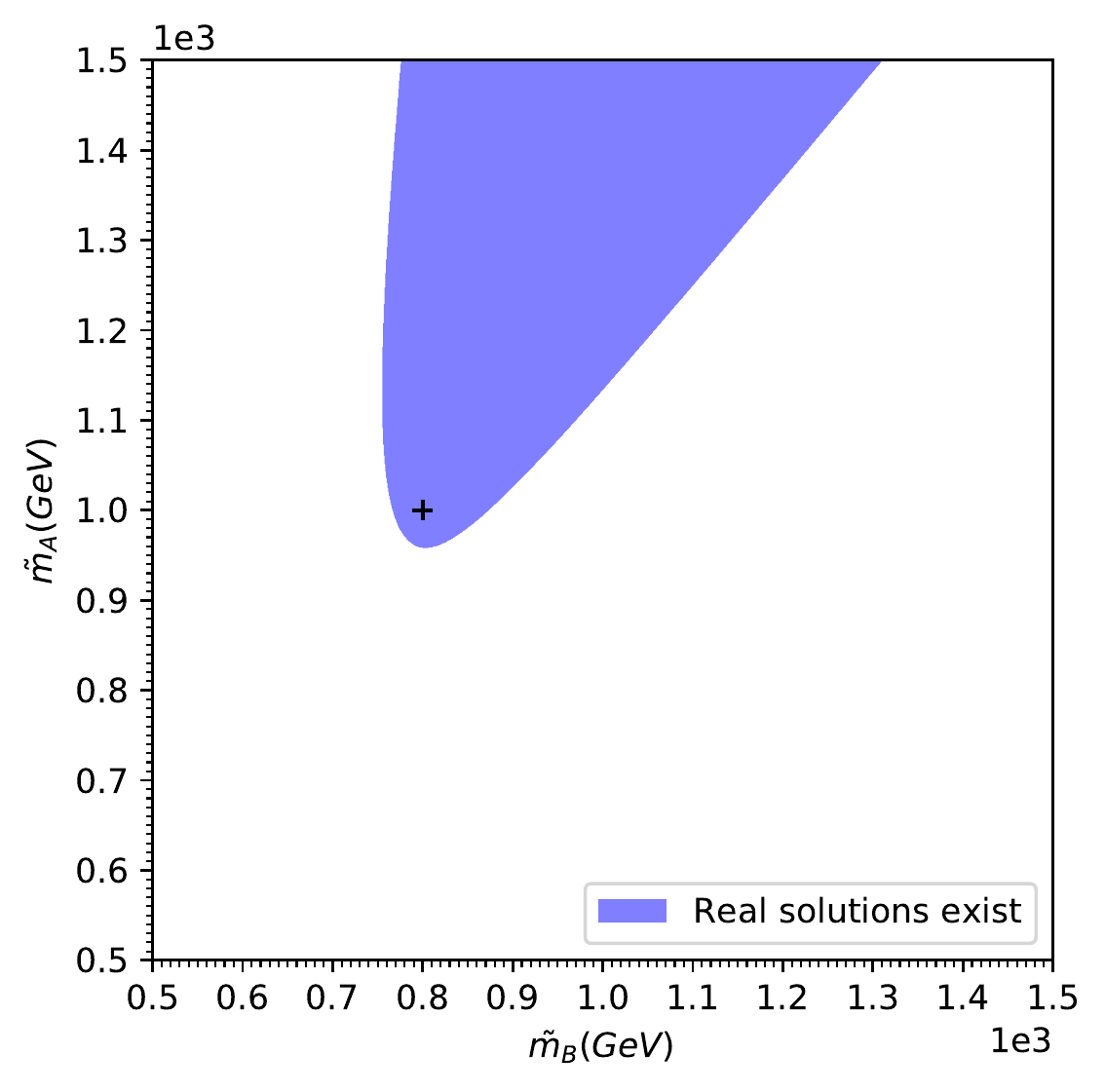}\\
 \includegraphics[width=.45\textwidth]{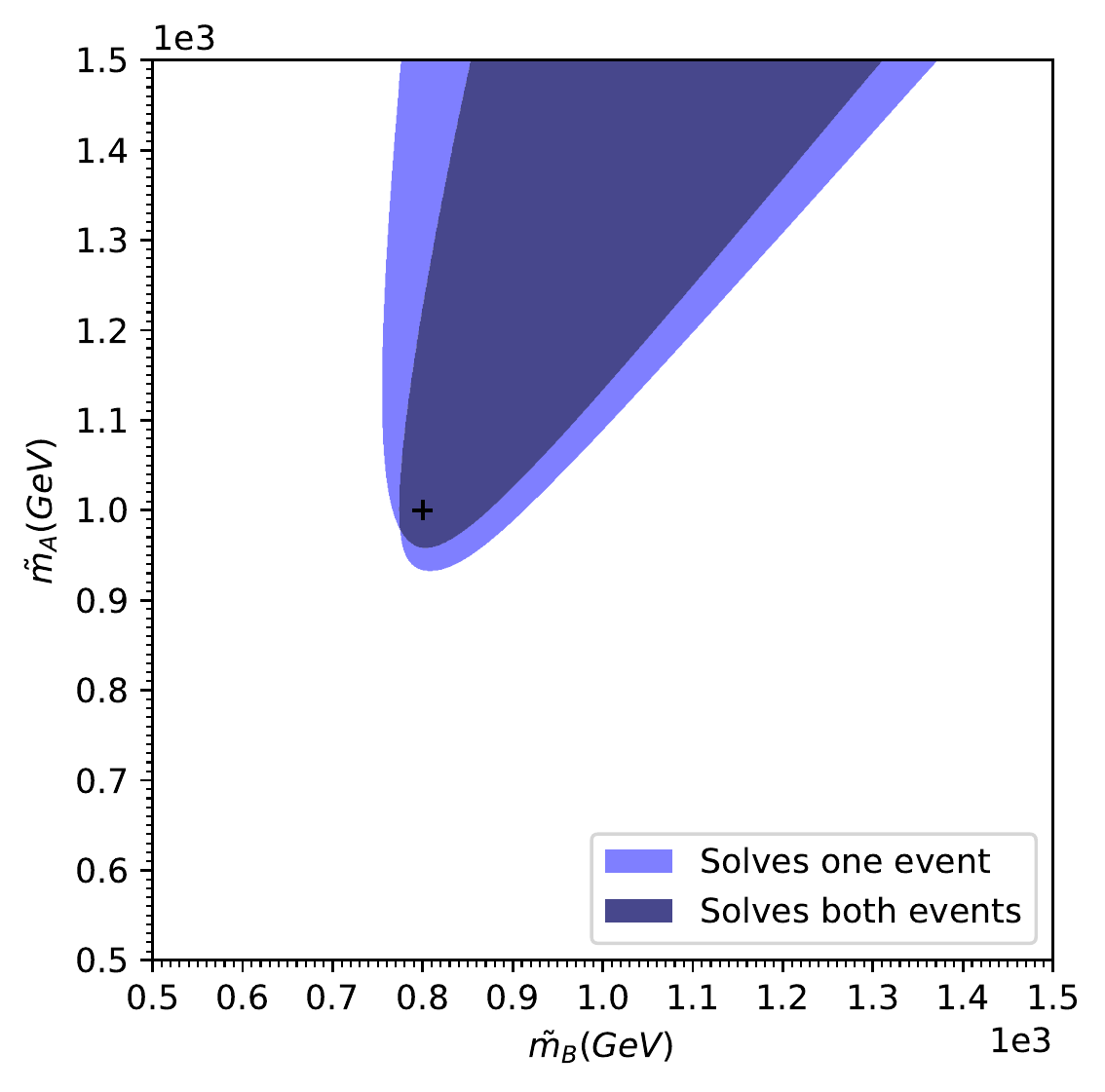}
 \caption{\label{fig:superimposedtongues}
 Top row: the same as Fig.~\ref{fig:700tongues} ($\tmC=\mC=700\GeV$), only now the allowed region is uniformly shaded (in light blue), 
 regardless of the number of real solutions. Bottom row: superposition of the allowed regions from the two events.
 White areas are incompatible with both events, light-shaded areas solve one event but not the other, while
 the dark-shaded area solves both events.}
\end{figure}

This procedure is
pictorially illustrated in Fig.~\ref{fig:superimposedtongues}, where we use the same two events 
as in Figs.~\ref{fig:700tongues}-\ref{fig:800tongues} above.
In the top row, we replot the result from Fig.~\ref{fig:700tongues} for $\tmC=\mC=700\GeV$, 
only now we remove the distinction between the green and yellow regions (with 2 and 4 real solutions, respectively)
and uniformly shade the allowed region in light blue. Then in the plot shown in the bottom row, we superimpose these two allowed regions, 
thus obtaining a new partition of the $(\tmB,\tmA)$ plane into three possible regions:
where neither event is solvable (white areas), where only one of the events is solvable but not the other (light blue areas) and 
where both events are solvable (dark blue areas). By construction, the dark blue region where both events are solvable is 
smaller than each of the individual allowed regions seen in the plots in the top row.
This demonstrates the benefit from adding more events to the discussion, i.e., increasing the statistics.
At the same time, a pessimist might point out that a) the benefit does not seem to be that great, since the improvement 
(the light blue areas in the bottom panel of Fig.~\ref{fig:superimposedtongues}) is relatively minor when 
compared to the overall size of the remaining allowed region (the dark blue area); and b) that 
the remaining dark blue allowed region still seems to extend out to infinity in the ``northeast" direction.
A crucial question, therefore, becomes how much further the allowed region will shrink once we add 
all of our remaining events in the data sample. We will answer this question in the next two subsections
\ref{subsec:mbma} and \ref{subsec:mc}, where for clarity of the presentation we split the discussion into two parts:
in Sec.~\ref{subsec:mbma} we focus on the issue of measuring $\mA$ and $\mB$ for a given
value of $\tmC$ (which throughout that subsection will be fixed to be the true mass $\mC$)
and then in Sec.~\ref{subsec:mc} we shall tackle the question of measuring $\mC$ itself.
The reason for this separation is twofold: first, it is difficult to present and visualize our results in the full
three-dimensional mass parameter space $(\tmA, \tmB, \tmC)$, and second, as we shall see below, 
the three-dimensional mass parameter space exhibits a direction of relatively low sensitivity,
which can be parametrized by the value of $\tmC$ (see Appendix~\ref{app:flatdirection} for details). As a result, the
measurement of $\mC$ will turn out to be much more challenging than the measurements
of $\mA$ and $\mB$. 

\begin{figure}
 \centering
  \includegraphics[width=.45\textwidth]{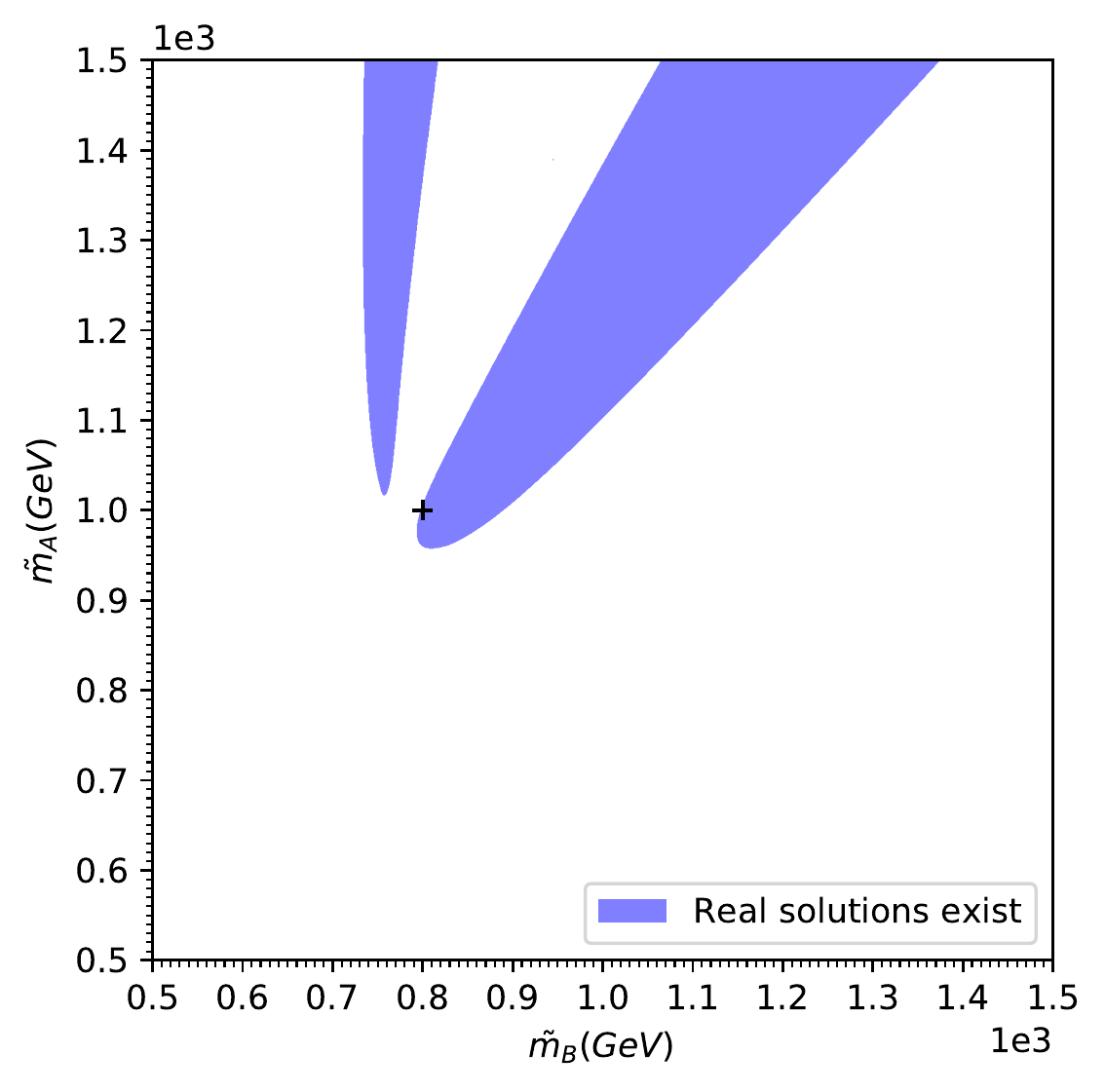}
 \hskip 5mm
 \includegraphics[width=.45\textwidth]{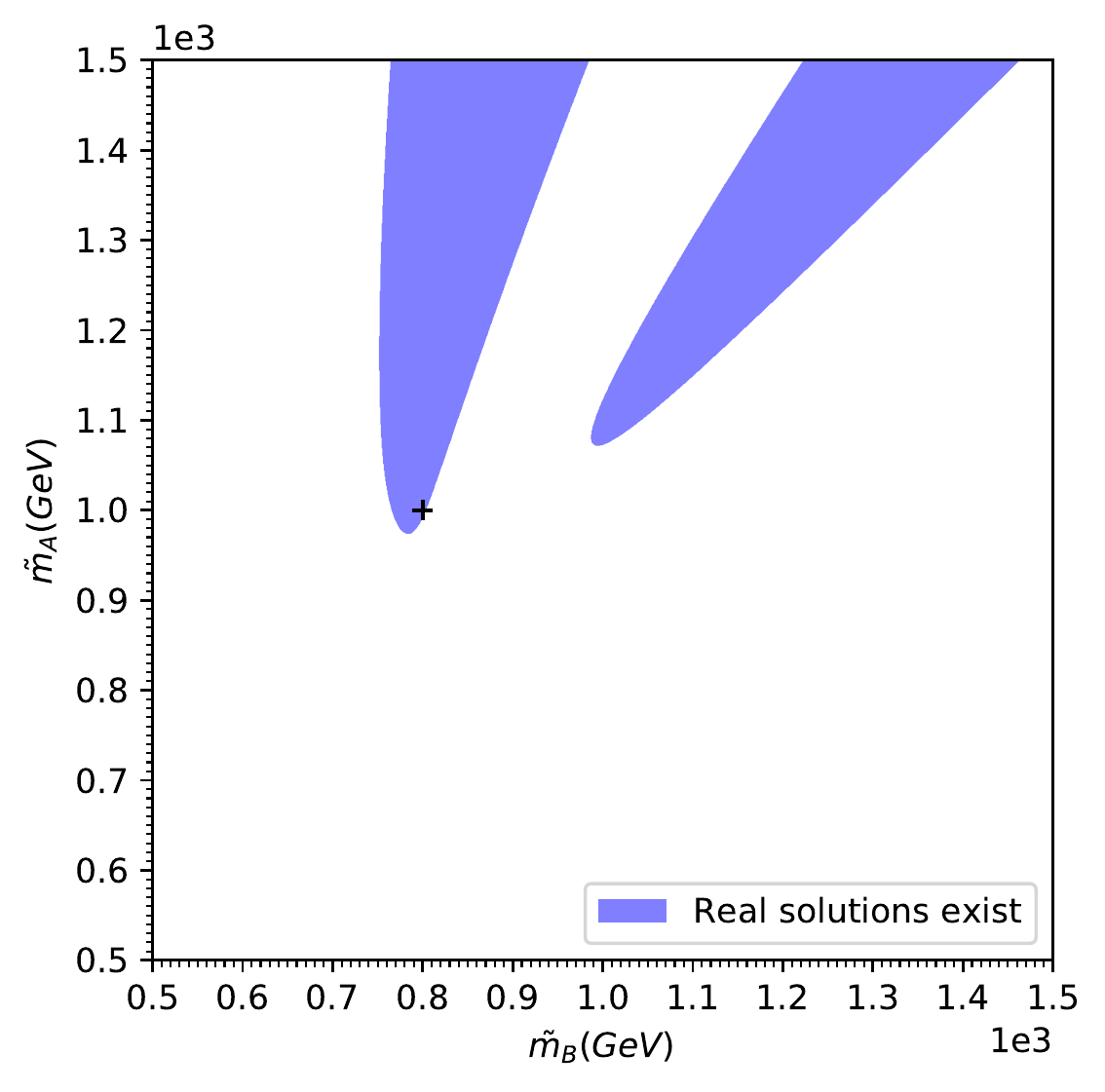}\\
  \includegraphics[width=.45\textwidth]{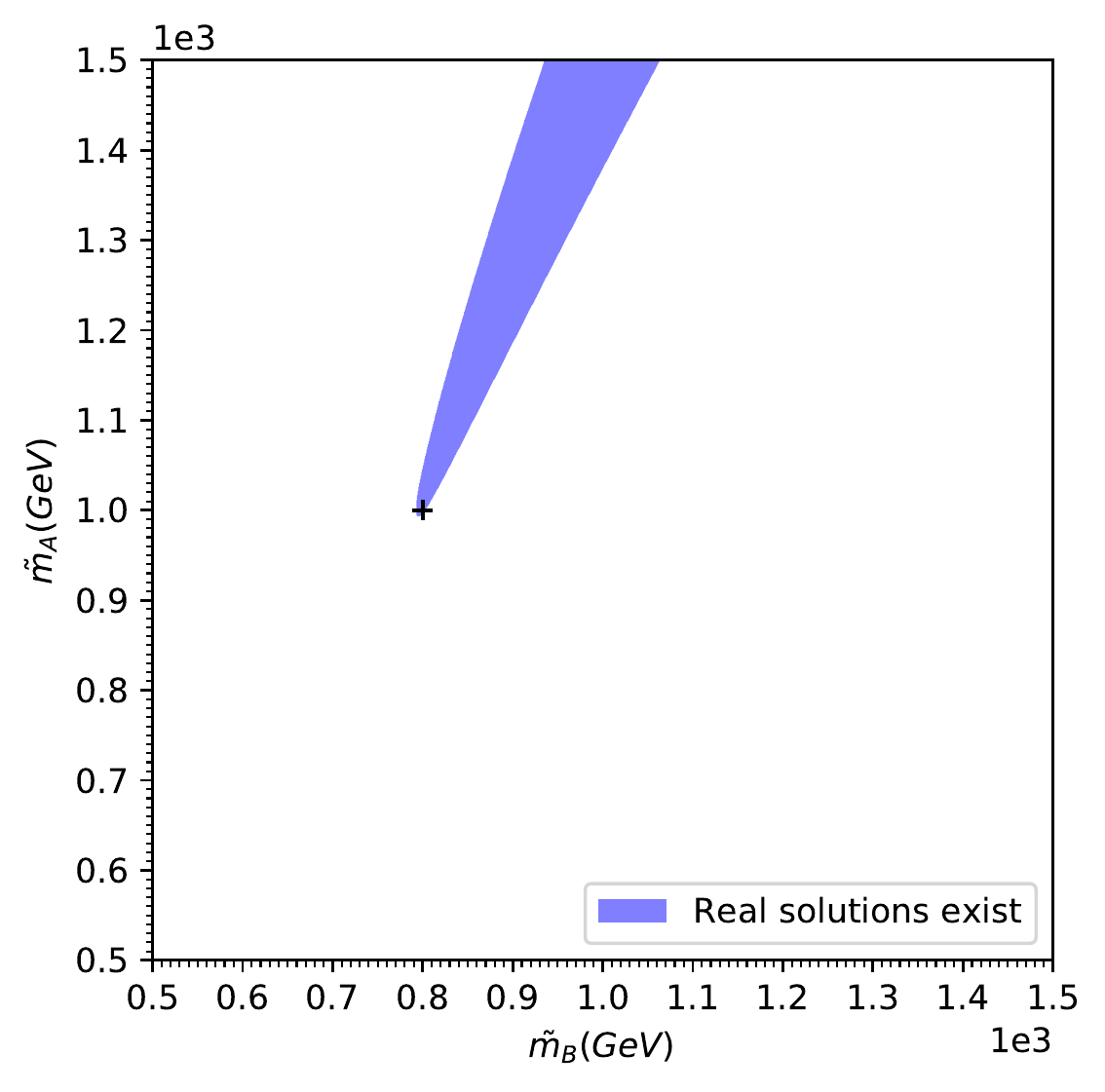}
 \hskip 5mm 
 \includegraphics[width=.45\textwidth]{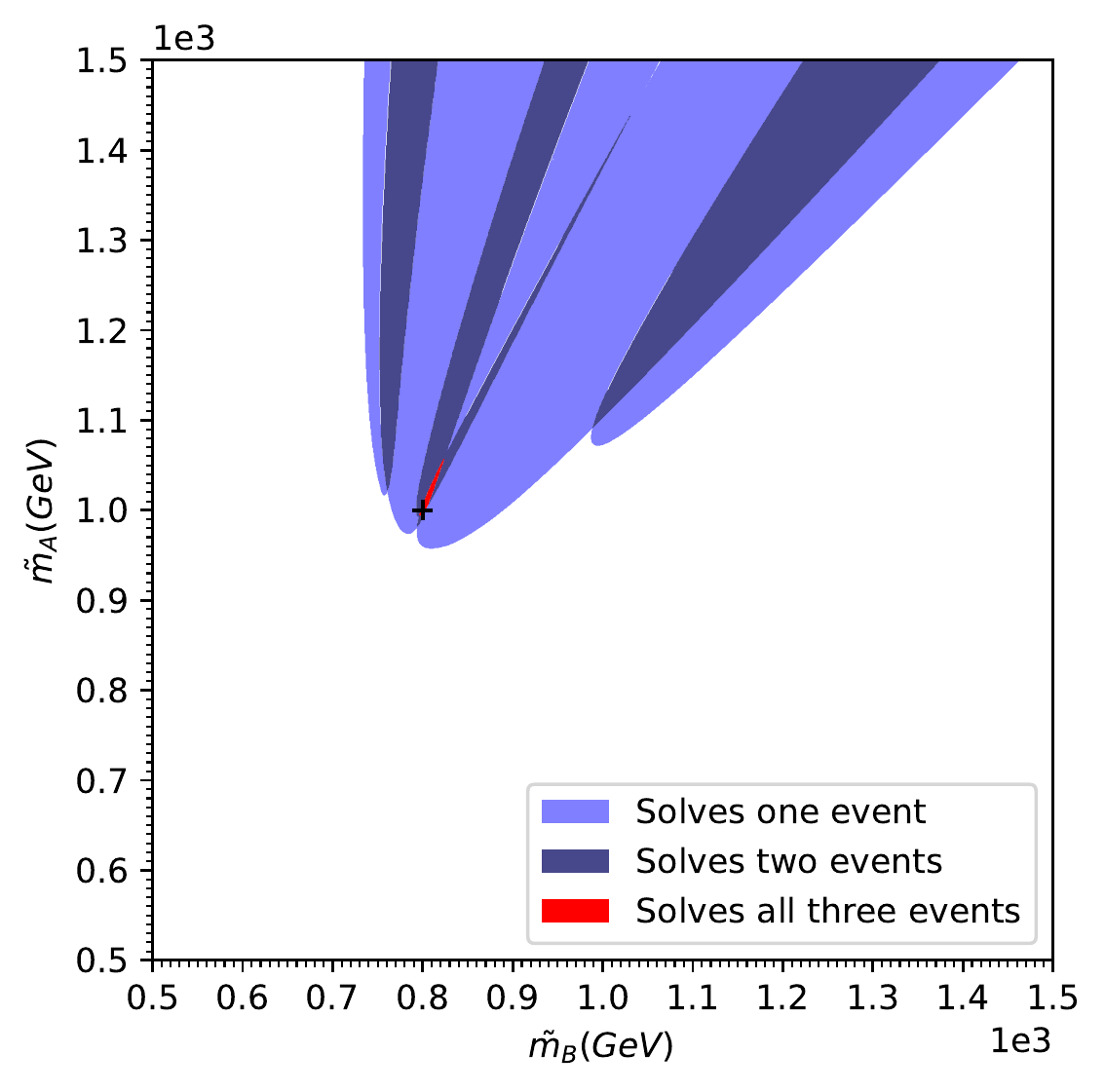}
 \caption{\label{fig:extreme3or2} The same as Fig.~\ref{fig:superimposedtongues}, 
 only now we superimpose the allowed regions for three suitably chosen events 
 (shown in the top left, top right and bottom left panels). The cross marks the true mass point $(\mB,\mA)$.
 The three events were selected to be complementary to each other in the sense that when taken
 together they would rule out a large portion of the parameter space.
 In the bottom right panel white areas are incompatible with all three events, 
 blue light-shaded areas solve one event but not the other two, blue dark-shaded areas solve two events, but not the third, 
 while the narrow red-shaded sliver extending away from the cross solves all three events.}
\end{figure}

However, before proceeding to show results with the full event sample, it is useful to 
consider one more simple exercise, illustrated in Fig.~\ref{fig:extreme3or2}, where
we repeat the superposition of allowed regions done in Fig.~\ref{fig:superimposedtongues}, still
with just a handful of events, in this case three. The important new twist here is that 
the three events in Fig.~\ref{fig:extreme3or2} were not chosen at random, 
as was done in Fig.~\ref{fig:superimposedtongues}, but were more carefully selected. 
The idea was to pick events which are maximally incompatible with each other, 
and would therefore rule out the largest amount of mass space by themselves.
The top left, top right and bottom left panels in Fig.~\ref{fig:extreme3or2} 
show the allowed regions for the three selected events.
There are three notable features of these individual allowed regions.
\begin{itemize}
\item For all three events, the cross marking the true mass point $(\mB,\mA)$ lies on the boundary of the allowed region.
This answers the question posed towards the end of the previous section, demonstrating that 
a kinematic boundary may pass through the true mass point. 
We shall have a lot more to say about that in the next Section~\ref{sec:extremeevents}.
\item The shapes of the three allowed regions are very different. For the events in the top row, we see 
the appearance of two apparently disjoint branches --- however, those are actually connected to each other
at large values of $\tmB$ and $\tmA$ beyond the plot range.
\item More importantly, the locations of the individual allowed regions happen to be different, so that when they
are superimposed in the bottom right panel, there is only a very narrow sliver of allowed mass space left (the red-shaded area).
Note the different location of the cross (the true mass point) for the three events:
in the top left (top right) panel the cross is on the upper (lower) boundary of the allowed region, so that when the 
two events are superimposed, they will leave only the ``southwest-to-northeast" direction as viable.
On the other hand, the cross in the lower left panel is right at the tip of the allowed region,
thus eliminating the ``southwest" portion, and leaving only the red-shaded area extending to the northeast of the true mass point.
\end{itemize}

The lesson from Figs.~\ref{fig:superimposedtongues} and~\ref{fig:extreme3or2} is that some events are better 
at ruling out mass parameter space than others. Unfortunately, since {\em a priori} we do not know the values of the
true masses, i.e., the location of the cross in Figs.~\ref{fig:superimposedtongues} and~\ref{fig:extreme3or2},
we are unable to preselect events which are ``good at" eliminating parameter space, and the best we can do is go over the whole sample,
thus guaranteeing ourselves that at some point we will eventually hit on some ``good" events as well.
This is precisely what we intend to do in the next two subsections.

\subsection{Measuring $m_B$ and $m_A$ for a given trial mass $\tilde m_C$}
\label{subsec:mbma}

In this subsection we focus on a slice of the three-dimensional mass parameter space $(\tmA, \tmB, \tmC)$ at a constant $\tmC$,
i.e., we shall attempt to measure the values of $\mB$ and $\mA$ given a value for $\tmC$, which we shall take to be the true mass, $\mC$. 
However, this choice is inconsequential, and the results will be similar for any other choice of $\tmC$
(in the next subsection we shall return to the question of measuring $\mC$ itself).

To this end, we follow the procedure illustrated in Figs.~\ref{fig:superimposedtongues} and~\ref{fig:extreme3or2},
only this time we use the full event sample. With realistic detector resolutions and finite widths for the resonances,
even the true mass point is not expected to solve 100\% of the events. Its success rate will be further reduced in
the presence of background events. So, instead of successively restricting the allowed mass space 
with more and more events, which would eventually result in an empty set, 
a better approach is to look at the fraction of events that are solvable by a given test mass hypothesis,
with the hope that the true test mass exhibits the highest such fraction.
\begin{figure}[t]
 \centering
 \includegraphics[width=.49\textwidth]{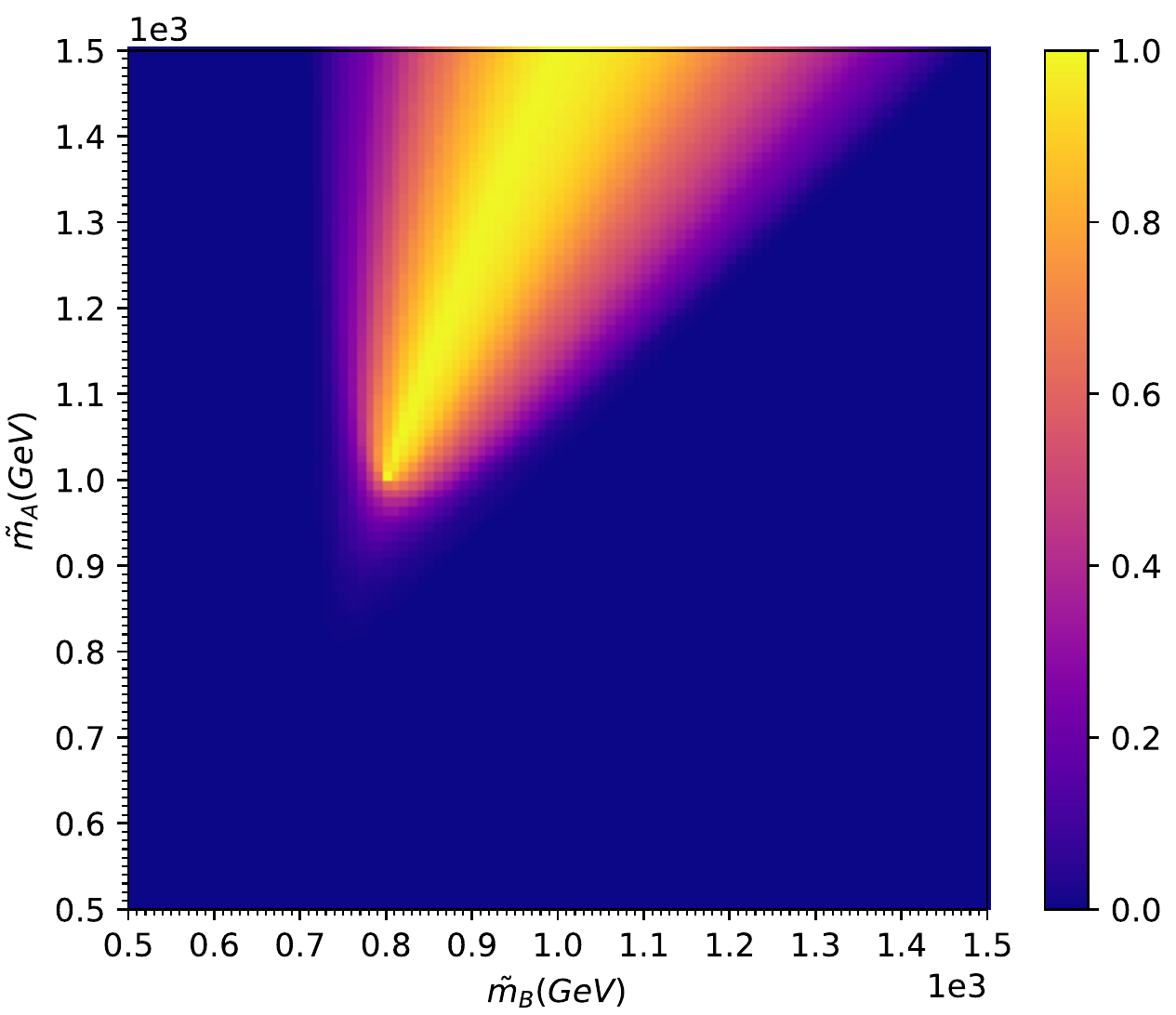}
 \includegraphics[width=.49\textwidth]{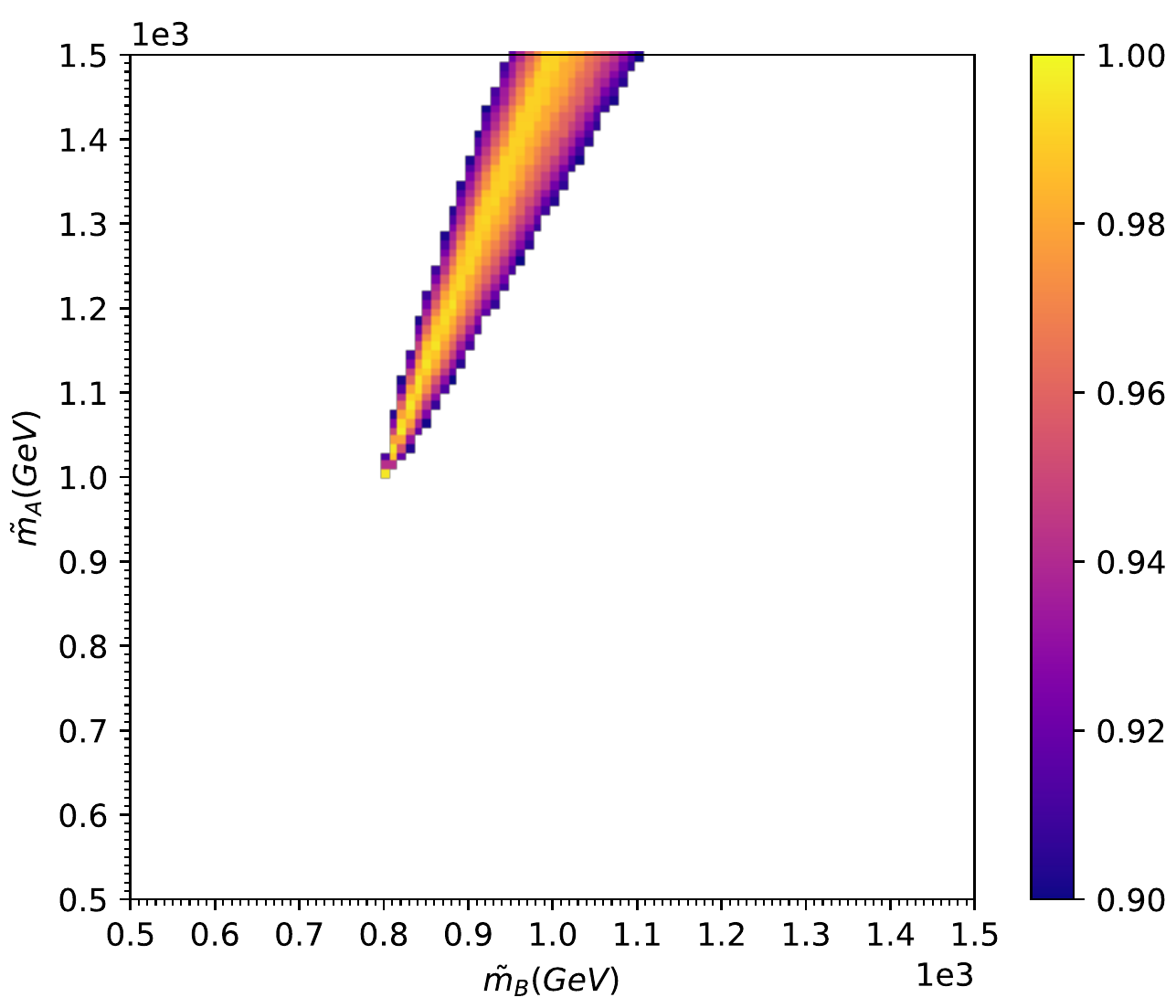}
 \caption{\label{fig:solvabilityhist} Solvability heat maps for fixed $\tmC=\mC=700\GeV$. 
 For each point in the $(\tmA, \tmB)$ plane we show the fraction of events which are solvable. 
 The plot on the right considers only mass points which solve over 90\% of the events (note the change of scale in the colorbar).}
\end{figure}
The result\footnote{From now on such plots will be referred to as ``solvability heat maps".} 
is plotted in Fig.~\ref{fig:solvabilityhist}, where we use only signal events and for now ignore the effects of detector resolution and finite particle widths.
The color represents the fraction of events which are solvable for the given mass hypothesis $(\tmA, \tmB)$.

Fig.~\ref{fig:solvabilityhist} confirms that, as expected, the true mass point $\mB=800\GeV$ and $\mA=1000\GeV$ 
has a 100\% solvability rate. However, in addition to the true mass point, there also seems to be a whole line of masses 
with nearly 100\% solvability. This ``solvability flat direction" is precisely the northeast-pointing red-shaded region previously encountered in Fig.~\ref{fig:extreme3or2}.
What we are now finding in Fig.~\ref{fig:solvabilityhist} is that this problematic region persists even after considering the full statistics in the sample.
This certainly presents a problem --- it suggests that in its current form the solvability method cannot uniquely determine the masses,
even in this simplified exercise where we only look at the two-dimensional $(\tmB,\tmA)$ plane.

One can think of several possible ways out. Perhaps the flat direction seen in Fig.~\ref{fig:solvabilityhist} 
is not exactly flat, but has a gentle slope, which would nevertheless pick out the true masses.
To test this, in the right panel of Fig.~\ref{fig:solvabilityhist} we change the color scale, zooming in on the events with maximal solvability (above 90\%).
We observe that the problematic direction is still pretty flat, and any existing gentle slope will be washed out
once we add the realistic effects of detector resolution and backgrounds.\footnote{The careful reader will note that
the flat direction appears to terminate at the true mass point, so that the true mass point is given not 
by the condition of maximal solvability, but by the sudden drop in the solvability rate along the flat direction.
This conjecture is correct \cite{Cheng:2008hk} and can be further motivated by ideas from Sections~\ref{sec:extremeevents} and \ref{sec:focus} below.}
We conclude that for any practical purposes the ``solvability flat direction" is there, and we need additional 
information in order to lift this degeneracy.

\begin{figure}
 \centering
 \includegraphics[width=.6\textwidth]{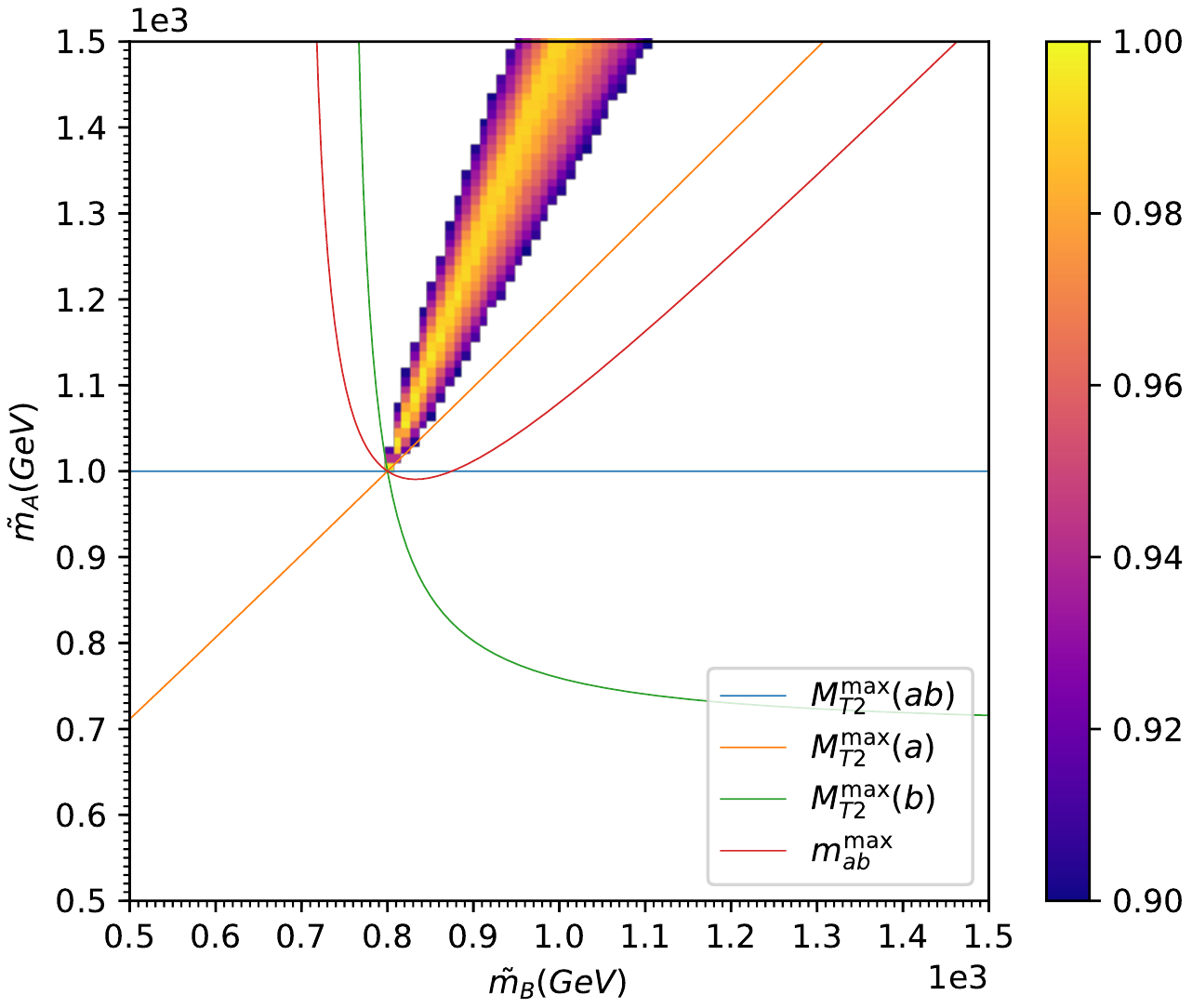}
 \caption[]{\label{fig:solvabilityandendpoints} Solvability heat map and constraints from measuring the kinematic endpoints (\ref{endpoints}):
 $M^{\mathrm{max}}_{T2}(ab)$ (blue), $M^{\mathrm{max}}_{T2}(a)$ (orange), $M^{\mathrm{max}}_{T2}(b)$ (green) and $m^{\mathrm{max}}_{ab}$ (red).}
\end{figure}

For example, we can augment the solvability method with constraints on the masses 
from kinematic endpoint measurements. The available measurements for this topology are \cite{Burns:2008va}:
\beq
\left\{M^{\mathrm{max}}_{T2}(ab),  M^{\mathrm{max}}_{T2}(a), M^{\mathrm{max}}_{T2}(b), m^{\mathrm{max}}_{ab}\right\}.
\label{endpoints}
\eeq
The first three are the upper kinematic endpoints for the three possible subsystem $M_{T2}$ variables in the event topology of Fig.~\ref{fig:feynmandiag},
where we have used the notation of Ref.~\cite{Cho:2014naa}. The last one is the upper kinematic endpoint in the distribution of 
the invariant mass of $\a{i}$ and $\b{i}$. The measurement of any one of these endpoints acts as an additional constraint on the masses.
In particular, Fig.~\ref{fig:solvabilityandendpoints} shows the effect of each individual kinematic endpoint measurement
on the $(\tmB,\tmA)$ parameter space from Fig.~\ref{fig:solvabilityhist}. Since $\tmC$ is still fixed at $\tmC=\mC=700\GeV$,
each kinematic endpoint measurement leads to a relation among $\tmB$ and $\tmA$ as given by the corresponding colored curve:
blue for $M^{\mathrm{max}}_{T2}(ab)$, orange for $M^{\mathrm{max}}_{T2}(a)$, green for $M^{\mathrm{max}}_{T2}(b)$, and red for $m^{\mathrm{max}}_{ab}$.
We see that all four curves intersect at the true mass point $(\mB,\mA)$, as they should.
More importantly, neither of the four curves is aligned with the solvability flat direction, and consequently,
any one of them can be used for lifting the degeneracy. We thus conclude that the solvability method is able to 
pinpoint the correct mass values in the $(\tmB,\tmA)$ plane (at fixed $\tmC=\mC$), 
but only when supplemented with a kinematic endpoint measurement.

\subsection{Measuring $m_C$}
\label{subsec:mc}

\begin{figure}
 \centering
 \includegraphics[width=.49\textwidth]{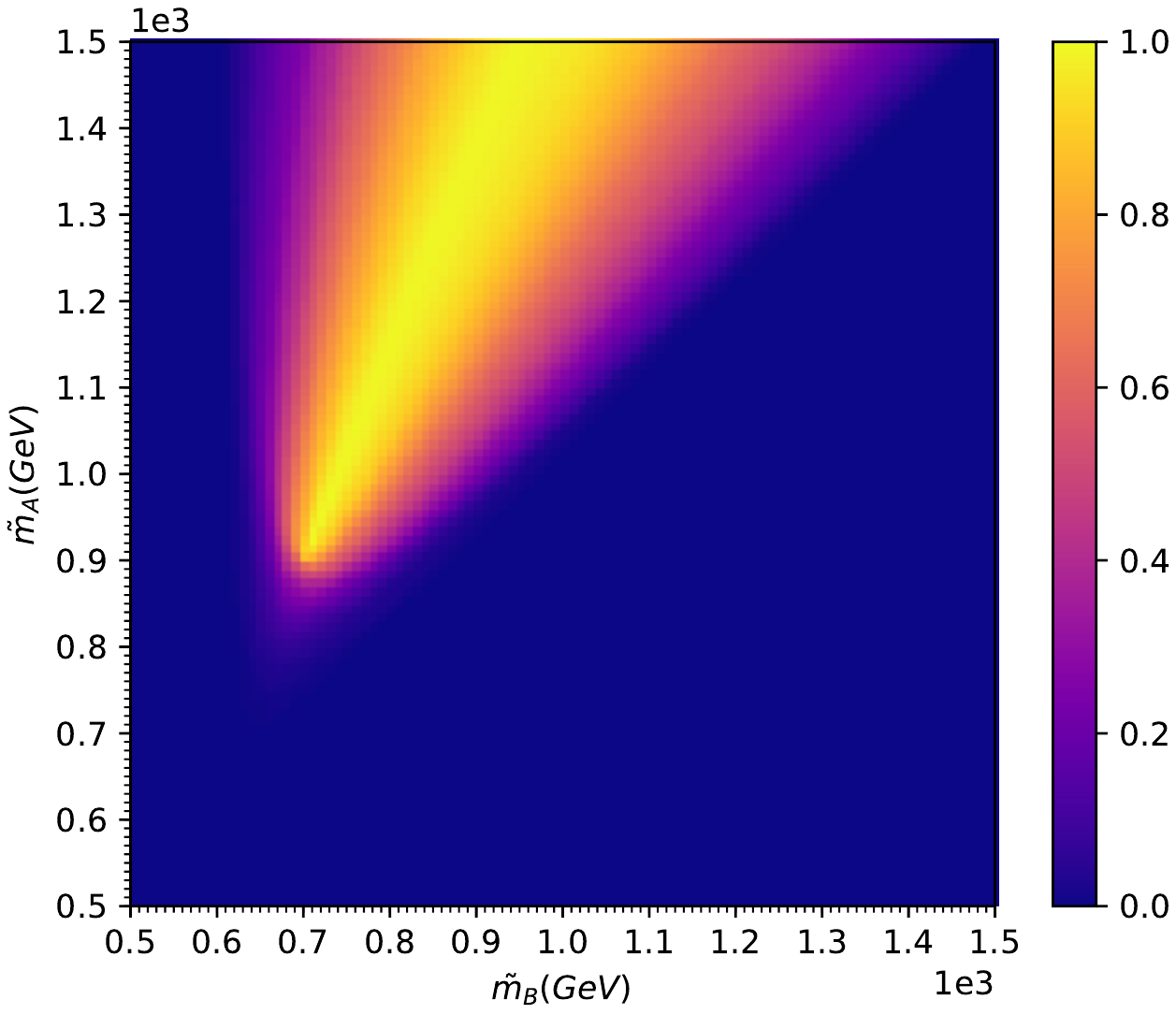}
 \includegraphics[width=.49\textwidth]{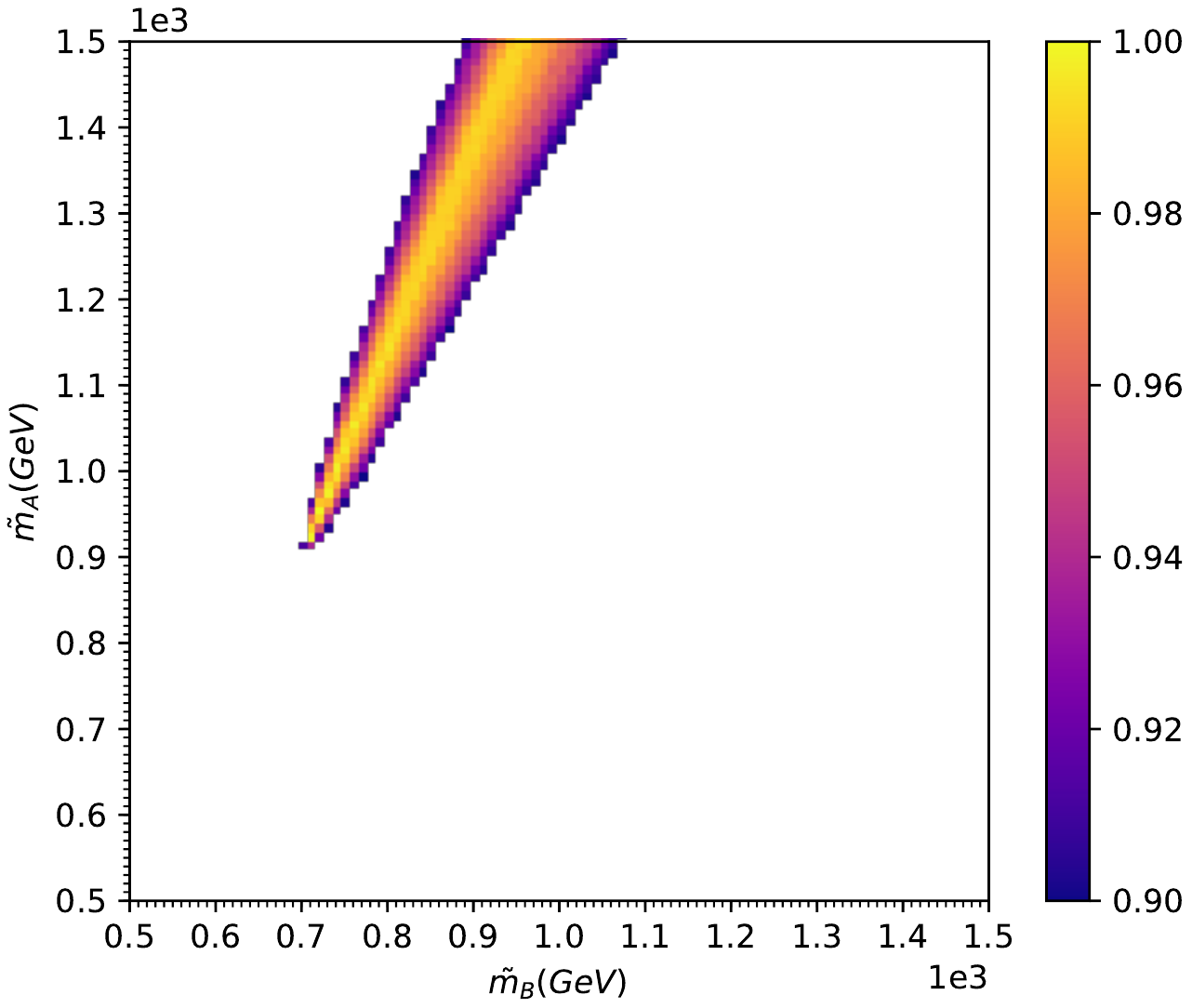}
 \caption{\label{fig:solvabilityhist600} Solvability heat maps as in Fig.~\ref{fig:solvabilityhist}, except here $\tmC=600\GeV<\mC$. }
\end{figure}

\begin{figure}
 \centering
 \includegraphics[width=.49\textwidth]{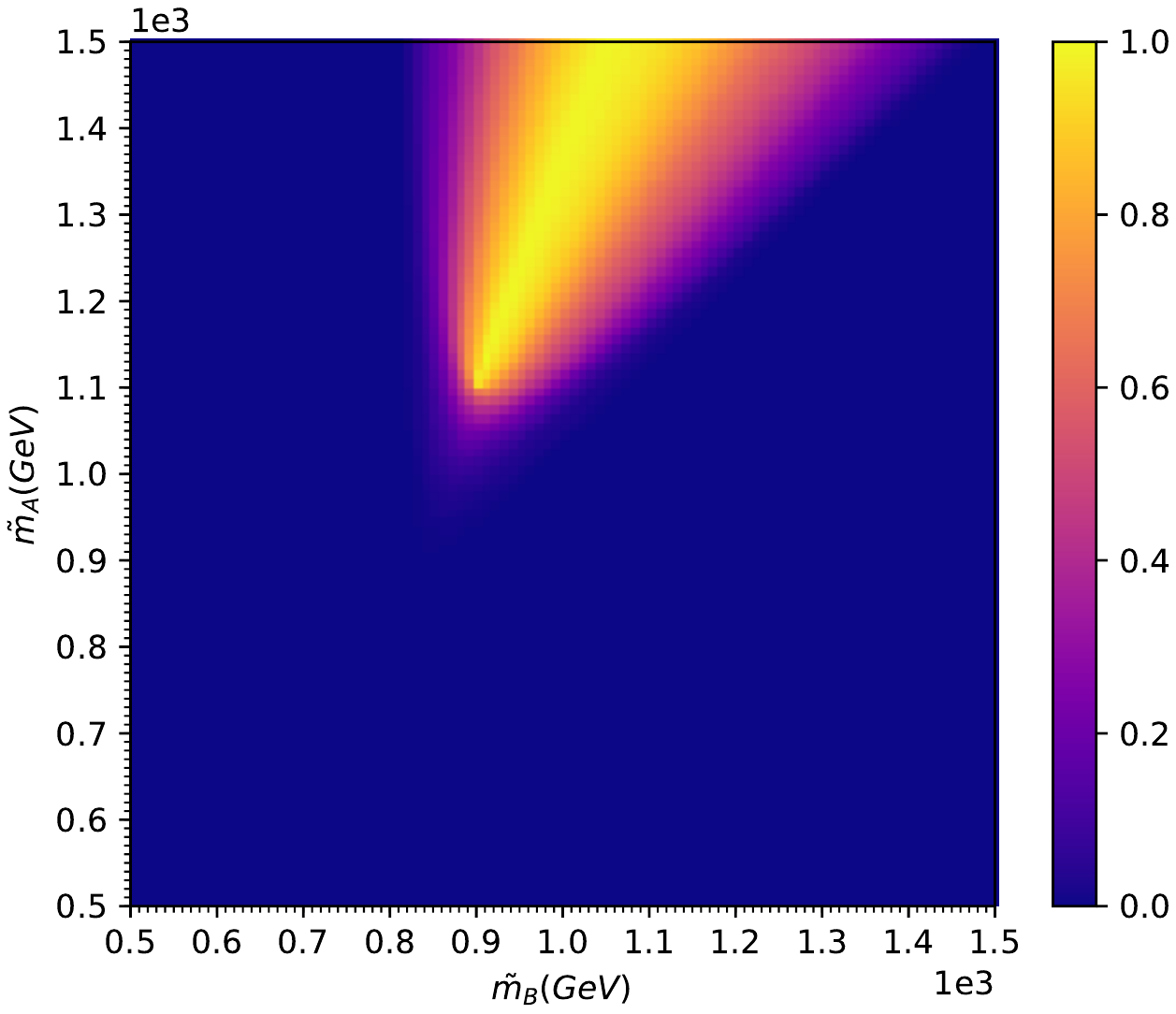}
 \includegraphics[width=.49\textwidth]{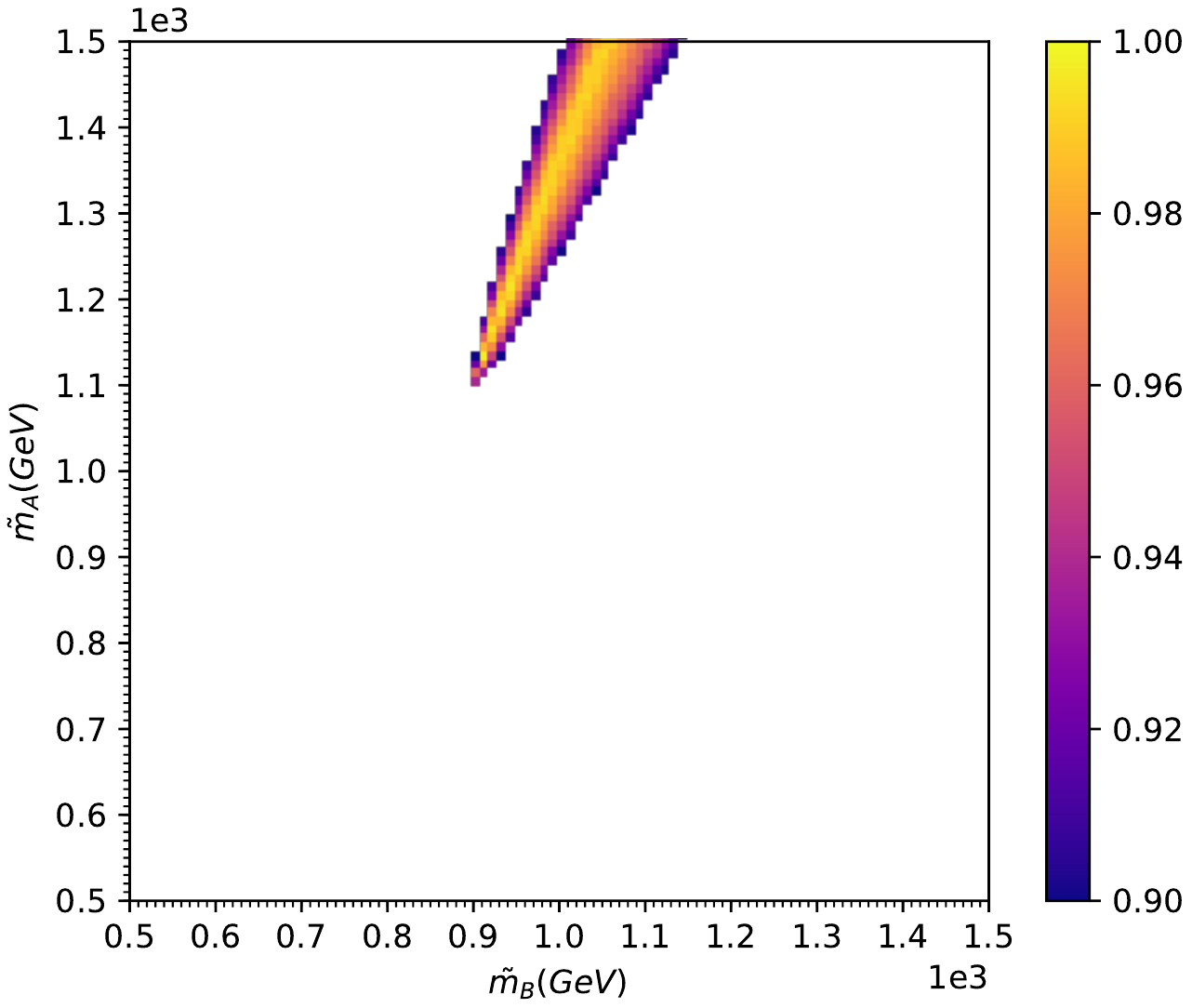}
 \caption{\label{fig:solvabilityhist800}Solvability heat maps as in Fig.~\ref{fig:solvabilityhist}, except here $\tmC=800\GeV>\mC$. }
\end{figure}

We now turn our attention to determining $\mC$. We begin by recreating the solvability heat maps from Fig.~\ref{fig:solvabilityhist} in Sec.~\ref{subsec:mbma},
only this time we choose different values of the trial mass parameter $\tmC$, away from the true value $\mC=700\GeV$. 
Results for $\tmC=600\GeV$ and $\tmC=800\GeV$ are shown in Figs.~\ref{fig:solvabilityhist600} and~\ref{fig:solvabilityhist800}, respectively. 
They confirm the trend observed in Figs.~\ref{fig:700tongues}-\ref{fig:800tongues}, namely, that as we increase (decrease) the value of $\tmC$, the preferred 
values of $\tmB$ and $\tmA$ increase (decrease) as well. Unfortunately, the solvability heat maps in Figs.~\ref{fig:solvabilityhist600} and~\ref{fig:solvabilityhist800}
look qualitatively very similar to the heat map in Fig.~\ref{fig:solvabilityhist} for $\tmC=\mC$. Therefore, it appears very difficult to extract the correct value of $\mC$
solely on the basis of these two-dimensional heat maps in the $(\tmB,\tmA)$ plane; some additional input is needed.

\begin{figure}[t]
 \centering
 \includegraphics[width=.49\textwidth]{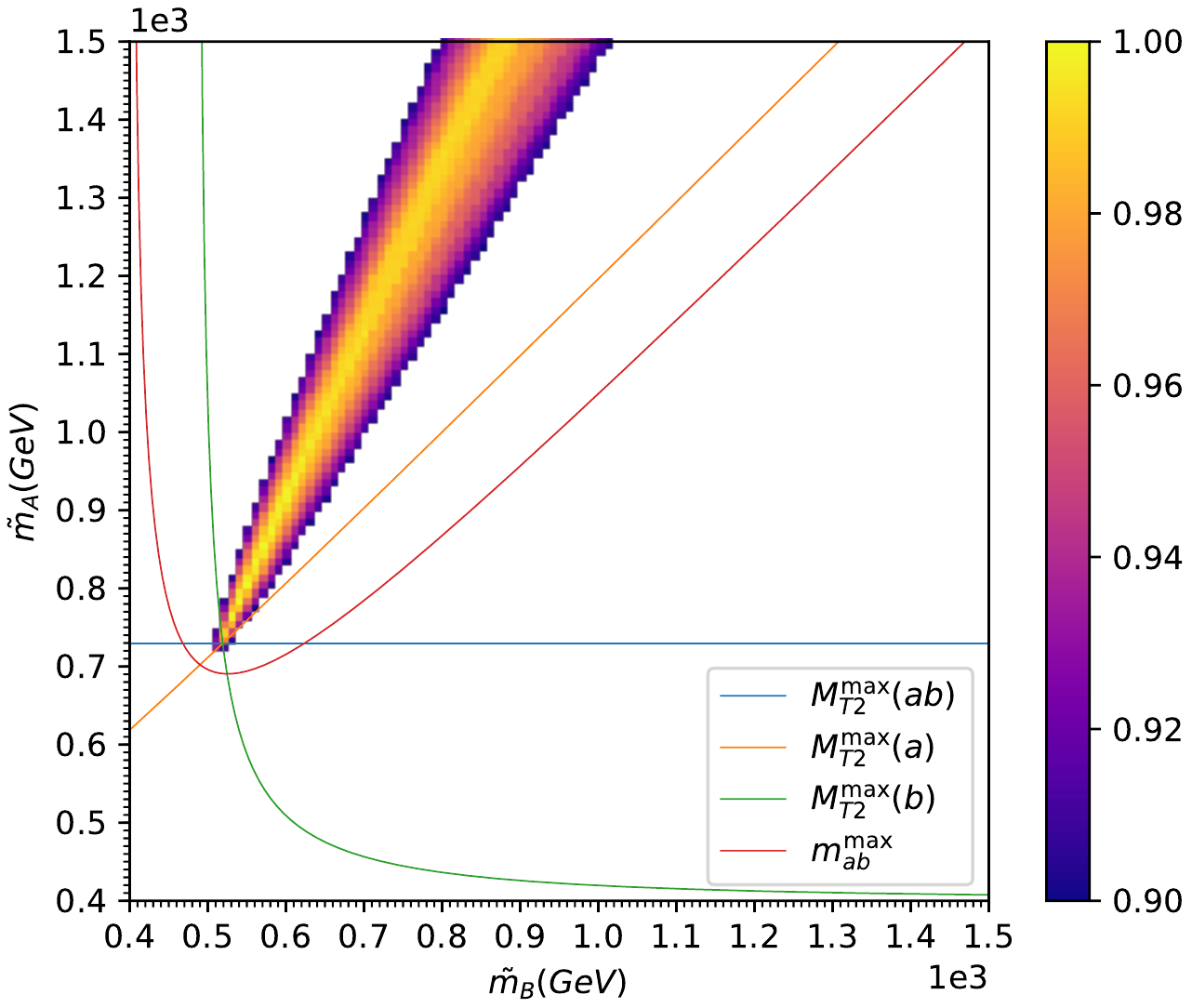}
 \includegraphics[width=.49\textwidth]{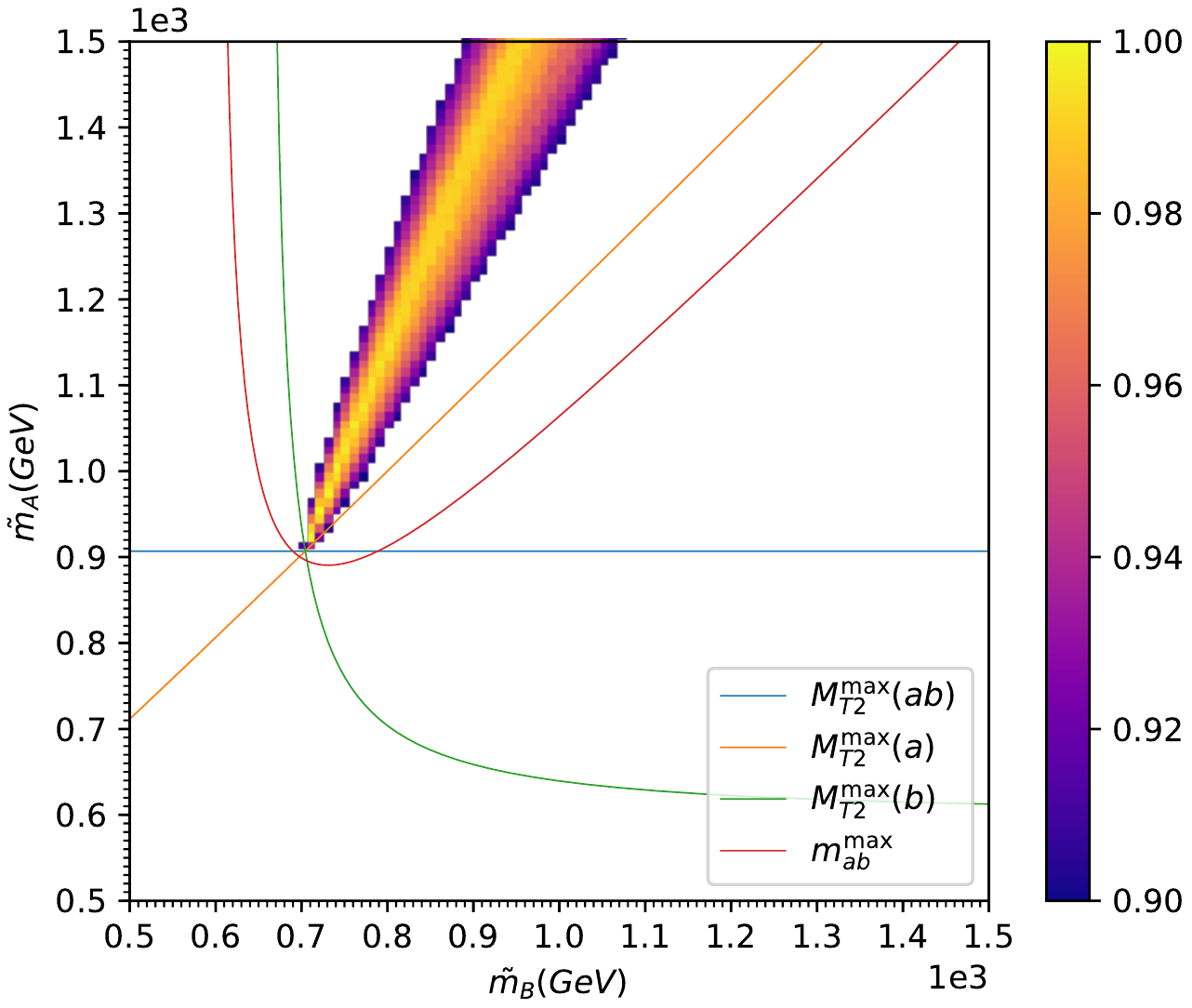}
 \\
 \includegraphics[width=.49\textwidth]{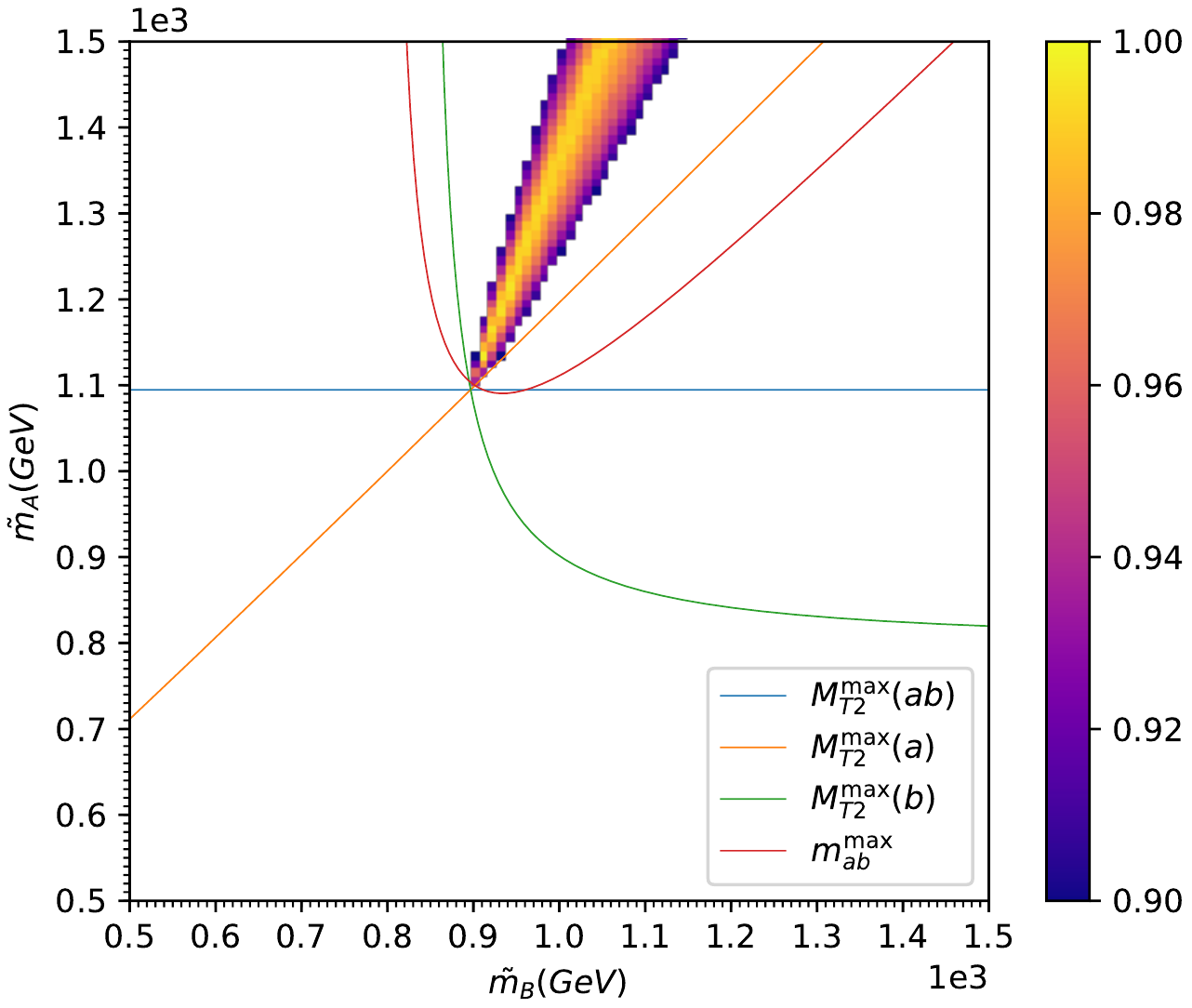}
 \includegraphics[width=.49\textwidth]{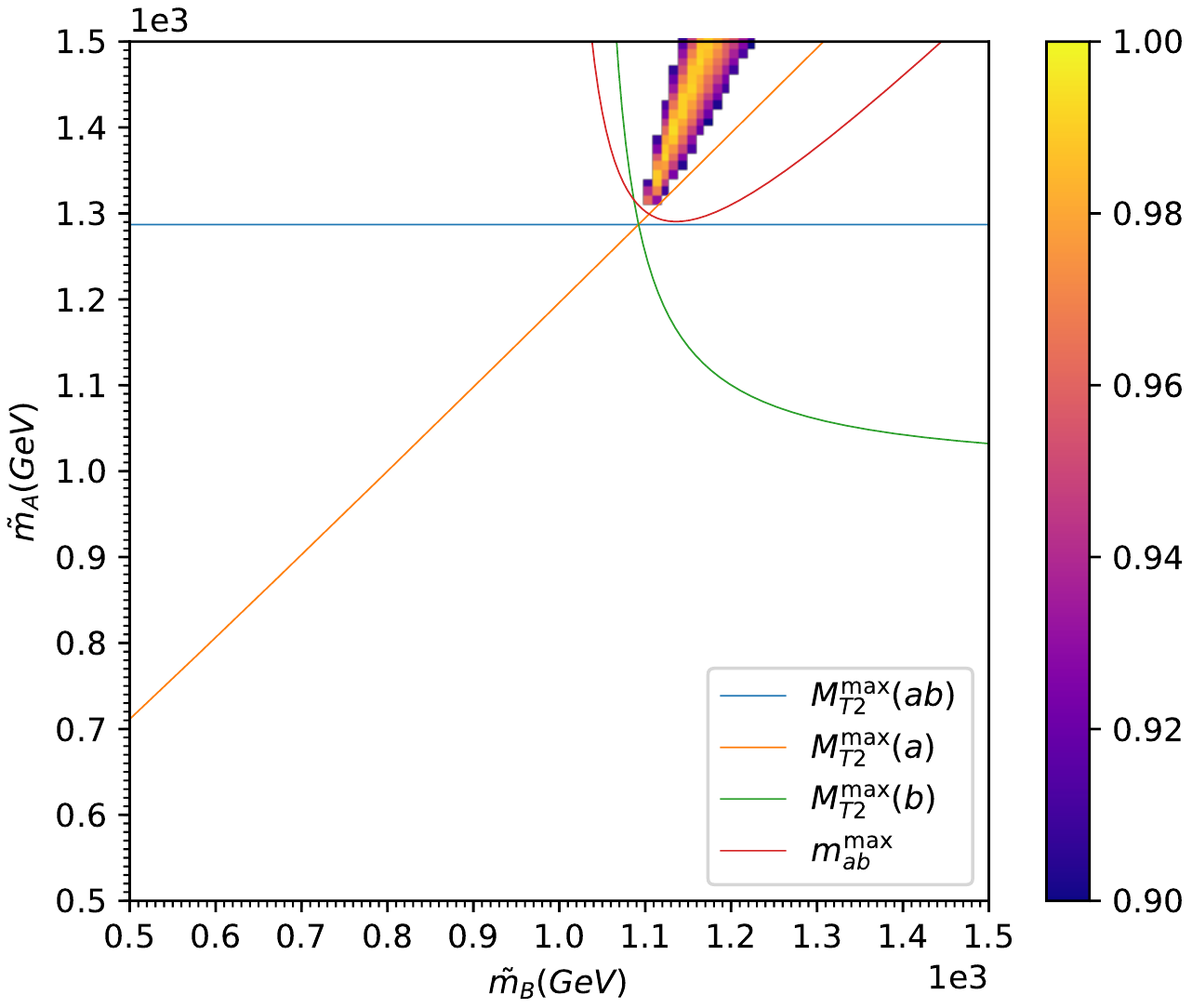}
 \caption{\label{fig:solvabilityandendpointstmCdiff}
 The same as Fig.~\ref{fig:solvabilityandendpoints}, but for different values of $\tmC$, not equal to the true mass $\mC=700$ GeV. 
 The top two plots have $\tmC<\mC$: $\tmC=400\GeV$ (top left) and $\tmC=600\GeV$ (top right).
 The bottom two plots correspond to $\tmC>\mC$: $\tmC=800\GeV$ (bottom left) and $\tmC=1000\GeV$ (bottom right).
Note that for $\tmC<\mC$ ($\tmC>\mC$) it is the $m^{max}_{ab}$ constraint (the $M^{max}_{T2}(ab)$ constraint)
which misses the region of high solvability.}
 \vspace{10mm}
\end{figure}

The measurements of the kinematic endpoints (\ref{endpoints}) might just be this missing input. 
Fig.~\ref{fig:solvabilityandendpoints} demonstrated that with the correct value of $\tmC$,
the measurements (\ref{endpoints}) are consistent among themselves, as well as with the solvability heat map. 
It is worth checking whether this consistency is retained when we use erroneous values of $\tmC$, i.e., values away from the true mass.
This test is performed in Fig.~\ref{fig:solvabilityandendpointstmCdiff}, which shows
analogues of Fig.~\ref{fig:solvabilityandendpoints} for
$\tmC = 400\GeV$ (upper left panel), $\tmC = 600\GeV$ (upper right panel), $\tmC = 800\GeV$ (lower left panel) and $\tmC = 1000\GeV$ (lower right panel).

There are several interesting features present in Fig.~\ref{fig:solvabilityandendpointstmCdiff}.
First, we note that the blue, orange and green curves corresponding to the three $M_{T2}$ kinematic endpoints 
always cross at a single point, and therefore the three $M_{T2}$ measurements are consistent with each other,
regardless of the chosen value of $\tmC$. This is due to the fact that these measurements are not independent, but
obey the relation \cite{Burns:2008va} 
\beq
\left[M^{\mathrm{max}}_{T2}(b)\right]^2 =
M^{\mathrm{max}}_{T2}(ab) \left[ M^{\mathrm{max}}_{T2}(ab) - M^{\mathrm{max}}_{T2}(a) \right],
\label{MT2endpointRelation}
\eeq
which is in fact why the fourth measurement of $m^{\mathrm{max}}_{ab}$ needed to be added to the set (\ref{endpoints}).
On the other hand, the red curve corresponding to the invariant mass endpoint $m^{\mathrm{max}}_{ab}$ is {\em not} consistent 
with the others, as it always misses the common crossing point for the three $M_{T2}$ measurements, which is an indication that the 
chosen value for $\tmC$ was incorrect. The mismatch grows as $\tmC$ moves further and further away from the true value $\mC$.

Fig.~\ref{fig:solvabilityandendpointstmCdiff} also reveals a discrepancy between the high solvability region, 
on the one hand, and the various kinematic endpoint measurements, on the other. 
In particular, at the lower values of $\tmC$ below the true mass $\mC$,
the red curve for $m^{\mathrm{max}}_{ab}$ tends to miss the solvable region, 
while for values of $\tmC$ which are too high, it is the blue $M^{max}_{T2}(ab)$ curve which fails to come into contact with the high solvability region.
Therefore, by plotting the maximum fraction of solvable events {\em along} these 
constraint curves in the $(\tmB,\tmA)$ plane, as a function of $\tmC$,
we can identify the true $\mC$ as the point where the maximum solvable fraction is highest. 
This is the technique proposed by H.-C.~Cheng and Z.~Han in Ref.~\cite{Cheng:2008hk} 
(for an illuminating discussion on the relation between solvability and kinematic endpoints, edges and kinks, 
see \cite{Barr:2009jv,Lester:2013aaa}).

In conclusion of this section, we have seen that the solvability method, supplemented with one or more 
kinematic endpoint measurements, is in principle capable of determining all three unknown masses $\mA$, $\mB$ and $\mC$.
However, as revealed by the preceding discussion in Sections~\ref{subsec:mbma} and \ref{subsec:mc}, 
the method is not as sensitive in finding the true value of $\mC$ as it is in finding the true values of $\mB$ and $\mA$, given $\mC$. 
This is indicative of a ``flat direction" of low sensitivity along $\tmC$. This is not unexpected --- it is known 
that mass differences can be measured much better than the overall mass scale, as has been 
demonstrated with specific studies for the case of pair production \cite{Cho:2007qv,Barr:2007hy,Cho:2007dh,Matchev:2009fh,Matchev:2009ad}
and a single decay chain \cite{Gjelsten:2004ki,Gjelsten:2005aw,Gripaios:2007is,Debnath:2016gwz,Betancur:2017kqe}.

\section{Extreme Events and Kinematic Boundaries in Mass Space}
\label{sec:extremeevents}

With the intuition developed in the previous two sections, we are now ready to present our main idea. First we shall introduce some terminology.

\subsection{Definitions}
\label{sec:definitions}

As illustrated in the solvability plots in Figs.~\ref{fig:700tongues}-\ref{fig:800tongues}, for any given event, 
the kinematic constraints (\ref{eqn:met}) and (\ref{eqn:on-shell}) partition 
the three-dimensional parameter space $(\tmA, \tmB, \tmC)$ into regions allowing 0, 2 or 4 real solutions for the invisible momenta.
We shall call the two kinematic\footnote{In the sense that they are derived from kinematic constraints.} boundary surfaces 
separating those mass space regions {\bf degeneracy boundaries}, since {\em on} those boundaries there exists a pair of 
degenerate real solutions. For completeness, we shall also give a special name to the boundary separating the 
regions with zero and two solutions and call it a {\bf solvability boundary}, 
since the event is solvable on one side of the boundary and not solvable on the other.
In our event topology, the kinematic boundaries of an event are two dimensional {\em surfaces} in the three dimensional mass space $(\tmA, \tmB, \tmC)$. 
Therefore, when we take a slice of the parameter space at fixed $\tmC$ as in Figs.~\ref{fig:700tongues}-\ref{fig:800tongues},
we correspondingly obtain {\bf degeneracy curves} in the $(\tmB,\tmA)$ plane, some of which (the ones 
which delineate the region with zero solutions) are {\bf solvability curves}.

\begin{figure}
 \centering
 \includegraphics[width=.45\textwidth]{11_24_mnu=700.pdf}
 \hskip 5mm
 \includegraphics[width=.45\textwidth]{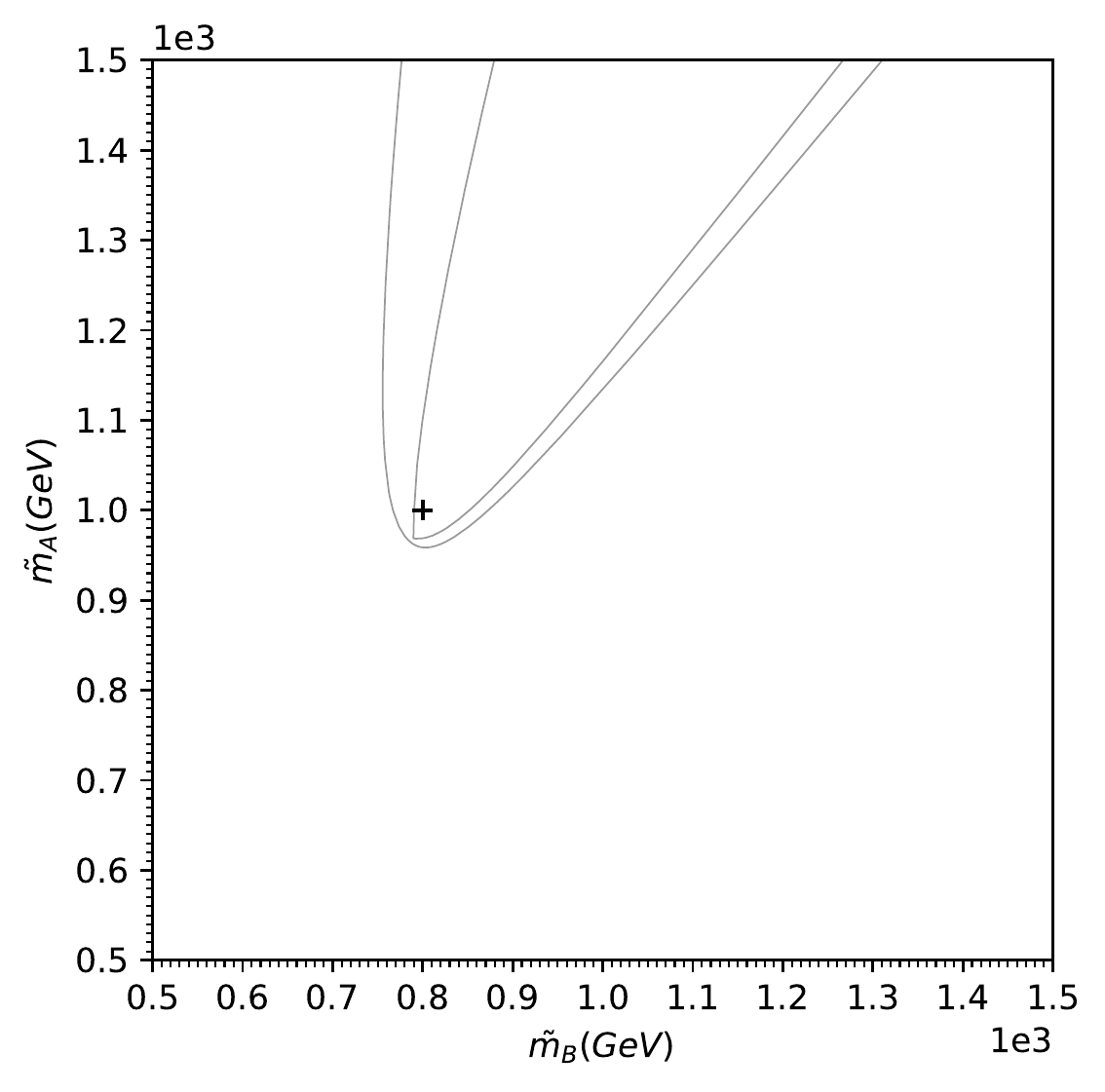}
 \caption{\label{fig:sampleextremenesscurve} Illustration of degeneracy curves. We recycle the same event 
 used to make the right panels in Figs.~\ref{fig:700tongues}-\ref{fig:800tongues}. 
 On the left we show the solvability plot for $\tmC=\mC=700\GeV$, while on the right we show the corresponding degeneracy curves.
 As before, the cross marks the true mass point $(\mB,\mA)$.}
\end{figure}

Fig.~\ref{fig:sampleextremenesscurve} illustrates the notion of degeneracy curves, using the event from the right panels
in Figs.~\ref{fig:700tongues}-\ref{fig:800tongues}. In the left panel of Fig.~\ref{fig:sampleextremenesscurve}, 
we reproduce the solvability plot from Fig.~\ref{fig:700tongues} which was made for $\tmC=\mC=700\GeV$.
Again we can see the three different regions --- with 0, 2 and 4 real solutions (colored white, green and yellow, respectively).
Then in the right panel of Fig.~\ref{fig:sampleextremenesscurve} we ignore the interiors of those regions and only plot the 
degeneracy curves for that event. As before, the cross marks the true mass point $(\mB,\mA)$.
We observe that the cross does {\em not} lie on any of the degeneracy curves --- indeed, 
there is no reason to think that the true masses will necessarily lead to degenerate real solutions for the momenta.
Since the event used in Fig.~\ref{fig:sampleextremenesscurve} was chosen at random, its properties are generic 
and follow this expectation.

At the same time, Fig.~\ref{fig:extreme3or2} also revealed the presence of events in the sample for which the 
cross {\em will} lie on a degeneracy curve. Now is a good time to give a name to such special events, 
since they will be the main topic of discussion in this section. In general, whenever a given trial mass point 
$(\tmA,\tmB,\tmC)$ lies on a degeneracy boundary of an event, the event will be said to be an {\bf extreme} event \emph{for that mass point}. 
Extreme events for the true mass point $(\mA,\mB,\mC)$ will be referred to simply as extreme events. 
Note that {\em a priori} we do not know which events in the event sample are extreme in the latter sense, 
since we do not know the true masses. Therefore, in order to analyze extreme events, we either have to cheat and use
Monte Carlo information, or develop some alternative strategies for recognizing such events in the data, 
see Section~\ref{subsec:extremecharacterization} below.

\begin{figure}
 \centering
 \includegraphics[width=.45\textwidth]{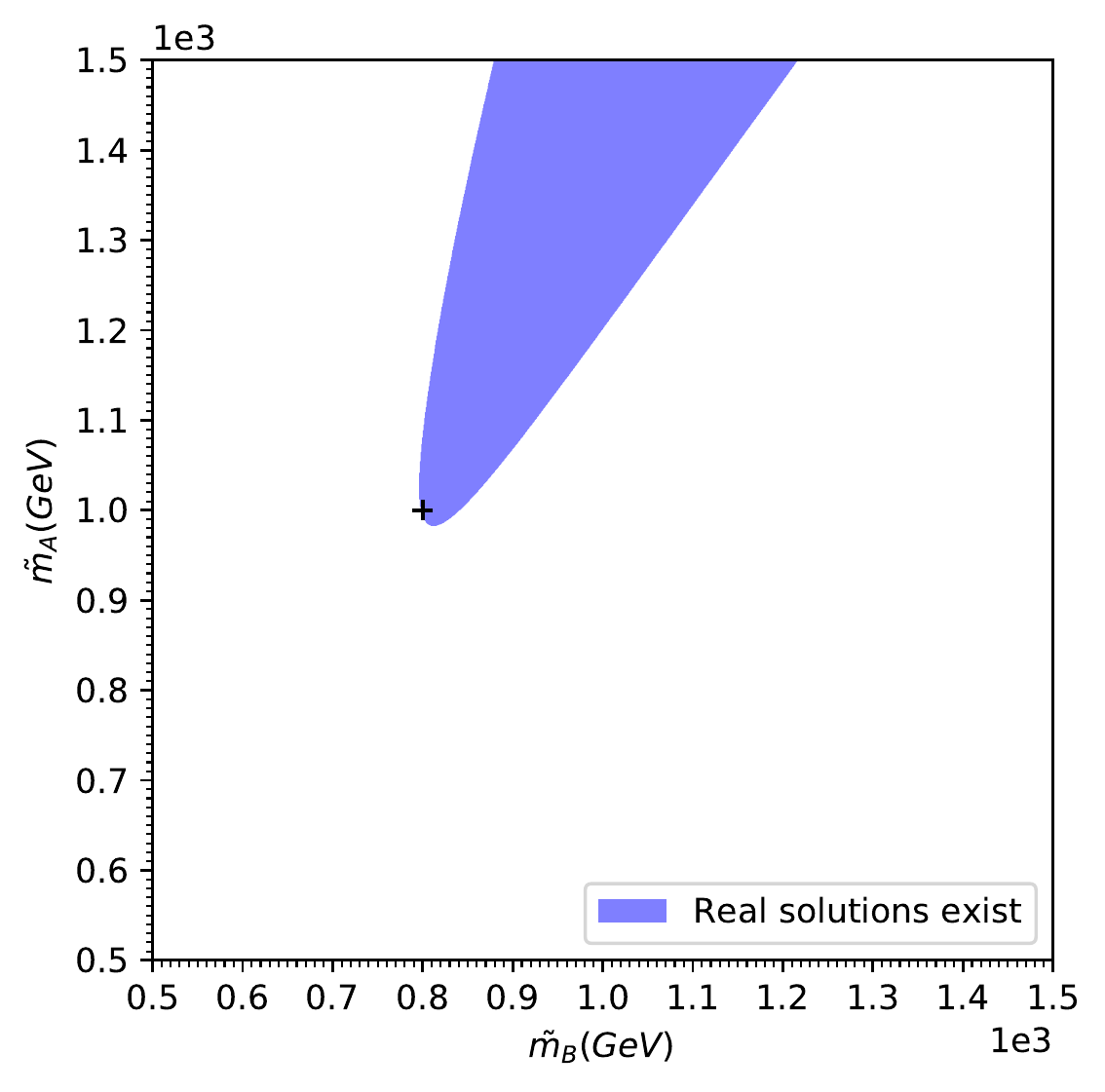}
 \includegraphics[width=.45\textwidth]{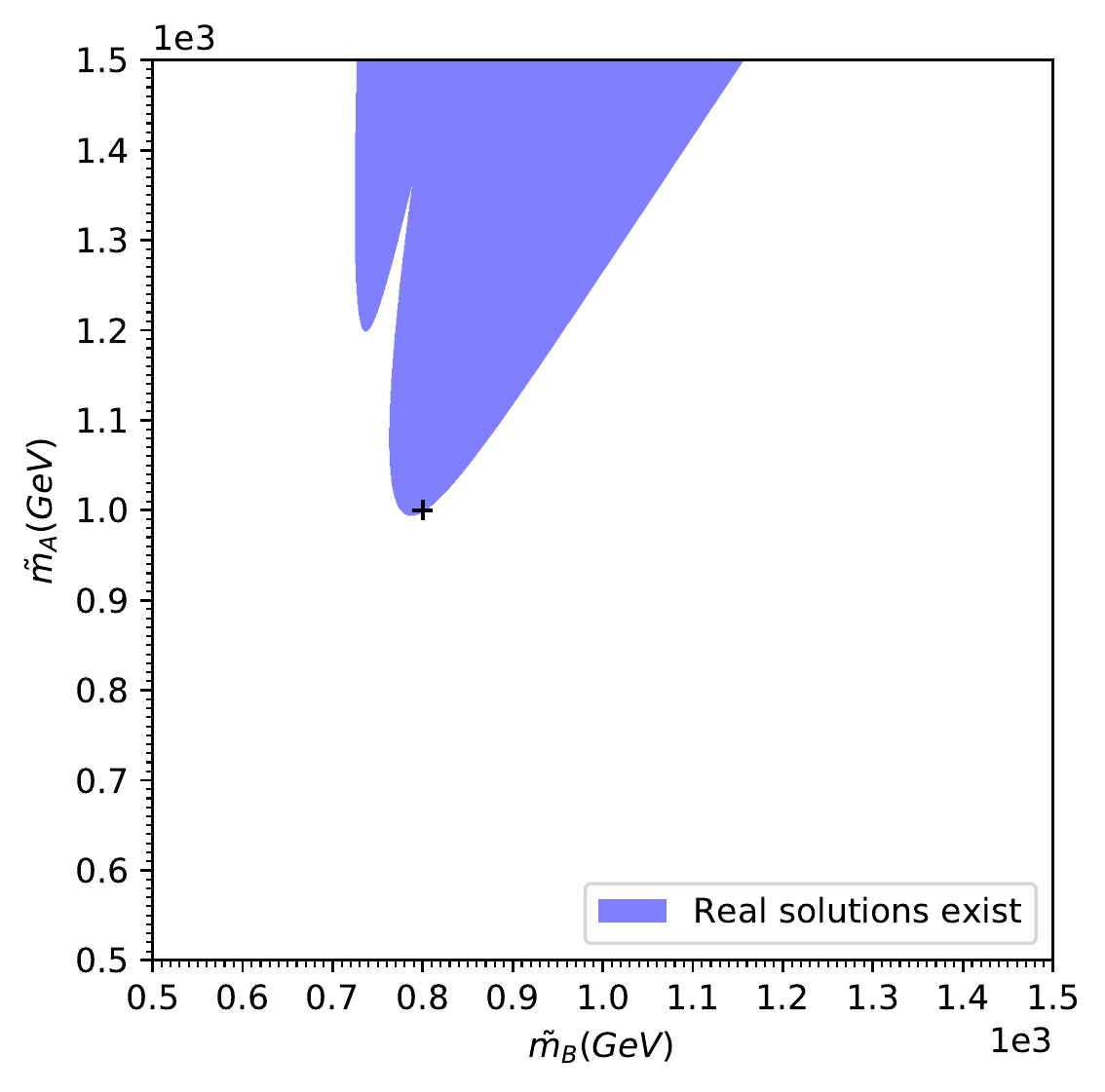}
 \\
 \includegraphics[width=.45\textwidth]{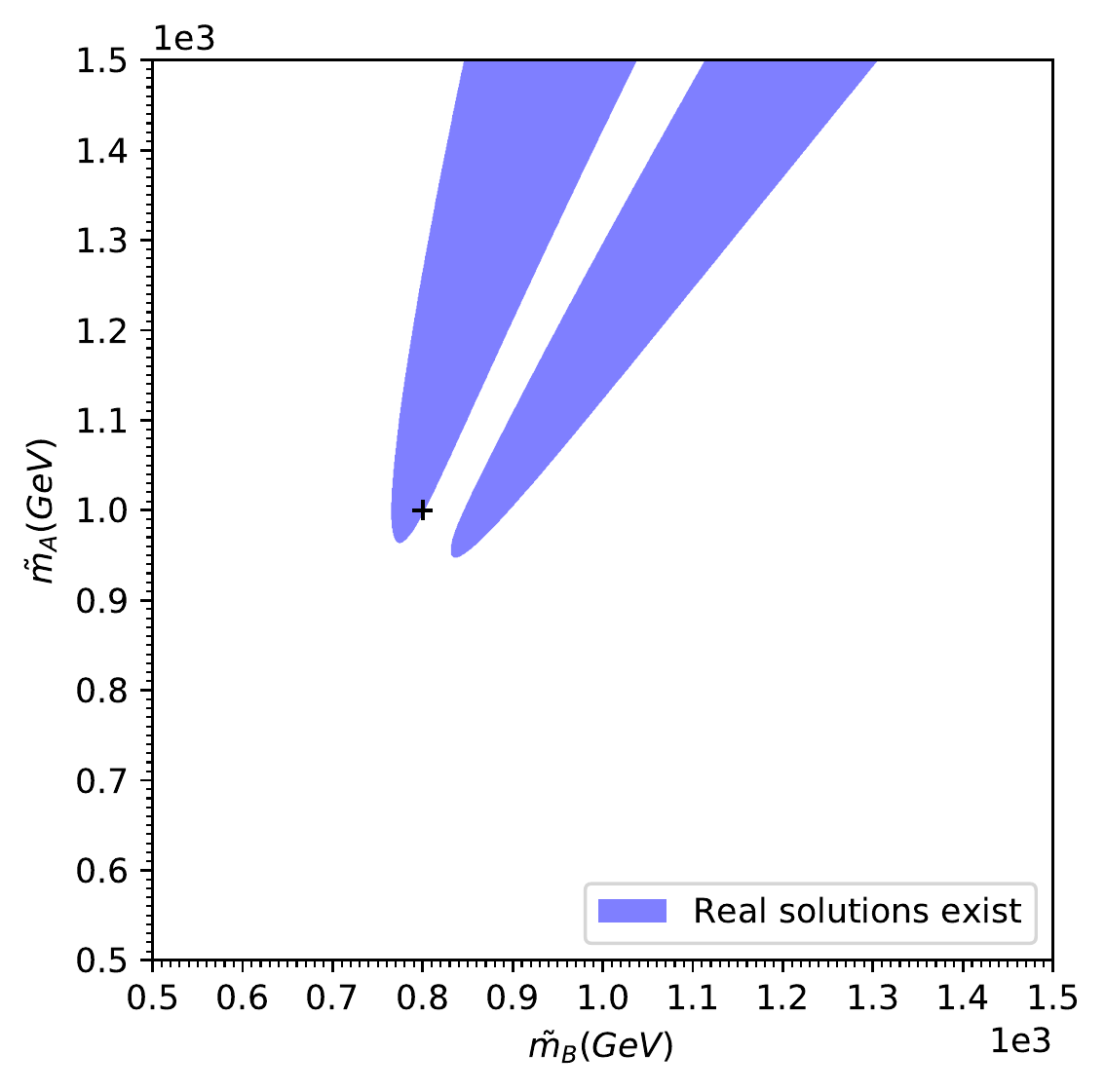}
 \includegraphics[width=.45\textwidth]{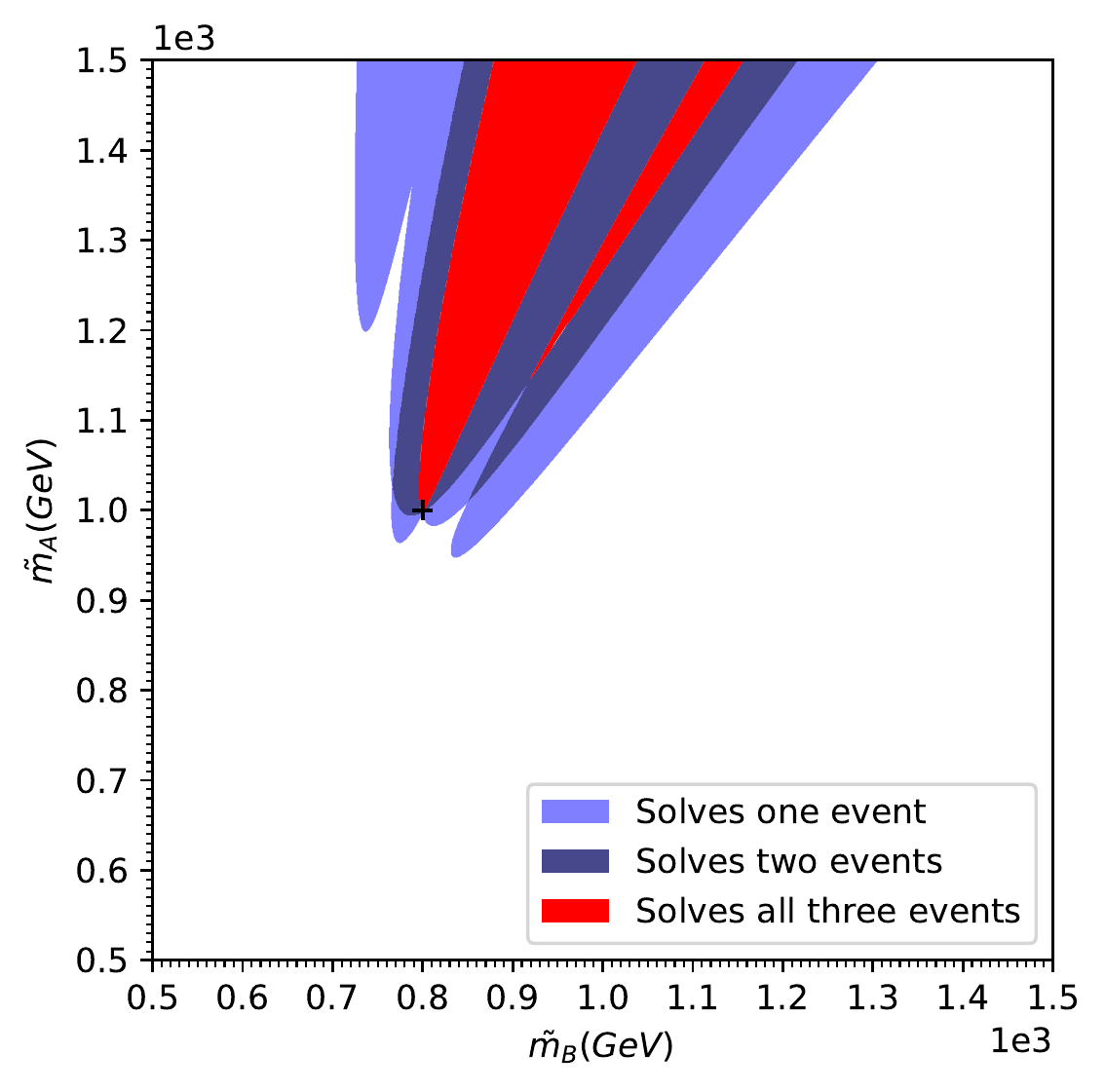}
 \caption{\label{fig:extreme-solvnosol}
 The same as Fig.~\ref{fig:extreme3or2}, but for three other randomly chosen extreme events, 
 i.e., the events were required to be extreme, but not necessarily complementary to each other.}
\end{figure}

As demonstrated in Fig.~\ref{fig:extreme3or2}, the primary motivation for studying extreme events is 
that they are very efficient in ruling out incorrect mass hypotheses. This is especially important 
in the region {\em in the immediate vicinity of the true mass point}.
Indeed, by definition, any event which is not extreme, will leave the whole neighborhood 
around the true mass point as viable. In that sense, extreme events are also {\em extremely} valuable, 
as they provide the only chance to probe the mass space close to the true mass point.
This point is illustrated in Fig.~\ref{fig:extreme-solvnosol}, which repeats the exercise 
from Fig.~\ref{fig:extreme3or2} with three {\em other}, randomly selected, extreme events.
Note that while the events were required to be extreme in the sense that the true mass point marked with the cross lies on a degeneracy boundary,
they were {\em not} required to be complementary to each other, as was previously done in Fig.~\ref{fig:extreme3or2}. 
As a result, their superposition now leaves a much larger red-shaded area in which all three events are solvable 
(compare to Fig.~\ref{fig:extreme3or2}). Even so, notice that the true mass point is on the very tip of the
allowed red-shaded region. In other words, three of the four possible directions away from the true mass 
point are ruled out, leaving only a relatively narrow funnel in the north-northeast direction. 
This is to be contrasted with the result from Fig.~\ref{fig:superimposedtongues}, for example, 
where the immediate vicinity of the true mass point is viable in any direction.

The observant reader would have noticed that each of the three events in Fig.~\ref{fig:extreme-solvnosol} happened to be extreme because
the true mass point lies on a solvability boundary as opposed to the other degeneracy boundary (the one between the regions with 2 and 4 real solutions). 
This is somewhat accidental --- in the data sample we do find extreme events of both types. If we were to categorize them,
we would find that the extreme events on a solvability boundary outnumber the others by a factor of roughly $5:1$.
This is consistent with the outcome in Fig.~\ref{fig:extreme-solvnosol} --- given three extreme events, the chances that  
they all have the true mass point on a solvability boundary are approximately $\left(\frac{5}{6}\right)^3\approx 58\%$.
One might question the usefulness of the remaining $1/6$ of extreme events which do not involve a solvability boundary. 
Indeed, they do not help to \emph{shave} the allowed mass space near the true mass point. 
Nevertheless, as we shall see in the next section, both types of extreme events will
contribute on an equal footing to the mass measurement method proposed in Section~\ref{sec:focus},
therefore from now on we shall stop making the distinction between them.

\subsection{Characterizing extreme events}
\label{subsec:extremecharacterization}

As mentioned earlier, the kinematic constraints (\ref{eqn:met}) and (\ref{eqn:on-shell}) allow four (in general complex) 
solutions for the invisible momenta $q_i$ of particles $\C{i}$. 
The four solutions are continuous functions of $(\tmA, \tmB, \tmC)$ and any
non-real solutions among them appear in complex conjugate pairs \cite{Gripaios:2011jm}. 
On the degeneracy boundaries, where the number of real solutions changes, 
the equations have a pair of degenerate real solutions.

We can therefore characterize extreme events (for a given choice of masses $\tmA$, $\tmB$ and $\tmC$) 
by finding the condition for a particular solution $(\qC{1}, \qC{2})$ to be degenerate. 
It can be shown that a solution $(\qC{1}, \qC{2})$ is degenerate iff
\beq
{\mathcal E}(\pa{i}, \pb{i}, \qC{i}) \equiv \Ea{1}\Eb{1}\EC{1}\Ea{2}\Eb{2}\EC{2} \left[(\vec{V}_1 \times \vec{V}_2)\cdot \hat{z}\right] = 0,
\label{eqn:tripleproduct}
\eeq
where $\hat z$ is the unit vector along the $z$-axis, 
the $E$'s refer to the energy components of the corresponding 4-momenta, and the three-vectors $\vec{V}_i$ are defined as
\beq
\vec{V}_i = (\vec{v}_{a_{i}} \times \vec{v}_{b_{i}}) + (\vec{v}_{b_{i}} \times \vec{v}_{\ms{C}_{i}}) + (\vec{v}_{\ms{C}_{i}} \times \vec{v}_{a_{i}}).
\eeq
Here the 3-vectors $\vec{v}$ are the velocities\footnote{Not to be confused with the spatial components of the 4-velocities.} of the particles, i.e., $\vec{v}=\vec{p}/E$:
\beq
\vec{v}_{a_{i}} = \frac{\vec{p}_{a_{i}}}{\Ea{i}};\quad\vec{v}_{b_{i}} = \frac{\vec{p}_{b_{i}}}{\Eb{i}};\quad\vec{v}_{\ms{C}_{i}} = \frac{\vec{q}_i}{\EC{i}}.
\eeq

Eq.~\eqref{eqn:tripleproduct} is useful in understanding the kinematic configurations of extreme events. 
As discussed later in Section~\ref{subsec:densityvendpoint}, extreme events correspond to kinematic endpoints, 
therefore eq.~\eqref{eqn:tripleproduct} (and its analogues for other topologies) can be useful in understanding 
and deriving various endpoints. Eq.~\eqref{eqn:tripleproduct} also provides an efficient way of generating a pure sample of
extreme events for dedicated studies of such events\footnote{Any simulated event (for which we know the invisible momenta) 
can be converted into an extreme event by appropriately rotating the decay of $\B{2}$ about its direction 
of motion to satisfy eq.~\eqref{eqn:tripleproduct}.}. Of course, the quantity ${\mathcal E}$, being a scalar, 
conveniently quantifies the level of extremeness of the event: the smaller the magnitude of ${\mathcal E}$, the 
more extreme an event is.

One downside of the parametrization (\ref{eqn:tripleproduct}) is that it refers to the particular solution $(\qC{1}, \qC{2})$,
so that in order to use this criterion in practice, one has to solve the constraint equations first. Since they may lead to up to 4 
real solutions, one must test the condition (\ref{eqn:tripleproduct}) for each real solution and 
then use the smallest value of ${\mathcal E}$ thus obtained in order to judge whether the event is extreme or not.

\subsection{Extreme events are more common than you think}
\label{subsec:extremeeventdensity}

Having motivated extreme events through their useful and interesting properties, the 
crucial question to address next is, how frequent are they in a realistic data sample? 
The pessimist from Sec.~\ref{sec:superposition} could again raise an objection that 
the volume of a two-dimensional surface in three-dimensional space is zero, and therefore,
the chances that an event will land exactly on a degeneracy boundary and thus become extreme
are vanishing as well. However, this argument contains a hidden assumption --- that the distribution
of events throughout the three-dimensional space is a normal, well-behaved function, i.e., the event density has no singularities. 
In this subsection we shall argue that the hidden assumption is false, and that in fact, 
in the narrow width approximation, the density of events is expected to be singular 
{\em precisely} on the degeneracy boundaries, thus greatly increasing the odds of encountering extreme events in the sample.
The gist of our argument goes back to the algebraic singularity method of I.-W.~Kim \cite{Kim:2009si}. As a specific example,
phase space singularities on kinematic surfaces were recently explored in the case of a single SUSY-like decay chain in
\cite{Agrawal:2013uka,Debnath:2016gwz,Debnath:2018azt}. However, our main argument here is very general, 
thus we shall try to keep the discussion independent of the underlying topology.

Consider the space of the $N_{p\,\cup\,q}$  components
of the 4-momenta of all final state particles, visible and invisible. 
Out of these, let $N_{p}$ be the total number of visible momentum components $\{p\}$
and $N_{q}$ be the total number of invisible momentum components $\{q\}$, 
where trivially $N_{p\,\cup\,q}=N_{{p}}+N_{{q}}$. On-shell events generated from a given set of true masses will obey 
a certain number, $C_{{q}}$, of kinematic constraints involving the invisible momenta,
plus possibly also a certain number of constraints $C_{{p}}$ involving {\em only the visible} momenta $\{{p}\}$ and no unknown parameters.
As a result of the kinematic constraints, the final state momenta will lie on a hypersurface of
dimension $N_{{p}\,\cup\,{q}}-C_{{q}}-C_{{p}}$, which is
embedded in the original $(N_{{p}\,\cup\,{q}})$-dimensional space.
This hypersurface represents the full phase space for this process, and
events will be more or less evenly distributed in the full phase space.

Now the crucial question is, how are events distributed in the allowed $(N_{{p}}-C_{{p}})$-dimensional {\em visible} subspace? 
For this purpose, we need to project the full $(N_{{p}\,\cup\,{q}}-C_{{q}}-C_{{p}})$-dimensional phase space 
onto the $(N_{{p}}-C_{p})$-dimensional allowed visible subspace \cite{Kim:2009si}.
The dimensionality of the resulting projection is 
\beq
\min\left(N_{{p}\,\cup\,{q}}-C_{{q}}-C_{{p}}, N_{{p}}-C_{{p}}\right).
\eeq 
In this paper, we shall restrict our attention to the case where $N_{{p}\,\cup\,{q}}\,-\,C_{{q}}\,-\,C_{{p}} = N_{{p}}\,-\,C_{{p}}$, 
i.e., the dimensionality of the full phase space equals the dimensionality of the allowed visible subspace.
Since $N_{{p}\,\cup\,{q}}=N_{{p}}+N_{{q}}$, an equivalent way to state the same condition is simply
$N_{{q}}=C_{{q}}$, i.e., there are just enough constraints in order to solve for the invisible momenta $\{{q}\}$.
This was precisely the case in Section~\ref{sec:solvability}: we were able to solve for the invisible momenta in the event topology 
of Fig.~\ref{fig:feynmandiag}. Let us check: there were 4 visible final state particles ($N_{{p}}=4\times 4=16$),
2 invisible final state particles ($N_{{q}}=2\times 4=8$), two constraints from (\ref{eqn:met}) and six
constraints from (\ref{eqn:on-shell}) for a total of $C_{{q}}=2+6=8$.
There are also four mass shell constraints on the visible momenta ($C_{{p}}=4$).
Thus both the full phase space and the visible phase space are 12-dimensional:
$N_{{p}\,\cup\,{q}}-C_{{q}}-C_{{p}}=(16+8)-8-4=12$ and  $N_{{p}}-C_{{p}}=16-4=12$.

As already mentioned above, the probability density of events on the full phase space is well behaved,
i.e., there are no singularities or discontinuities. However, this is not the case with events in the visible phase space.
First we note that the mapping of the full phase space onto the visible phase space is not one-to-one (or invertible) ---
multiple points in the full phase space can have the same visible phase space projection, 
as evidenced by the existence of multiple solutions for the invisible momenta. To keep a simple visual analogy in mind, 
think of the projection of a two-dimensional hollow sphere onto the equatorial plane.  
Any image point in the equatorial plane has two possible preimages, one in the Northern hemisphere and one in the Southern hemisphere.

Let $\mathcal{P}_{\mathrm{full}}$ be the density of events in the full phase space, and let $\mathcal{P}_{\mathrm{vis}}$ 
be the density of events in the visible phase space. Consider an infinitesimal volume $dV_{\mathrm{vis}}$ in the visible phase space
and the corresponding infinitesimal volumes $dV_{\mathrm{full},i}$ which can be mapped onto it from the full phase space.
Here the index $i$ labels the ``sheet" on which $dV_{\mathrm{full},i}$ is located\footnote{Since the visible phase space and the full phase space have equal dimensionalities, 
there can only be a finite number of such ``sheets".} --- for example, in the hollow sphere example above, 
the two values of $i=1,2$ would label the Northern and Southern hemispheres. The two event densities are related as
\beq
\mathcal{P}_{\mathrm{vis}}\ dV_{\mathrm{vis}} = \sum_i \mathcal{P}_{\mathrm{full},i}\ dV_{\mathrm{full},i},
\eeq
or equivalently,
\beq
\mathcal{P}_{\mathrm{vis}}= \sum_i \mathcal{P}_{\mathrm{full},i}\ \frac{dV_{\mathrm{full},i}}{dV_{\mathrm{vis}} }.
\eeq
As already mentioned, $\mathcal{P}_{\mathrm{full},i}$ is well behaved. 
However, at points where $dV_{\mathrm{vis}}/dV_{\mathrm{full},i}$ is vanishing (for one of the $i$ values), $\mathcal{P}_{\mathrm{vis}}$ will be singular \cite{Kim:2009si}. 
The vanishing of $dV_{\mathrm{vis}}/dV_{\mathrm{full},i}$ occurs at those locations where 
the tangent plane of the full phase space is perpendicular to the tangent plane of the visible phase space.
(In the hollow sphere example above, this singularity occurs on the equator, where the surface of the sphere is perpendicular to the equatorial plane.)
It is also easy to see that those locations are precisely where two different branches in the full phase space (i.e., corresponding to two different values of $i$) meet
(in the hollow sphere example, think of the Northern and Southern hemisphere merging at the equator).
In other words, the points where $\mathcal{P}_{\mathrm{vis}}$ is singular correspond to extreme events 
which have degenerate real solutions for the invisible momenta. 
Consequently, the dimensionality of the subspace spanned by extreme events is $N_{{p}}-C_{{p}}-1$, due to the additional ``degeneracy condition".

\subsection{The connection between extreme events and kinematic endpoints}
\label{subsec:densityvendpoint}

For events on the boundary of the visible phase space, the true mass point will lie on the solvability boundary in mass space. 
In other words, events on the boundary of the visible phase space lead to a particular class of extreme events, namely
the ones involving the solvability boundary. As already discussed at the end of Section~\ref{sec:definitions}, those are the large majority of extreme events.

The shape of the boundary of the visible phase space is dependent on the underlying mass spectrum and can therefore reveal information about it. 
A standard approach to extract this information is to measure kinematic endpoints in the distributions of variables suitably constructed from the visible momenta \cite{Barr:2010zj,Matchev:2019sqa}.
It is clear that the events for which the kinematic variable under consideration acquires its endpoint value, are necessarily events on the boundary of the visible phase space
\cite{Cheng:2008hk,Barr:2009jv,Lester:2013aaa}. Since those are extreme events, it follows that extreme events are responsible for the 
endpoints in the distributions of kinematic variables. The reverse, however, is not necessarily true --- not {\em all} of the extreme events are necessarily mapped onto the kinematic endpoint value.
Therefore, compared to endpoints in kinematic distributions, extreme events present an interesting and more general target for studies of event kinematics. 

\subsection{The distribution pattern of kinematic boundaries in mass space}
\label{subsec:extremeboundarydensity}

In Section~\ref{subsec:extremeeventdensity} we argued that the density of extreme events is singular. 
Here we shall use that result to show that if we consider the degeneracy boundaries in the mass space 
for a sample of events, the density of these boundaries will be singular at the true mass point. 

We shall first give a heuristic justification before building it into a more rigorous argument. 
Note that any event is extreme for some set of points in mass space, namely, 
the ones on the degeneracy boundaries of that event. 
We do not know {\em a priori} whether any of those mass points happen to be the true masses or not. 
So let us then try to figure this out on a statistical basis. Let each event give a vote to all the mass points 
on its degeneracy boundaries, i.e., all masses for which it is extreme. 
We saw in Section~\ref{subsec:extremeeventdensity} that events which are extreme for the true mass 
have an enhanced (singular) probability density. On the other hand, events which are extreme for an 
incorrect test mass point do not necessarily have the same enhancement. By this logic, we can expect the true mass point
to pick up more votes in this procedure than an incorrect test mass choice.
``More votes" here corresponds to a higher density of degeneracy boundaries.

Let us rephrase this heuristic argument in a slightly more rigorous form. 
We shall do this by mapping the density of extreme events in the visible phase space to the density of degeneracy boundaries in the mass space.

In our event topology of Fig.~\ref{fig:feynmandiag} the mass space is 3 dimensional and the degeneracy boundaries are 2 dimensional surfaces. 
Their density, $\mathcal{P}^{db}_\mathrm{mass}(\tmA, \tmB, \tmC)$, 
will be defined as the fraction of degeneracy boundaries per unit 
$\sqrt[3]{\mathrm{volume}}$ in the mass space.\footnote{This is because the number of randomly distributed $d$-dimensional objects 
in a $D$-dimensional volume element scales as $\left(\mathrm{volume}\right)^{(1-d/D)}$. In our case $d=2$ dimensional surfaces are distributed in 
a $D=3$ dimensional volume element and their number scales as
$\left({\mathrm{volume}}\right)^{1/3}$.} 
It is given by
\beq
\mathcal{P}^{db}_{\mathrm{mass}}(\tmA, \tmB, \tmC) \sqrt[3]{d^3V_{\mathrm{mass}}} =  dV_1 \int\limits_{\substack{\mathrm{extreme}\\\mathrm{events}}}\mathcal{P}_\mathrm{vis}\, dV_{(N_{{p}}-C_{{p}}-1)},
\label{eqn:huh?}
\eeq
where the integral is over the $(N_{{p}}-C_{{p}}-1)$-dimensional subspace of extreme events 
for the given mass point $(\tmA, \tmB, \tmC)$. $dV_1$ represents a differential element in the visible phase space perpendicular to the space of extreme events. 
It corresponds to the change in the space of extreme events when the mass point varies within 
$d^3V_{\mathrm{mass}}$. It helps to think of the space of extreme events of a test mass as 
an $(N_{{p}}-C_{{p}}-1)$-dimensional curve which changes smoothly when the test mass changes.
Eq.~\eqref{eqn:huh?} can be rewritten as
\beq
\mathcal{P}^{db}_{\mathrm{mass}}(\tmA, \tmB, \tmC) = J \int\limits_{\substack{\mathrm{extreme}\\\mathrm{events}}}\mathcal{P}_\mathrm{vis}\, dV_{(N_{{p}}-C_{{p}}-1)},
\label{eqn:huh?-part2}
\eeq
where $J$ is a Jacobian factor. We can see that $\mathcal{P}^{db}_{\mathrm{mass}}$ will be singular iff the integral is divergent.

Note that the Jacobian factor and the integration volume for a given test mass are both independent of the true mass from which the event sample is generated. 
So, except possibly for some special test masses (special overall, not in relation to the true mass), 
we can expect their contribution to the variation of $\mathcal{P}^{db}_{\mathrm{mass}}$ to be well-behaved. 
Any local singularity in $\mathcal{P}^{db}_{\mathrm{mass}}$  can therefore only come from $\mathcal{P}_\mathrm{vis}$. 
As mentioned in Section~\ref{subsec:extremeeventdensity}, $\mathcal{P}_\mathrm{vis}$ is singular 
for extreme events of the true mass and for those events only. Thus the density of degeneracy boundaries 
in the mass space is singular at the true mass\footnote{The density is also singular at masses whose extreme events have an intersection of measure greater than 0 with the true extreme events.}.

\section{The Kinematic Focus Point Method}
\label{sec:focus}

\subsection{The basic idea}
\label{sec:idea}

Before proceeding further, let us condense the content of Section~\ref{subsec:extremeboundarydensity} into the following elevator pitch:
\begin{enumerate}
 \item Draw the degeneracy boundaries (in the mass space) for all events in the data sample. 
 As a reminder, the degeneracy boundary of an event contains masses for which the given event has degenerate real solutions for the invisible momenta.
 \item The density of these boundaries should peak at the true mass point.
\end{enumerate}

\begin{figure}[t]
 \centering
 \includegraphics[width=.45\textwidth]{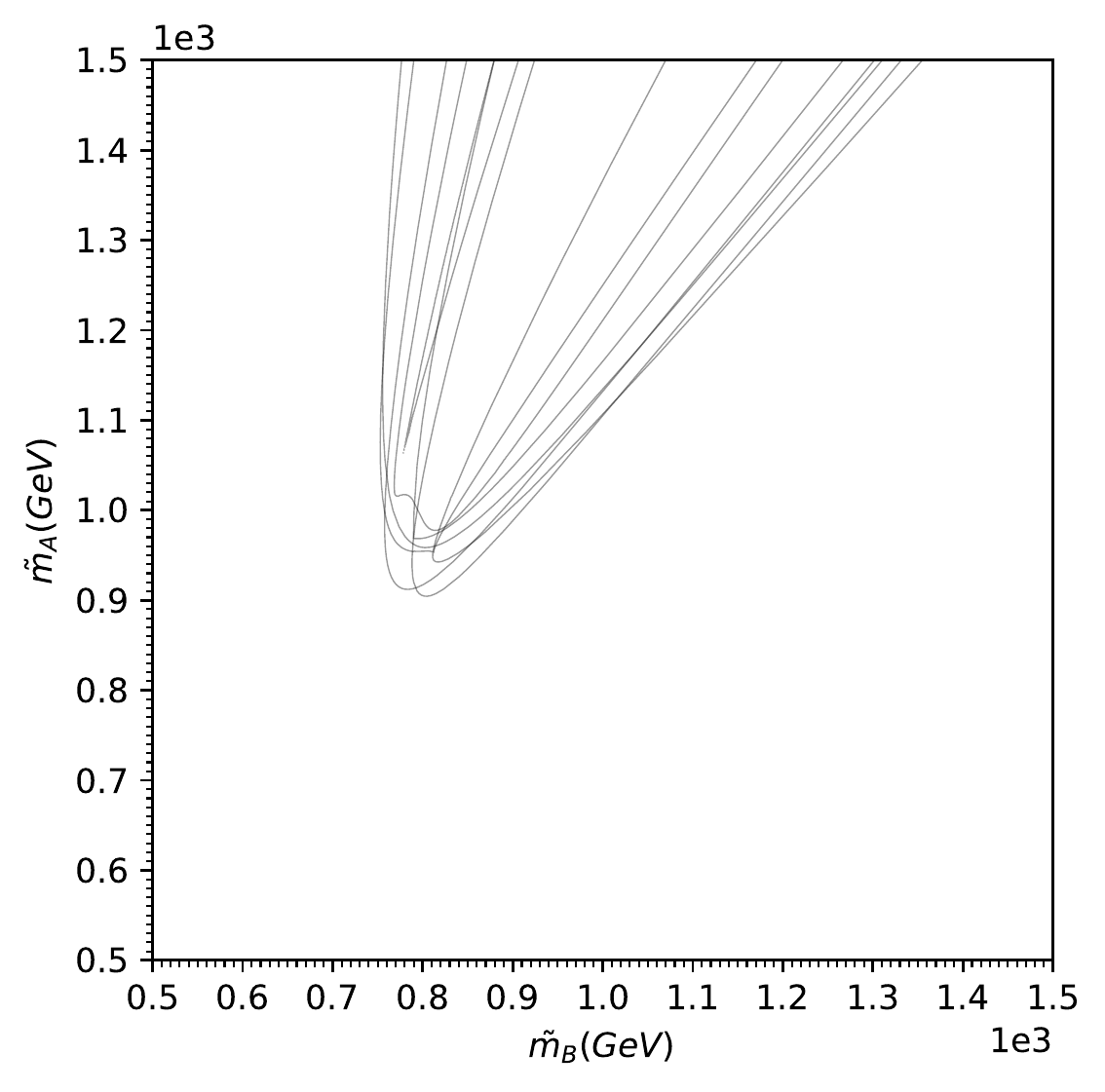}
 \includegraphics[width=.45\textwidth]{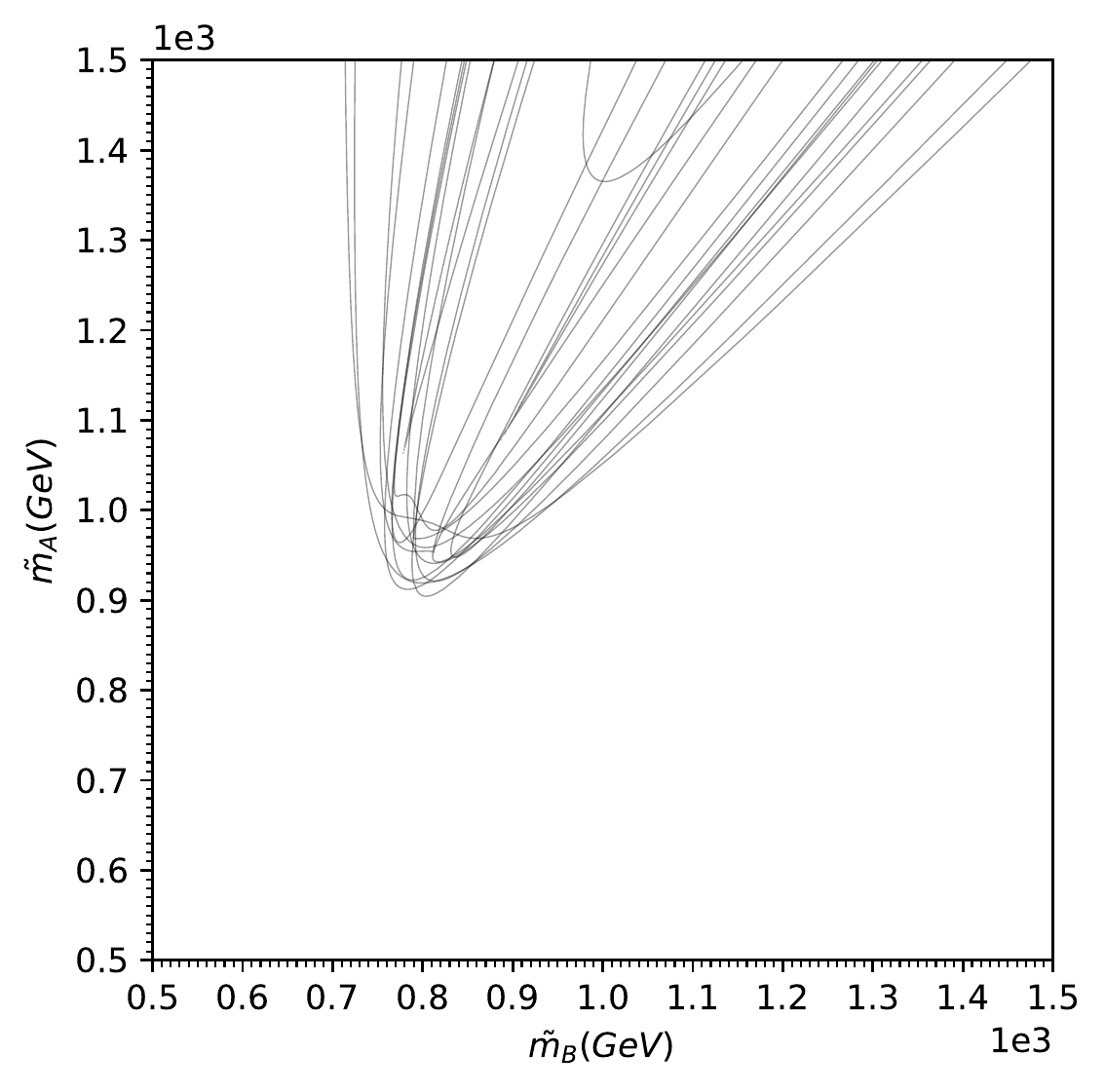}
 \includegraphics[width=.45\textwidth]{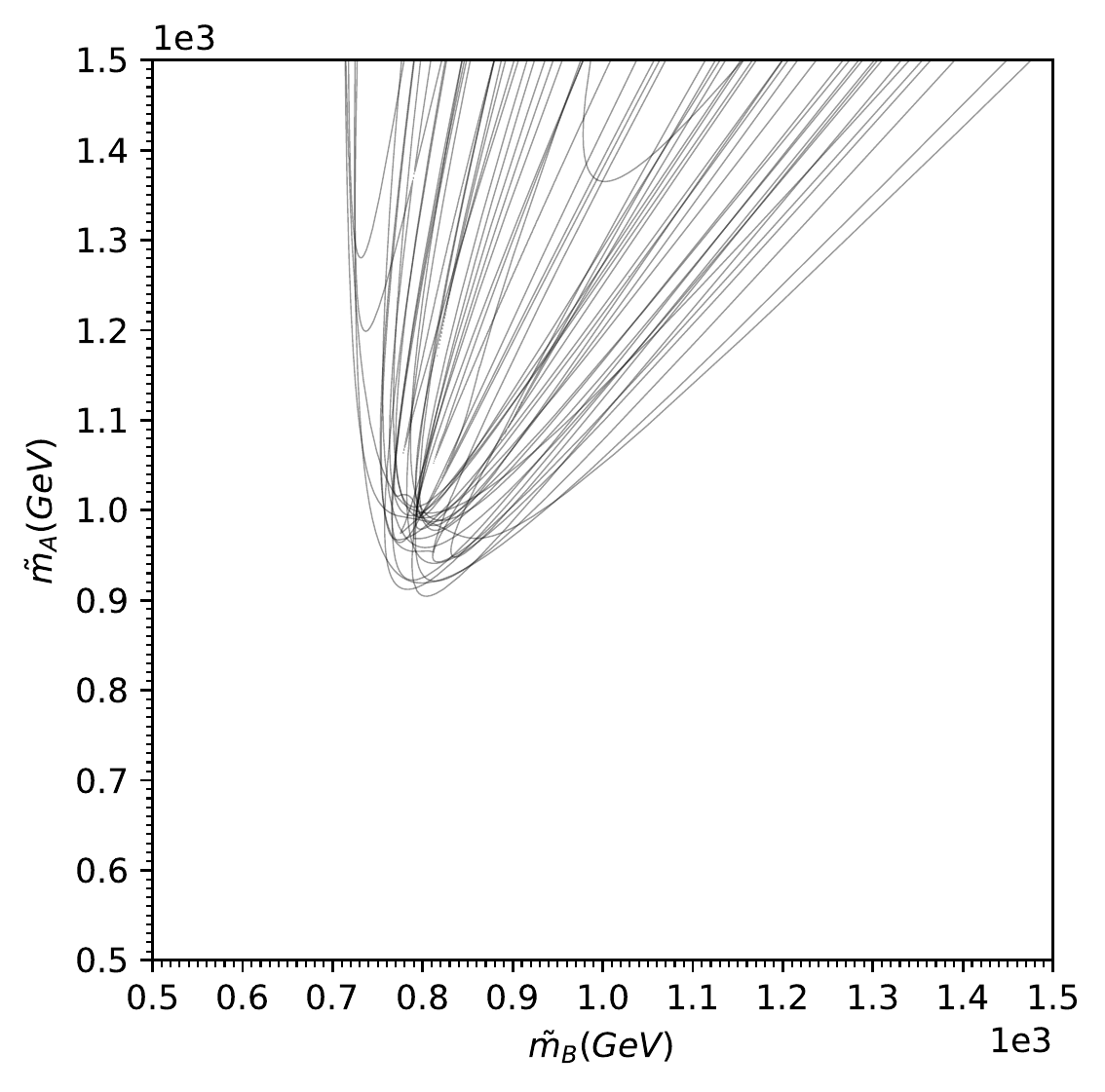}
 \includegraphics[width=.45\textwidth]{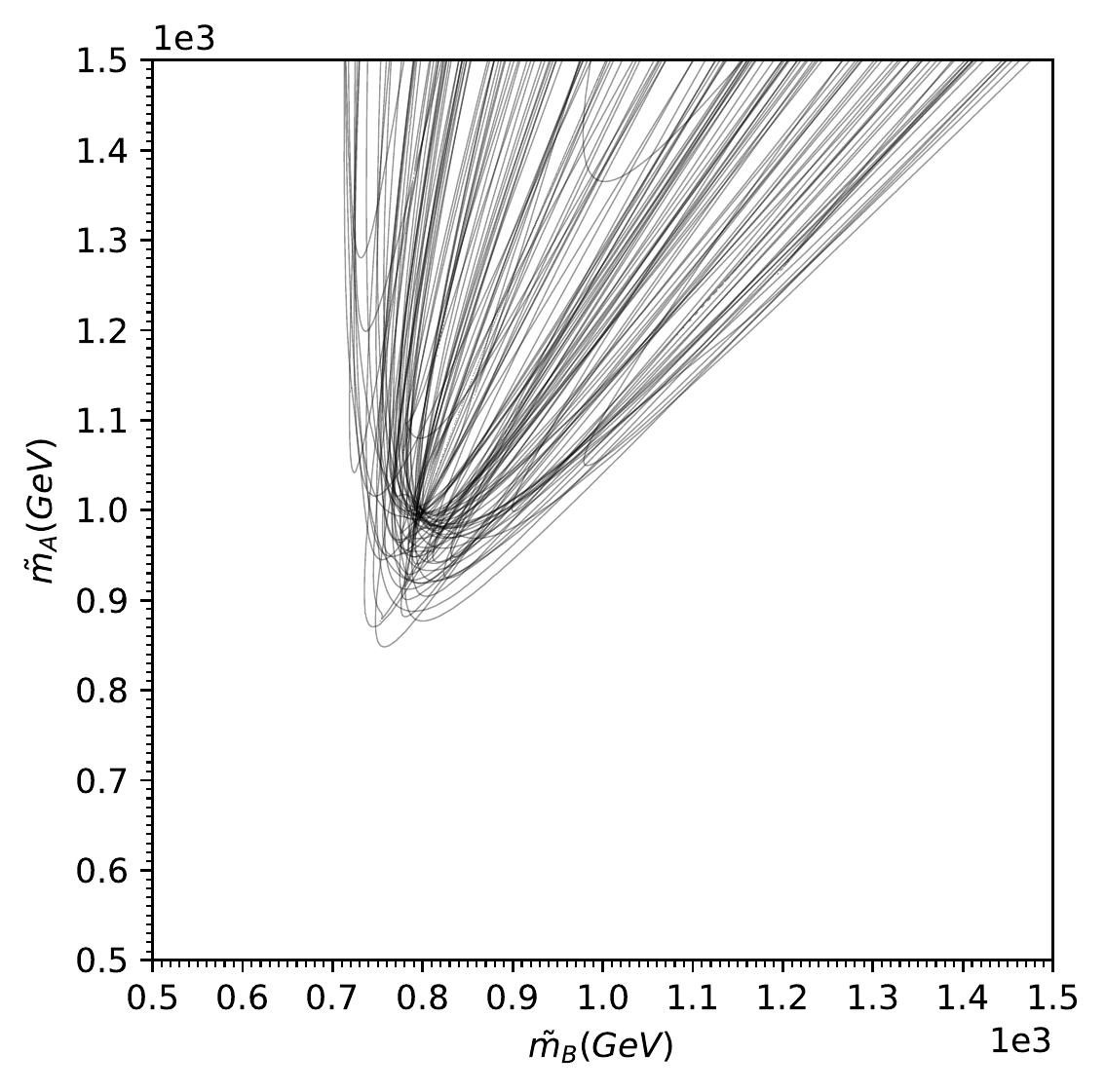}
 \caption{\label{fig:extremenesscurves} Degeneracy curves in the $(\tmB,\tmA)$ plane, for fixed $\tmC=\mC=700\GeV$.
 The plots are made with an increasing number of events: 5 events (top left), 10 events (top right), 
 20 events (bottom-left) and 50 events (bottom right). Each subsequent panel already contains the 
 events plotted previously, plus some new events.
 Note the gradual emergence of the focus point of boundary curves near the true mass $\mB=800$ GeV and $\mA=1000$ GeV. }
\end{figure}

\begin{figure}[t]
\centering
\includegraphics[width=0.49\textwidth]{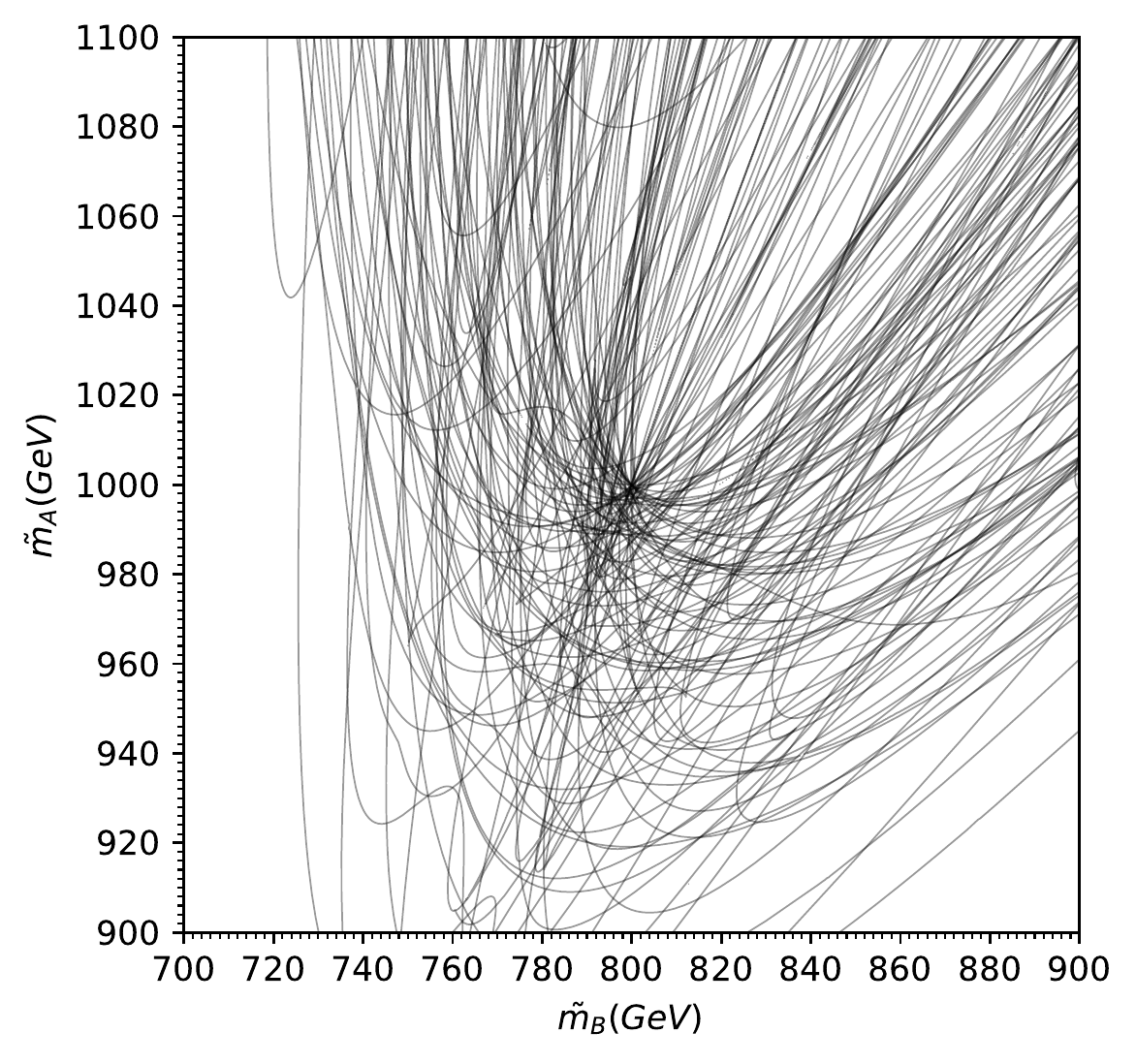}
\includegraphics[width=0.49\textwidth]{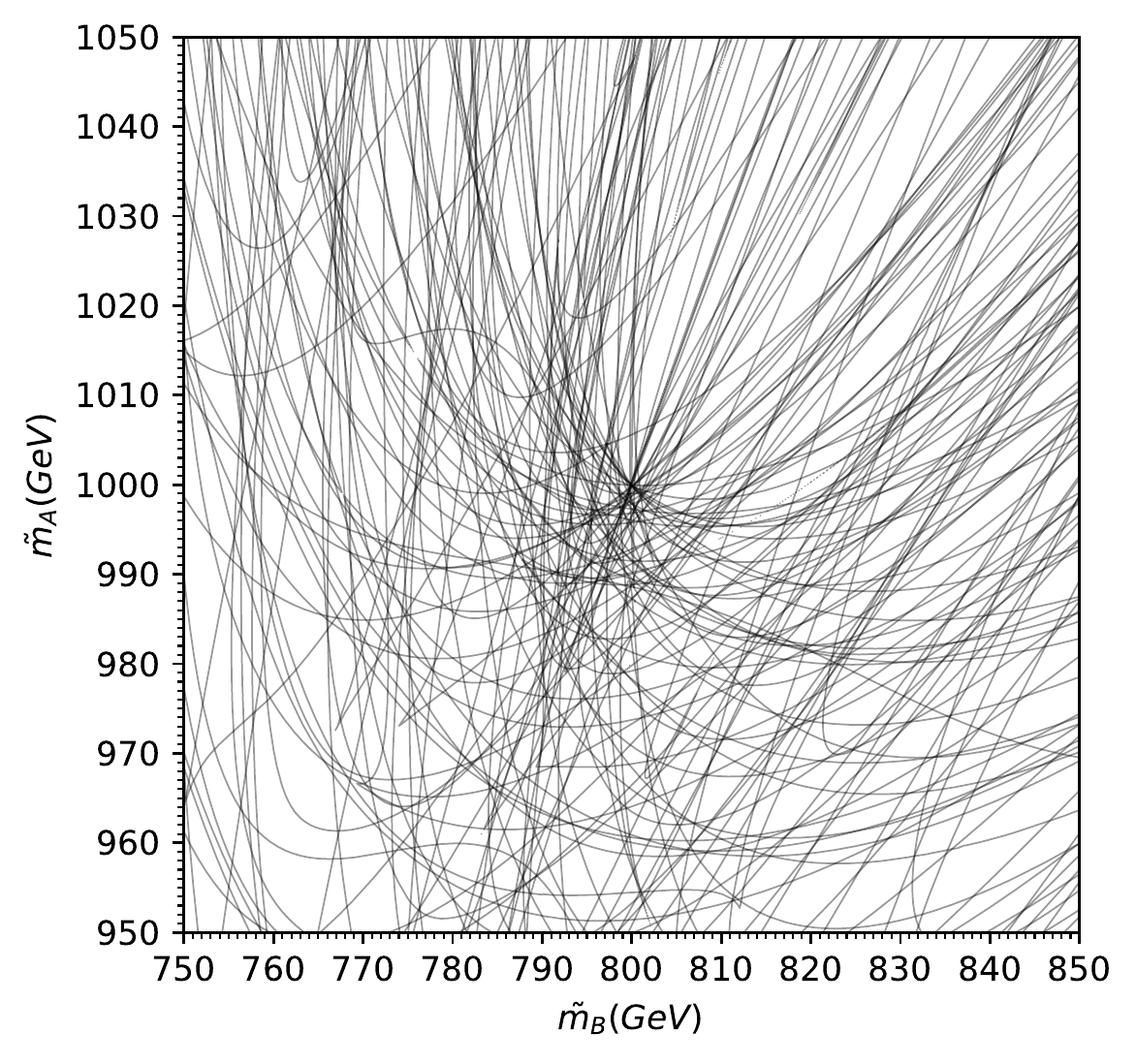}
\caption{\label{fig:alltongues700} 
The same as Fig.~\ref{fig:extremenesscurves}, but with 100 events and zoomed in by a factor of 5 (left panel) and a factor of 10 (right panel).}
\end{figure}

This lays out a straightforward mass determination method pictorially illustrated in Figs.~\ref{fig:extremenesscurves} and 
\ref{fig:alltongues700}.
We simply plot the degeneracy surfaces for {\em all} events in our sample, and look for a focus point where a maximal number of 
them either intersect or pass close by. Since it is difficult to visualize this in the full three dimensional mass space $(\tmA,\tmB,\tmC)$,
in Figs.~\ref{fig:extremenesscurves} and \ref{fig:alltongues700} we show results in the $(\tmA,\tmB)$ plane for fixed $\tmC=\mC=700\GeV$.
In Fig.~\ref{fig:extremenesscurves} we use the same plot range as in all of our previous figures, and gradually increase the number of events used for the plot
from 5 (top left panel) to 50 (bottom right panel). We observe that as we add more and more events, a focus point indeed
gradually emerges near the true mass values of $\mB=800$ GeV and $\mA=1000$ GeV.  Our method derives it name from this feature, 
since the degeneracy curves tend to ``focus" at the true mass point. The focus point is very easy to identify by eye, 
and in the next subsection we shall develop a procedure for quantifying the observed amount of ``focusing". 
Clearly, if we add even more events, the focus point should become even more pronounced, but unfortunately, the plot would become too cluttered. 
This is why when we further increase the number of events to 100 in Fig.~\ref{fig:alltongues700}, we zoom in on the region in the vicinity of the focus point.
The left (right) panel of Fig.~\ref{fig:alltongues700} uses a zoom factor of 5 (10) relative to the plot range in Fig.~\ref{fig:extremenesscurves}. 
We see that the focus point is very clear, and is in perfect agreement with the true values $\mB=800$ GeV and $\mA=1000$ GeV.
For example, if one were to estimate the values of $\mB$ and $\mA$ by simply eyeballing the plots in Fig.~\ref{fig:alltongues700}, 
the error would not be more than a couple of GeV! One should also keep in mind that the plots in Fig.~\ref{fig:alltongues700}, which were made only for
illustration, use only 100 events, while the actual LHC statistics is likely to be much higher.

\subsection{Measuring $m_B$ and $m_A$ for a given trial mass $\tilde m_C$}
\label{sec:fpmbma}

Hopefully by now the reader has been convinced by the money plots in Figs.~\ref{fig:extremenesscurves} and \ref{fig:alltongues700}
that the kinematic focus point method outlined in the previous Section~\ref{sec:idea} is a promising new technique for mass measurements.
In the remainder of this section, we shall discuss how to implement it in practice, as well as some practical challenges 
and how they might impact the sensitivity of the method.

Since the kinematic focus point method relies on the high density of degeneracy curves near the true mass point, we need to design a procedure
for quantifying this effect. Following the outline of Section~\ref{sec:fd}, 
let us first concentrate on the task of measuring $\mB$ and $\mA$ at a fixed value of $\tmC$, postponing 
the task of measuring $\mC$ until the next Section~\ref{sec:fpmc}. For this purpose, we need to plot one-dimensional degeneracy curves 
in the two-dimensional $(\tmB,\tmA)$ plane, as was done in Figs.~\ref{fig:extremenesscurves} and \ref{fig:alltongues700}.
The discussion from Sec.~\ref{subsec:extremeboundarydensity} showed that the relevant quantity in that case is the 
number of degeneracy curves per unit $\sqrt{\mathrm{area}}$, which is what we shall use in our analysis.
The area elements in the $(\tmB,\tmA)$ plane will be chosen to be $10\GeV\times10\GeV$ squares, 
and so the resulting unit for the density of degeneracy curves will be $(10\GeV)^{-1}$.
For the remainder of this subsection, the test mass $\tmC$ will be set to the true value of $700\GeV$.

\begin{figure}[t]
 \centering
 \includegraphics[width=.7\textwidth]{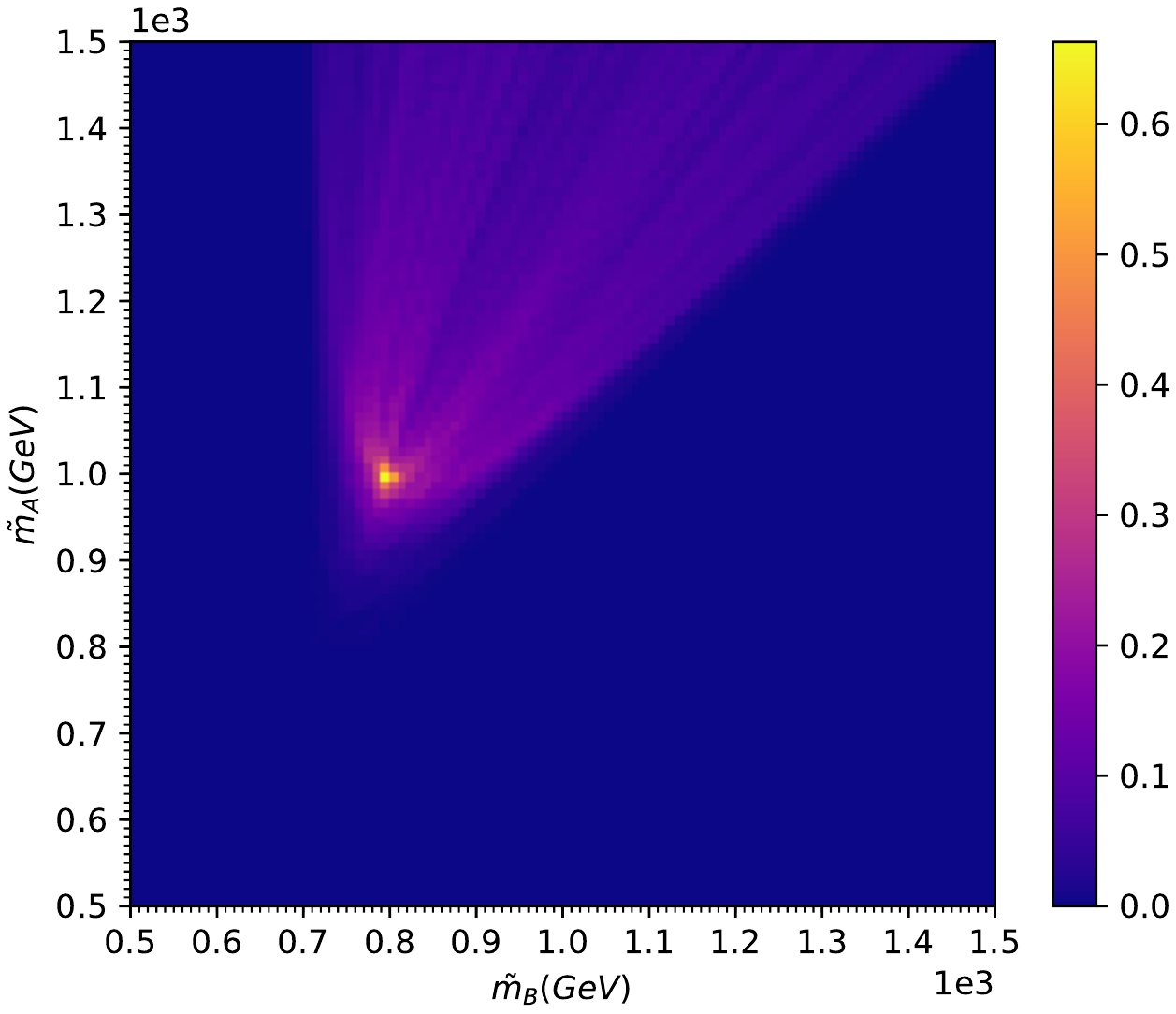}
 \caption{\label{fig:moneyplot1} A heat map of the density of degeneracy curves in the $(\tmB,\tmA)$ plane  for $\tmC=\mC=700\GeV$. 
 The density is displayed in units of fraction of events per $10\GeV$.
 Note the sharp peak at the true values $\mB=800\GeV$, $\mA=1000\GeV$ --- over $60\%$ of the signal events 
 have a degeneracy curve passing through a $10\GeV\times10\GeV$ square around the true mass point.}
\end{figure}

\begin{figure}[t]
 \centering
 \includegraphics[height=.25\textwidth]{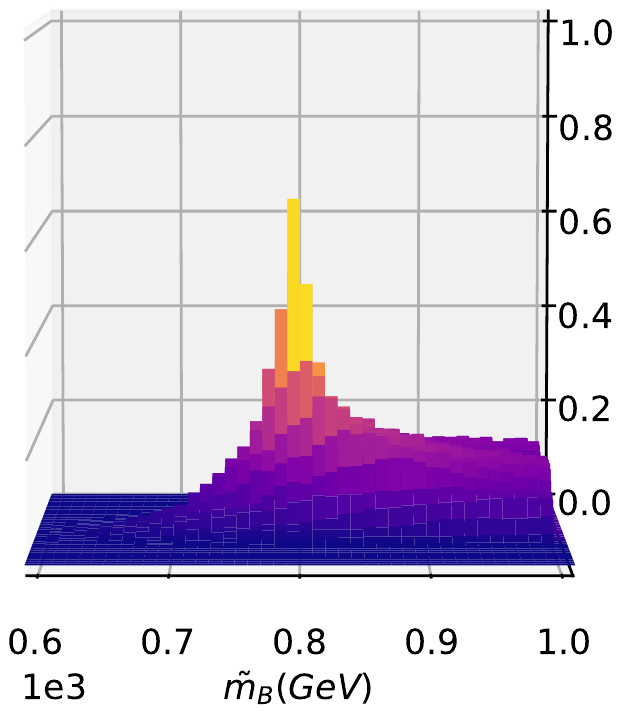}
 \includegraphics[height=.25\textwidth]{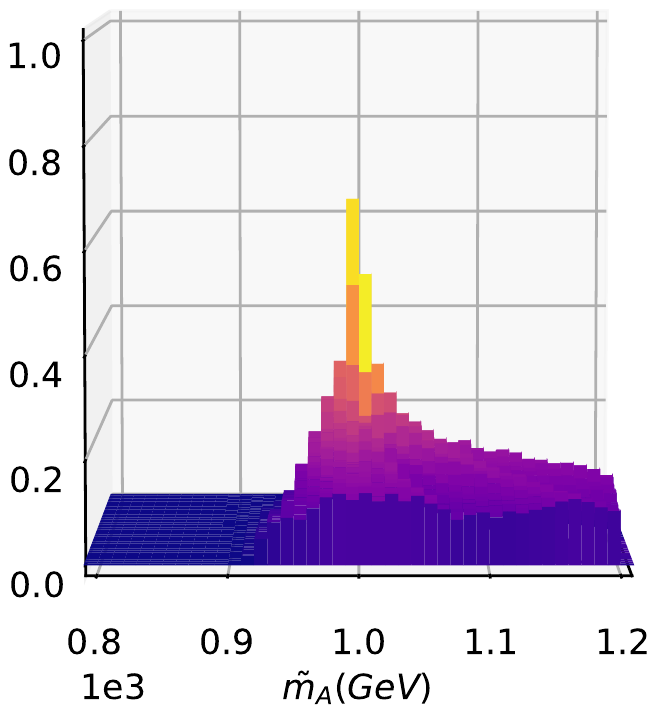}
 \includegraphics[height=.25\textwidth]{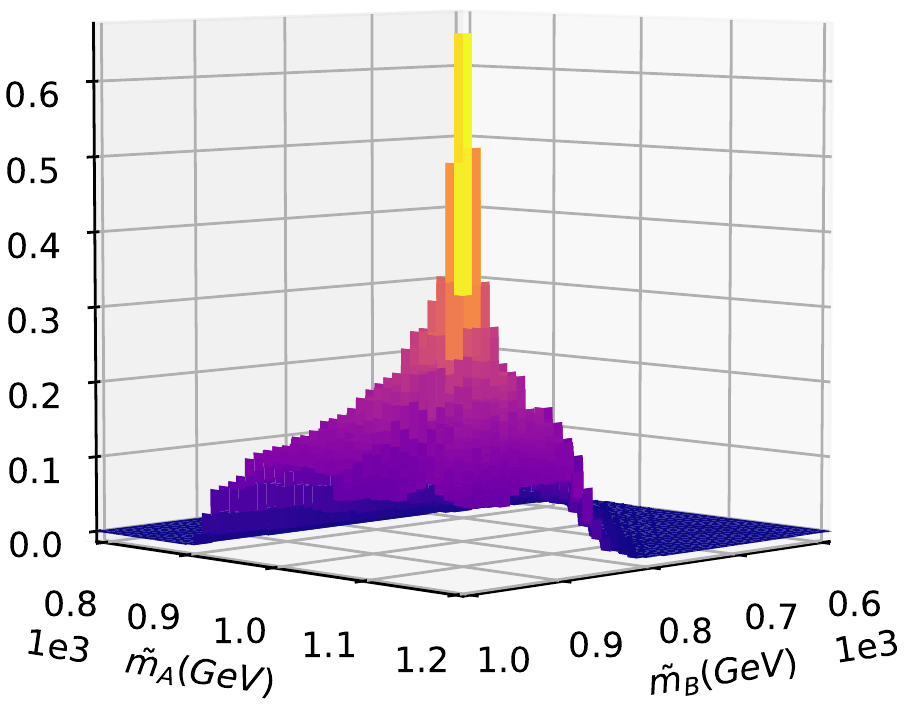}
 \caption{\label{fig:moneyplot2} The same as Fig~\ref{fig:moneyplot1}, 
 except densities are plotted as a 3D histogram instead of a heatmap. 
 The three figures show the 3D histogram from different angles. 
 The peak is at the true mass $\mB=800\GeV, \mA=1000\GeV$.}
 \end{figure}

We are now ready to study the density of degeneracy curves in the $(\tmB,\tmA)$ plane.
The result is shown as a heat map in Fig.~\ref{fig:moneyplot1} and as 3D histograms from different viewpoints in Fig.~\ref{fig:moneyplot2}. 
The density of degeneracy curves is displayed in units of fraction of events per $10\GeV$, i.e., instead of plotting the total number of curves passing through a given 
$10\GeV\times10\GeV$ square, we plot the fraction of events in the sample which have a degeneracy curve passing through that square.
By normalizing to a fraction, our results become insensitive to the statistics used to make the plots
(in this case, Figs.~\ref{fig:moneyplot1} and \ref{fig:moneyplot2} were made with 1000 events).
The figures show that the density has a sharp peak at the true mass ($\mB=800\GeV, \mA=1000\GeV$). 
The peak is actually very well pronounced: note that over $60\%$ of the events in the sample 
have one of their degeneracy curves passing through a $10\GeV\times10\GeV$ square around the true mass point.
This confirms that our method is really effective in finding $\mB$ and $\mA$, given $\tmC$.
Note the advantages of our method over the solvability method from Sec.~\ref{sec:fd} ---
first, we do not encounter the flat direction of $100\%$ solvability seen in Figs.~\ref{fig:solvabilityhist},  \ref{fig:solvabilityhist600} and \ref{fig:solvabilityhist800},
and as a result we do not need any additional experimental information (from kinematic endpoints or otherwise);
and second, we are measuring a sharp peak structure which is centered on the true values of the particle masses.

\subsection{Measuring $m_C$}
\label{sec:fpmc}

\begin{figure}[t]
 \centering
 \includegraphics[width=.49\textwidth]{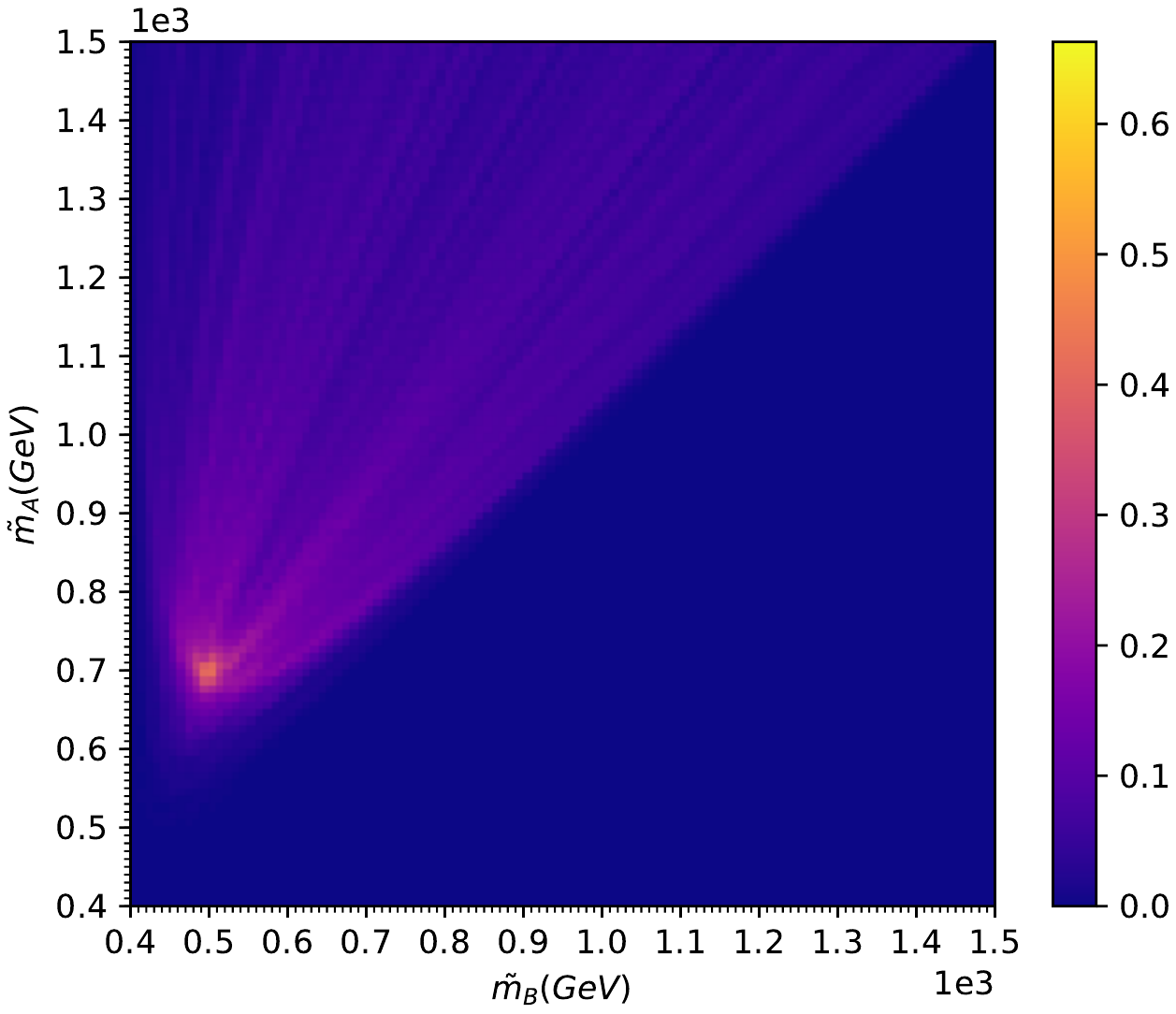}
 \includegraphics[width=.49\textwidth]{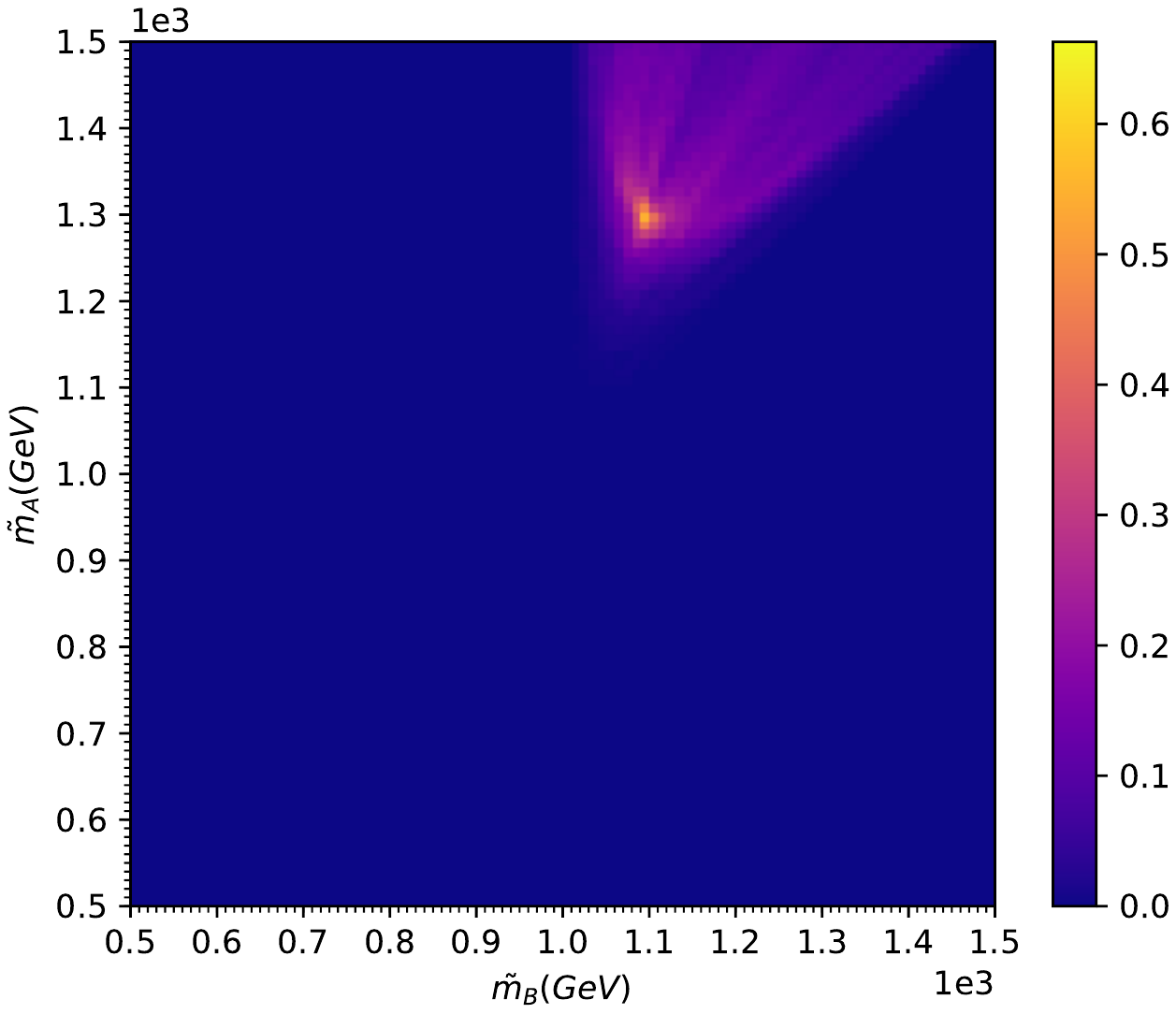}
 \caption{\label{fig:moneyplot3} The same as Fig.~\ref{fig:moneyplot1}, but for
 $\tmC=400\GeV$ (left panel) and $\tmC=1000\GeV$ (right panel). }
\end{figure}

As in Sec.~\ref{subsec:mc}, we now turn to the more difficult task of measuring $\mC$.
For this purpose, we repeat the exercise from the previous subsection, only this time with different values of $\tmC$.
Heat maps analogous to the one in Fig.~\ref{fig:moneyplot1} are shown in
the left and right panel of Fig.~\ref{fig:moneyplot3} for two representative values, $\tmC=400\GeV$ and $\tmC=1000\GeV$, respectively.
We observe that each of the two plots in Fig.~\ref{fig:moneyplot3} exhibits its own density peak in the $(\tmB,\tmA)$ plane. 
As expected, the location of the peak tracks the value of $\tmC$ --- for lower (higher) values of $\tmC$, the
peak is located at lower (higher) values of $\tmB$ and $\tmA$. From that point of view, there  
is no qualitative difference in the heat maps at different values of $\tmC$. 
However, there is a quantitative difference: the height of this peak, i.e, the maximum density in the $(\tmB,\tmA)$ plane 
for a given value of $\tmC$, is different for different values of $\tmC$. 
The maximum height is expected to occur at the true value of $\tmC=\mC$.
\begin{figure}[t]
 \centering
 \includegraphics[width=.6\textwidth]{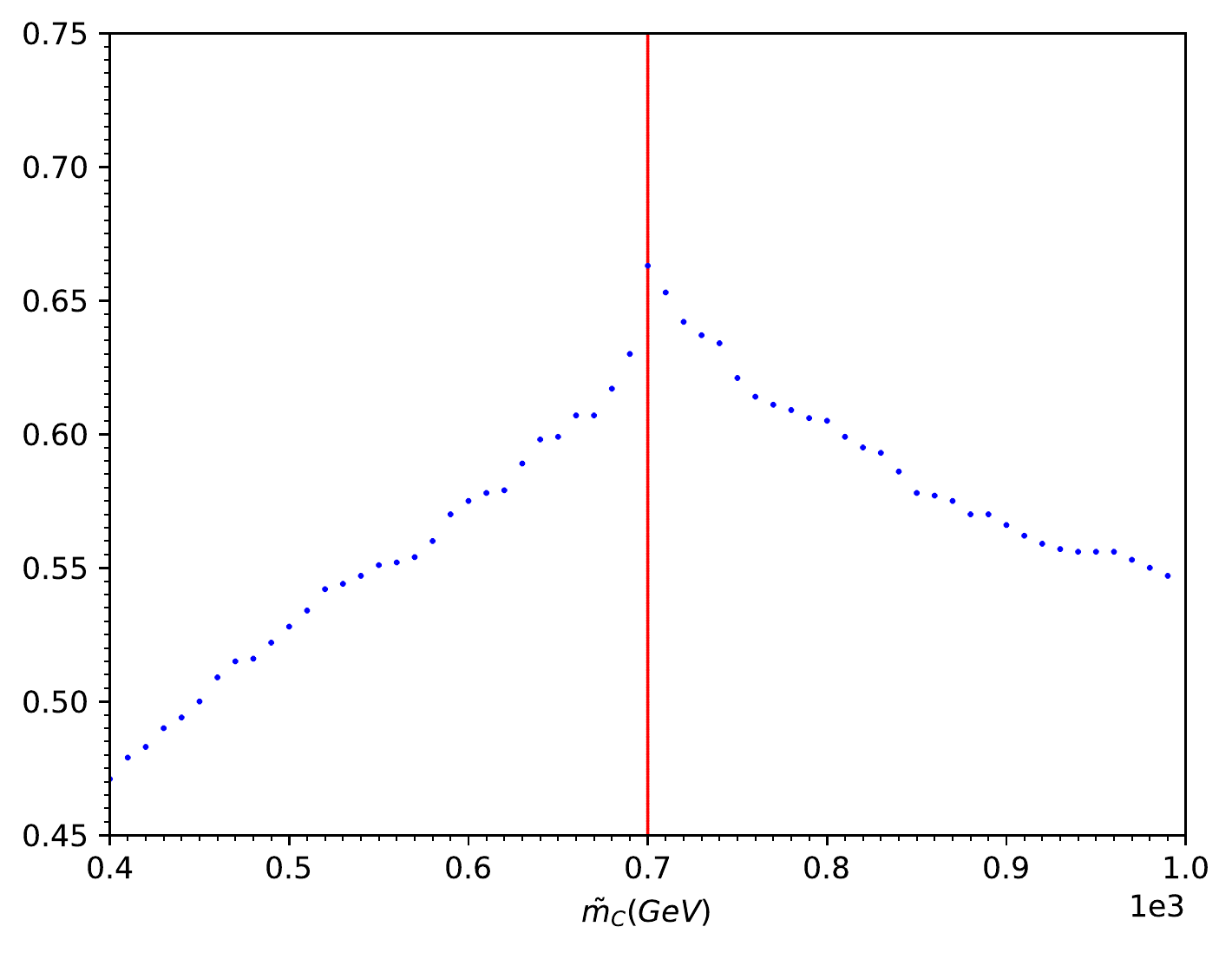}
 \caption{\label{fig:mCmeasurement} Plot of the maximum density of degeneracy curves (in units of fraction/($10\GeV$)) for different fixed values of $\tmC$. 
 For each value of $\tmC$, the maximization is done in the $(\tmB,\tmA)$ plane. The peak value in the plot is observed at  the true value of $\mC=700\GeV$ marked with the vertical red line. }
\end{figure}
While this expectation is consistent with the results in Figs.~\ref{fig:moneyplot1} and \ref{fig:moneyplot3},
we quantify it further in Fig.~\ref{fig:mCmeasurement}, which shows a plot of the maximum density 
of degeneracy curves in the $(\tmB,\tmA)$ plane (after fixing the value of $\tmC$), as a function of $\tmC$. 
As before, the density is plotted in units of fraction/($10\GeV$). Fig.~\ref{fig:mCmeasurement} nicely confirms that
the density peaks at the true value of $\mC=700$ GeV (marked with the vertical red line).
As already discussed in Sec.~\ref{subsec:mc}, the $\mC$ measurement is rather challenging --- 
in our case this is evidenced by the fact that the peak in Fig.~\ref{fig:mCmeasurement}
is not as sharp as the peak in Figs.~\ref{fig:moneyplot1} and \ref{fig:moneyplot3}. Nevertheless,
Fig.~\ref{fig:mCmeasurement} confirms that in principle our method can be used to find $\mC$ as well.

\subsection{The impact of the detector resolution}
\label{sec:resolution}

Until now, all of our studies were done at the parton level and in the narrow width approximation, 
without smearing from either detector resolution or particle width effects.
We shall now investigate the size of those effects. In this subsection we shall first
add the effect of the detector resolution. We shall consider the typical scenario where
the particles $\a{i}$ and $\b{i}$ in the event topology of Fig.~\ref{fig:feynmandiag}
are bottom quarks and leptons, respectively. Correspondingly, we shall apply smearing to the energies of the two $b$-jets 
according to typical LHC resolutions \cite{Khachatryan:2016kdb}: for jet $p_T$ up to 
$\{10,20,30,50,100,400,1000\}$ GeV, the $p_T$-dependent energy resolution parameter is
$40\%, 28\%, 19\%, 13\%, 10\%, 6\%, 5\%$. 

\begin{figure}[t]
\centering
\includegraphics[width=0.32\textwidth]{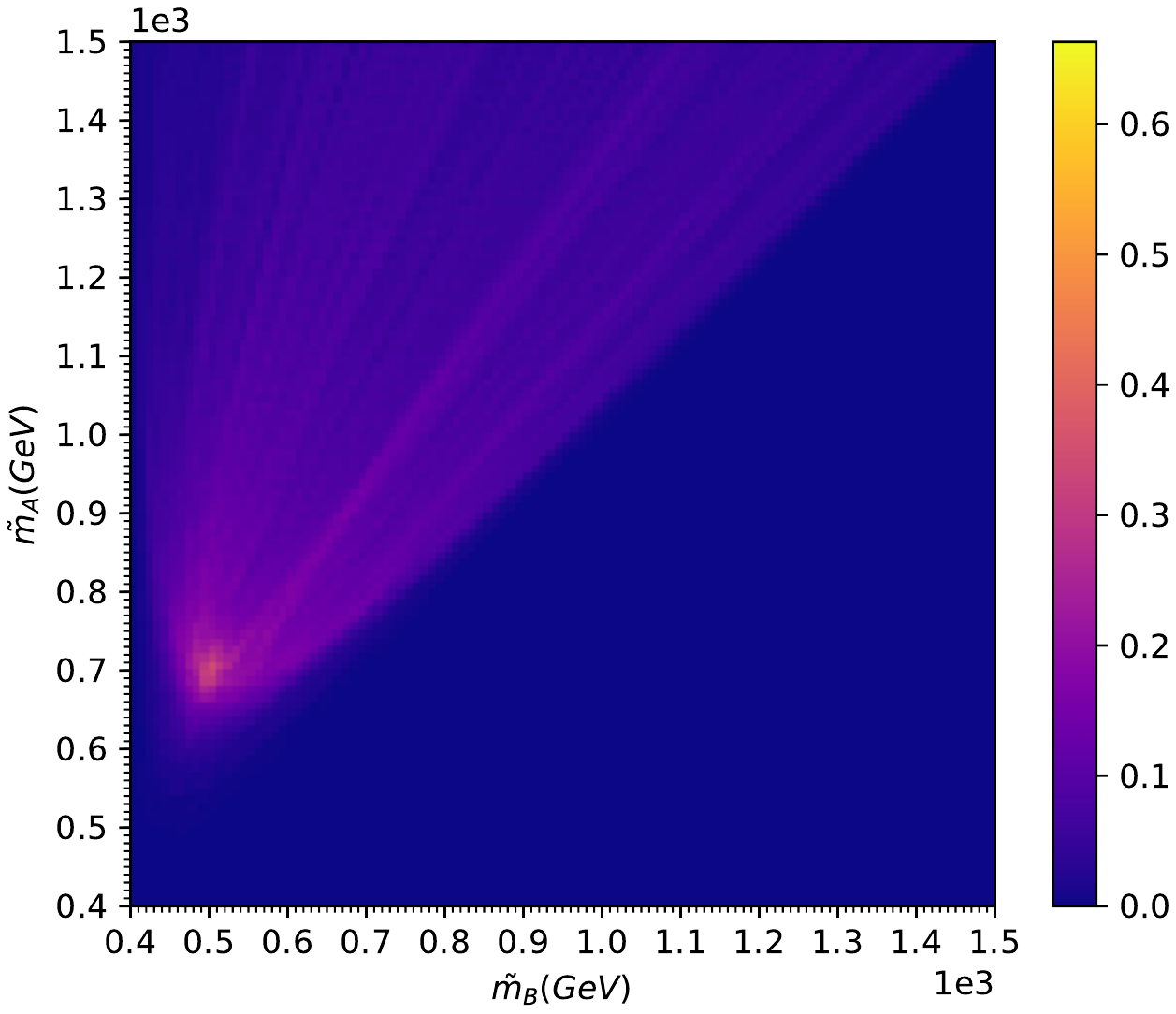}
\includegraphics[width=0.32\textwidth]{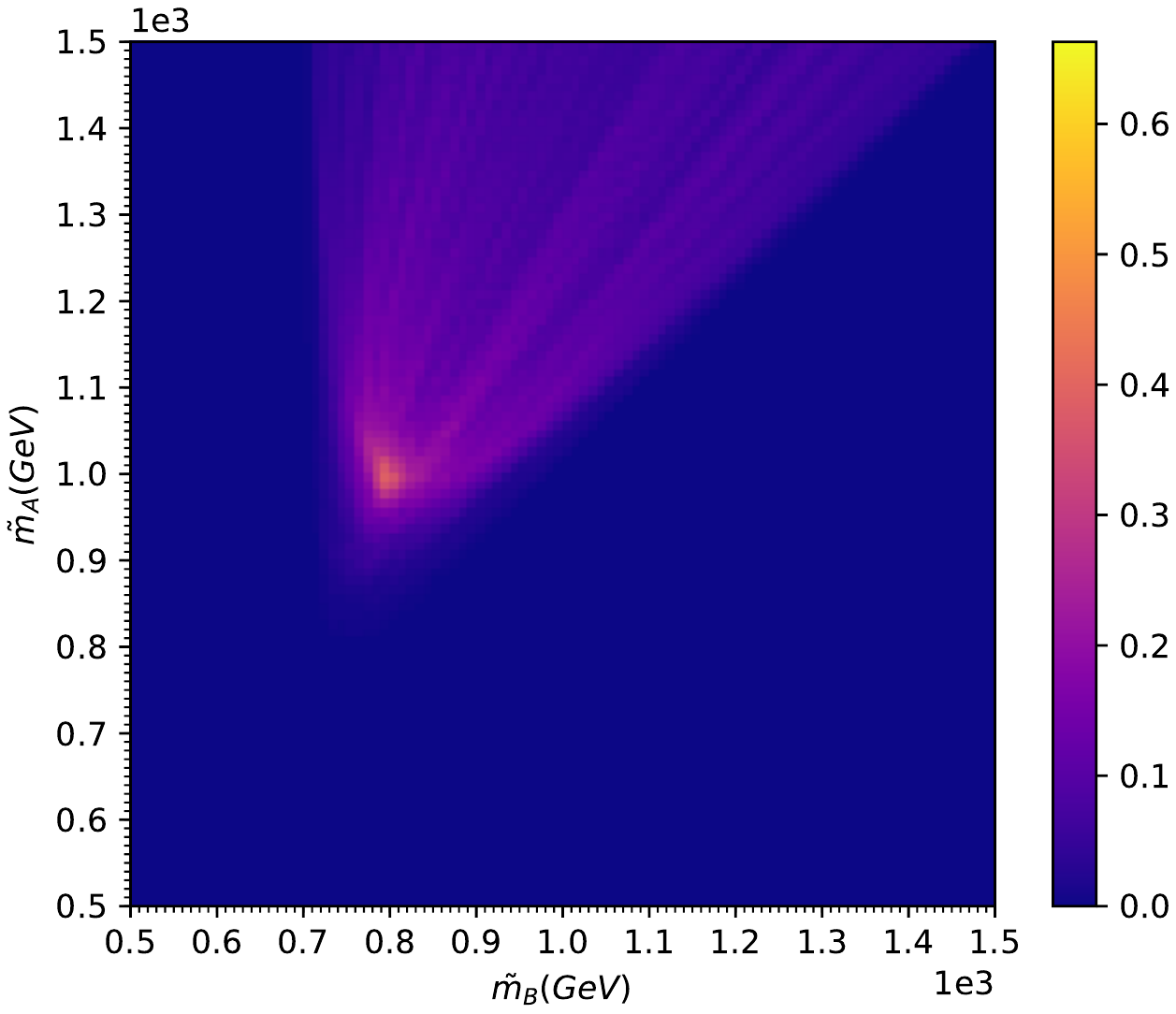}
\includegraphics[width=0.32\textwidth]{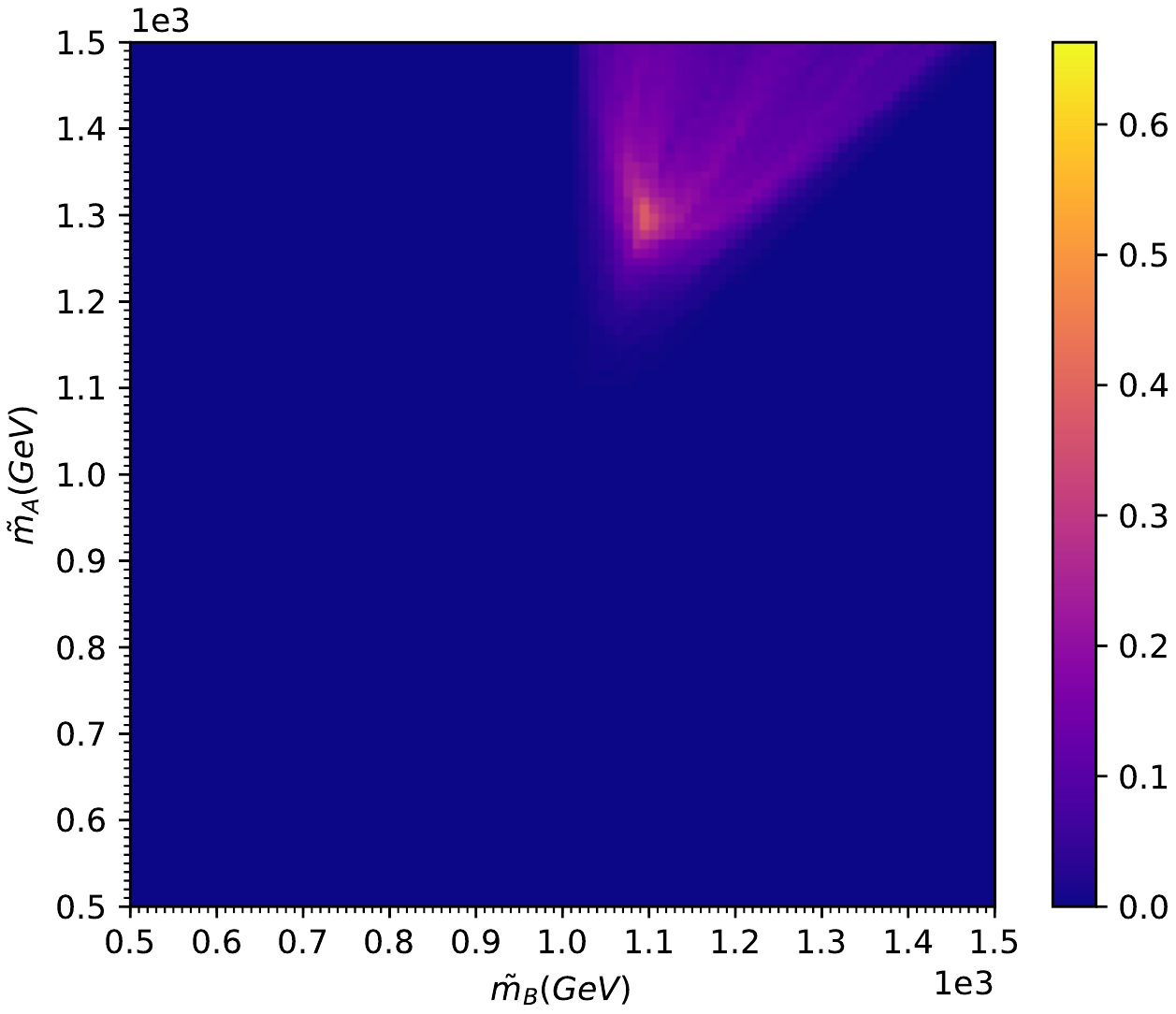}
\caption{\label{fig:focuszoomsmear} The same as Figs.~\ref{fig:moneyplot1} and \ref{fig:moneyplot3}, but including the effects of jet smearing.
The left (middle, right) panel shows results for fixed $\tmC=400\GeV$ ($\tmC=700\GeV$, $\tmC=1000\GeV$).}
\end{figure}

\begin{figure}[t]
 \centering
 \includegraphics[width=.6\textwidth]{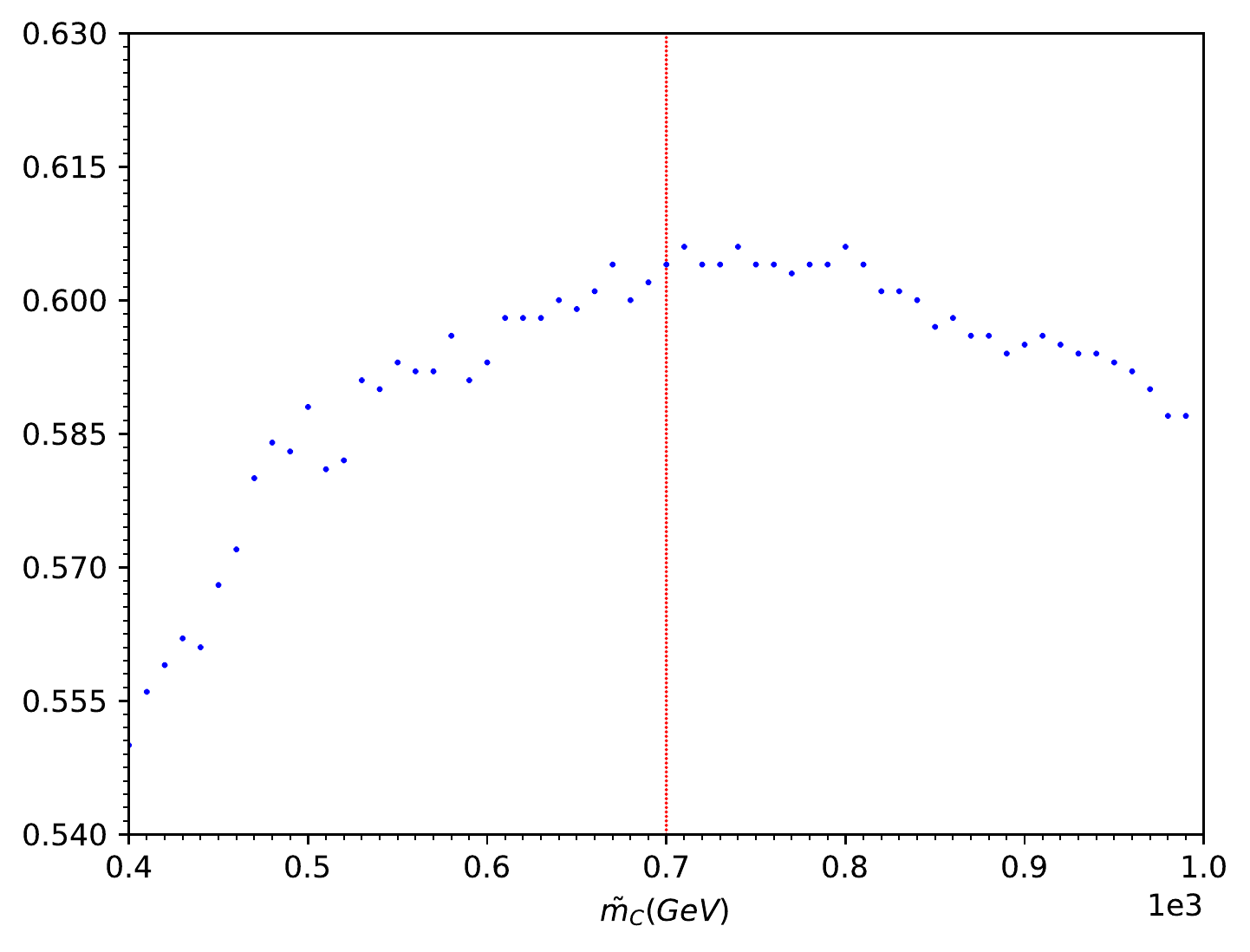}
 \caption{\label{fig:mCmeasurementsmear} The same as Fig.~\ref{fig:mCmeasurement}, but including the effects of jet smearing
 and counting events in a larger ($30\times30\GeV$) collection bin.}
\end{figure}

Figs.~\ref{fig:focuszoomsmear} and \ref{fig:mCmeasurementsmear} show the results analogous to
Figs.~\ref{fig:moneyplot1}, \ref{fig:moneyplot3} and \ref{fig:mCmeasurement}, only now with the detector resolution included.
As expected, the detector effects tend to wash out the peak structures a little bit.
%Nevertheless, the peaks are still discernible, and more importantly, are still located at the proper values of the masses.
Nevertheless, the peaks in Fig.~\ref{fig:focuszoomsmear} (which shows measurements of $m_A$ and $m_B$ for fixed $\tilde m_C$)
are still pronounced, and more importantly, the peak found for $\tilde m_C=m_C$ is still located at the proper values of the masses.
Fig.~\ref{fig:mCmeasurementsmear} shows the measurement of $m_C$ (in the presence of detector effects), which
was already a difficult task even at the gen-level, see Fig.~\ref{fig:mCmeasurement}.

\subsection{A Standard Model example: dilepton $t\bar{t}$ events}
\label{sec:ttbar}

Most of the mass measurement methods applicable to LHC events with missing transverse energy
were developed with some kind of new physics model in mind, e.g., supersymmetry \cite{Matchev:2019sqa}. 
However, experimenters who wish to test the kinematic focus point method in real data do {\em not}
have to wait for the discovery of any new physics --- the event topology of Fig.~\ref{fig:feynmandiag}
is already present in the LHC data in the form of Standard Model dilepton $t\bar{t}$ events, 
for which the available statistics is enormous. Therefore, it is worth supplementing our previous results with 
a study showcasing the method for the dilepton $t\bar{t}$ channel, and this is what we shall do in the current subsection.

\begin{figure}
 \centering
 \includegraphics[width=.49\textwidth]{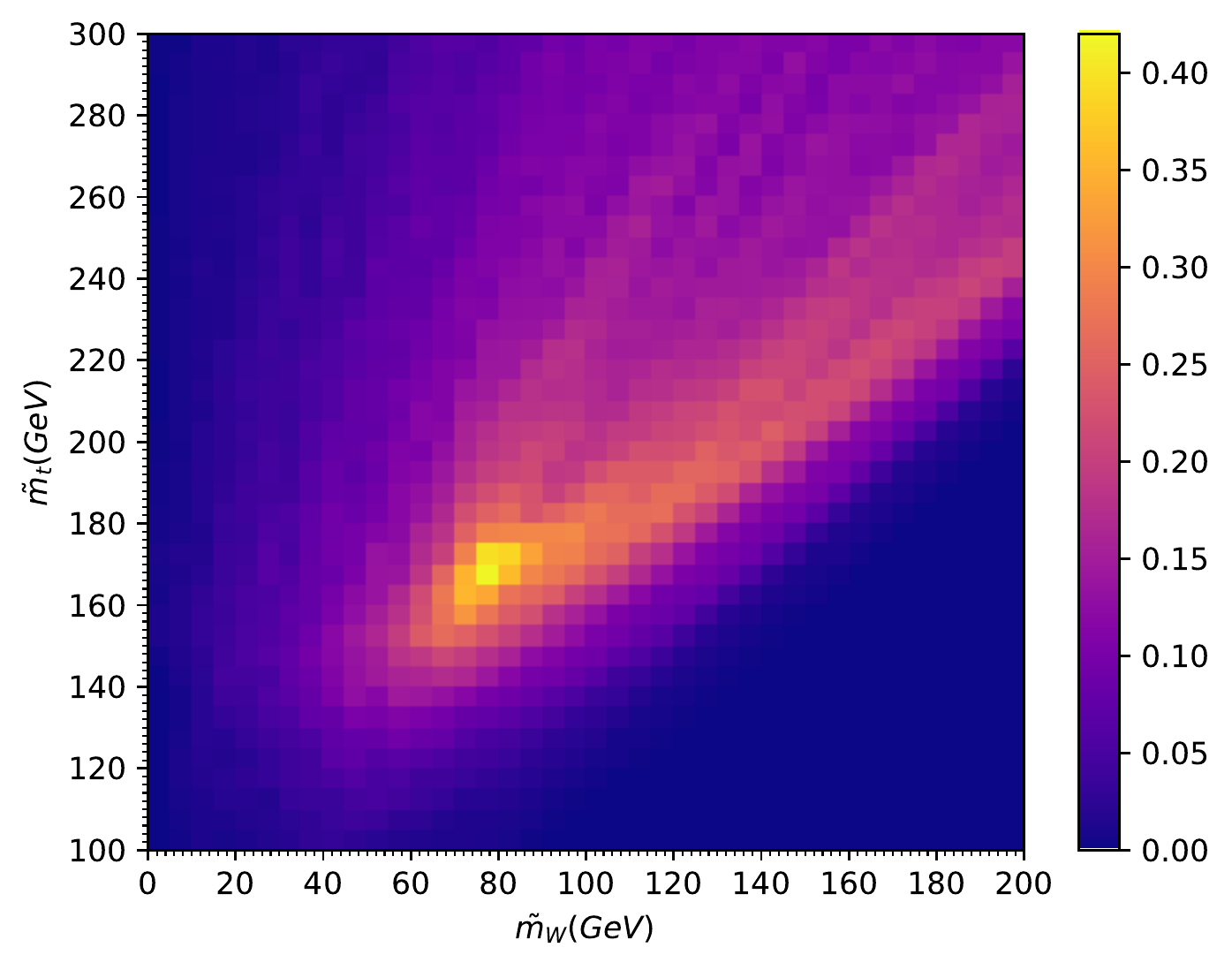}
 \includegraphics[width=.49\textwidth]{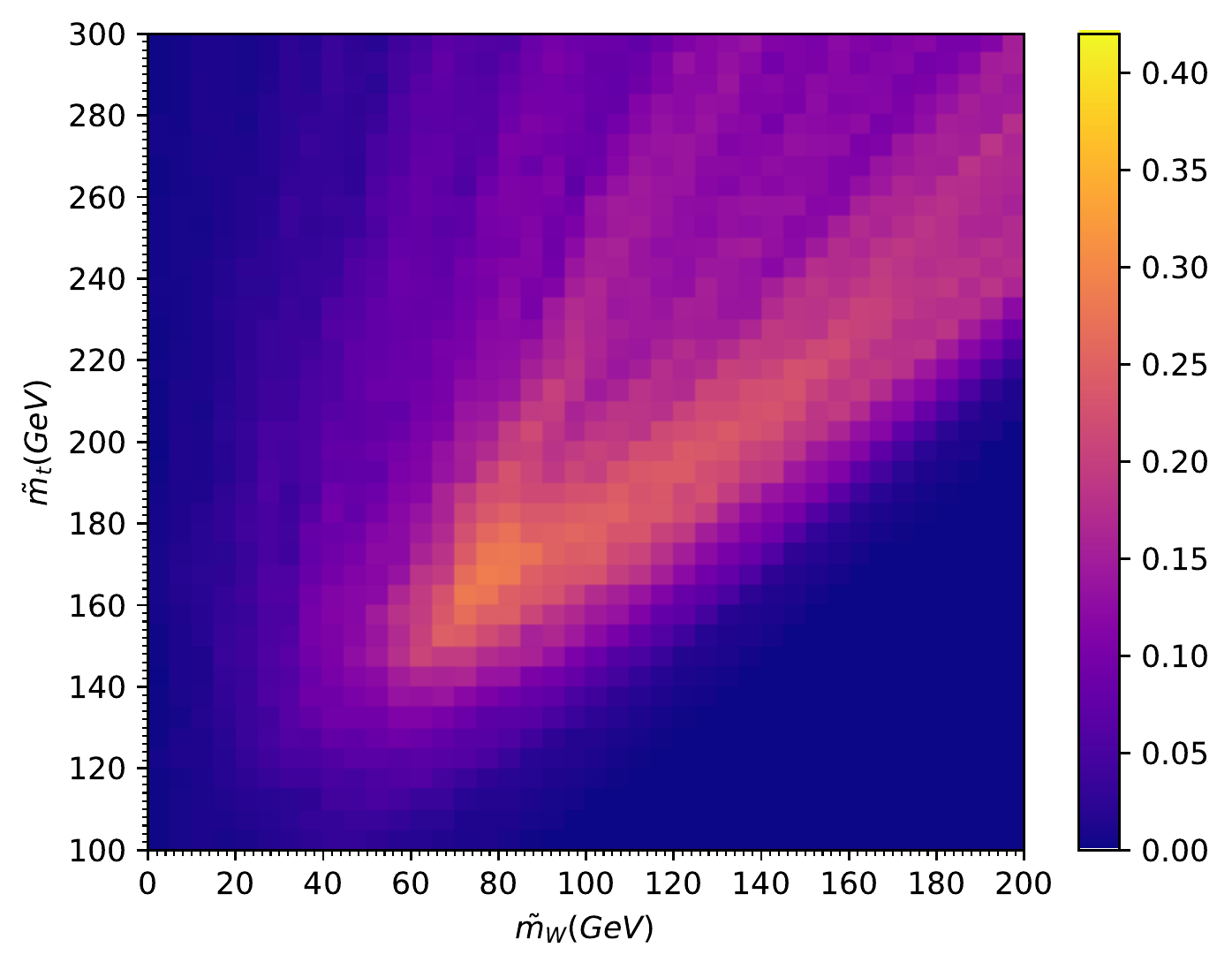}
 \caption{\label{fig:moneyplotttbar} Density of degeneracy curves for Standard Model dilepton $t\bar{t}$ events 
 in the $(\tilde m_{W}, \tilde m_t)$ plane for $\tilde m_\nu= m_\nu = 0$ GeV, 
 without detector smearing (left) and with detector smearing (right). Unlike Figs.~\ref{fig:moneyplot1},
 \ref{fig:moneyplot3} and \ref{fig:focuszoomsmear}, the density units here are fraction of events per $5\GeV$,
 i.e., the bins are $5\GeV \times 5\GeV$. }
\end{figure}

We generate Standard Model $t\bar{t}$ events with {\sc Madgraph} \cite{Alwall:2011uj}, 
using the proper widths for the top quark and the $W$ boson.
We select dilepton events in which both top quarks have decayed leptonically, and repeat the 
kinematic focus point analysis from Section~\ref{sec:fpmbma}, setting the trial neutrino mass $\tilde m_\nu$ to zero. 
We then consider the plane spanned by the trial masses for the $W$ boson ($\tilde m_W$) and for the top quark ($\tilde m_t$)
and compute the density of the degeneracy curves. The heat map analogous to Fig.~\ref{fig:moneyplot1}
is shown in Fig.~\ref{fig:moneyplotttbar}, where the left panel depicts the parton level result, while in the 
right panel we include the effects from the detector smearing as in Section~\ref{sec:resolution}.

As anticipated, Fig.~\ref{fig:moneyplotttbar} reveals a peak structure near the true masses $m_W=80\GeV$ and $m_t=173\GeV$.
The peak is especially pronounced at the parton level (left panel)  --- over $40\%$ of the events 
 have a degeneracy curve passing through a $5\GeV\times5\GeV$ square nearby. 
The detector resolution does tend to wash out the peak structure in the right panel, but a maximum is still visible.
This suggests that the kinematic focus point method could allow the 
simultaneous independent measurement of the top and $W$-boson masses
in the spirit of Refs.~\cite{Burns:2008va,Chatrchyan:2013boa}.

\section{Conclusions and Outlook}
\label{sec:conclusions}

In this paper we propose a new approach to mass measurements in events with missing energy, called the kinematic focus points method. 
The method derives from the study of the solvability of the kinematic constraints which are present in a given event topology 
\cite{Nojiri:2003tu,Kawagoe:2004rz,Cheng:2007xv,Cheng:2008mg,Cheng:2008hk,Cheng:2009fw,Webber:2009vm,Barr:2009jv}.
Using the dilepton $t\bar{t}$ topology as our example, we first critically examined the solvability method for measuring masses,
and its relation to the measurements of kinematic endpoints.
We then showed that any given event divides the relevant three-dimensional mass parameter space 
into regions with a certain number of (pairs of) real solutions for the unknown invisible momenta. 
The surface boundaries between those regions are where the solutions become degenerate, 
and we called those surfaces ``degeneracy boundaries". We illustrated and studied the shape of the degeneracy boundaries,
as well as their distribution throughout the mass parameter space. In particular, we singled out a special
class of events, extreme events, for which the true mass point lies on a degeneracy boundary.

Our main results were:
\begin{itemize}
\item Extreme events are very efficient in restricting the allowed mass parameter space.
\item Extreme events are abundant in realistic data samples, due to singularities in the phase space distribution.
\item This abundance of extreme events can be harnessed by drawing the degeneracy boundaries in the mass parameter space over the full event sample. 
The boundaries tend to focus at the true mass point and this property can be used to identify the true masses.
\item our study of the solvability mass measurement method revealed a flat direction of nearly 100\% 
solvability, which can be lifted only by adding additional information.
\end{itemize}
The kinematic focus points method has similarities to traditional mass-bump methods, thus allowing for
data driven estimation of the relevant backgrounds.

The method also represents an improvement over the traditional kinematic endpoint techniques.
\begin{figure}
 \centering
 \includegraphics[width=.49\textwidth]{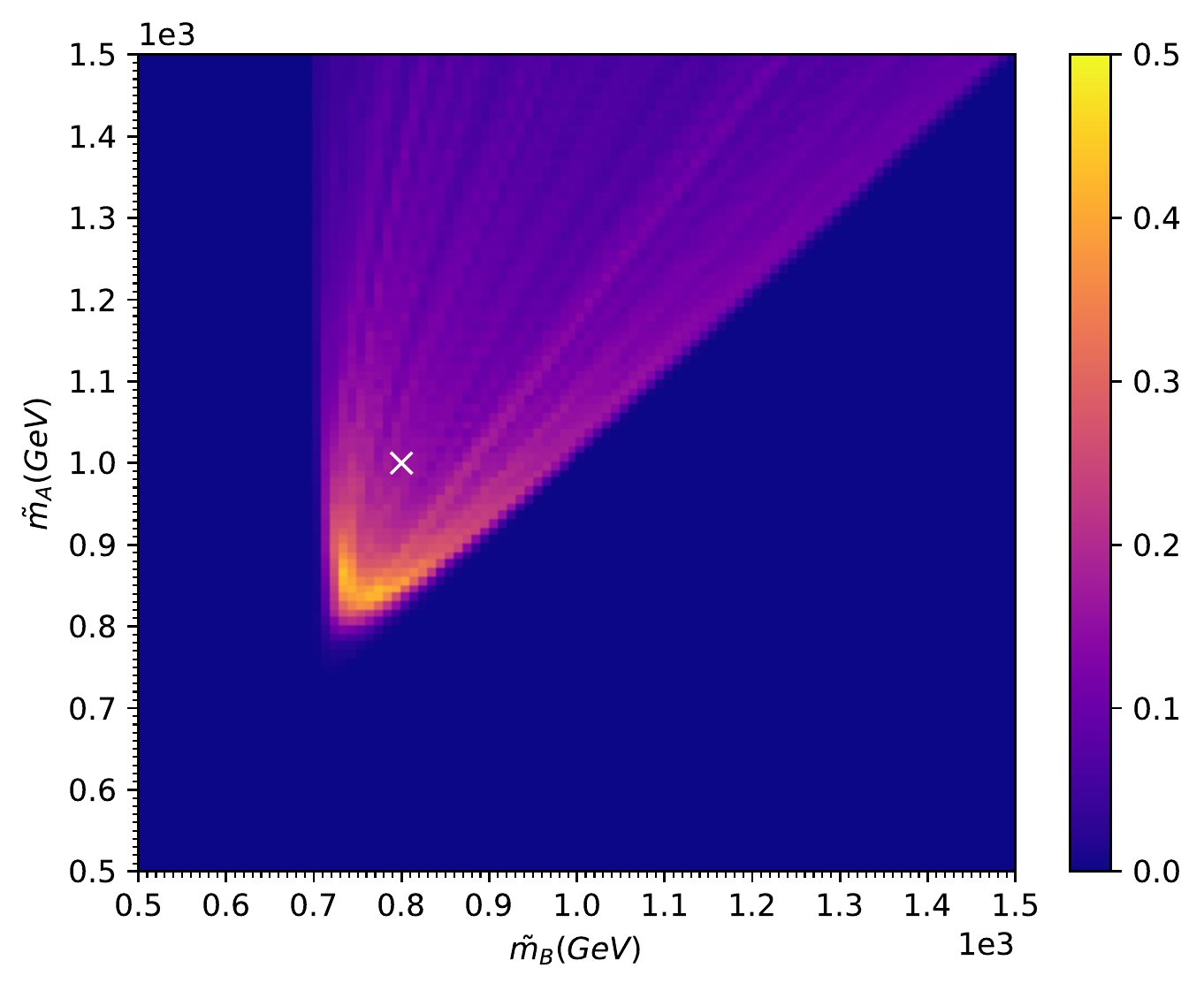}
 \includegraphics[width=.49\textwidth]{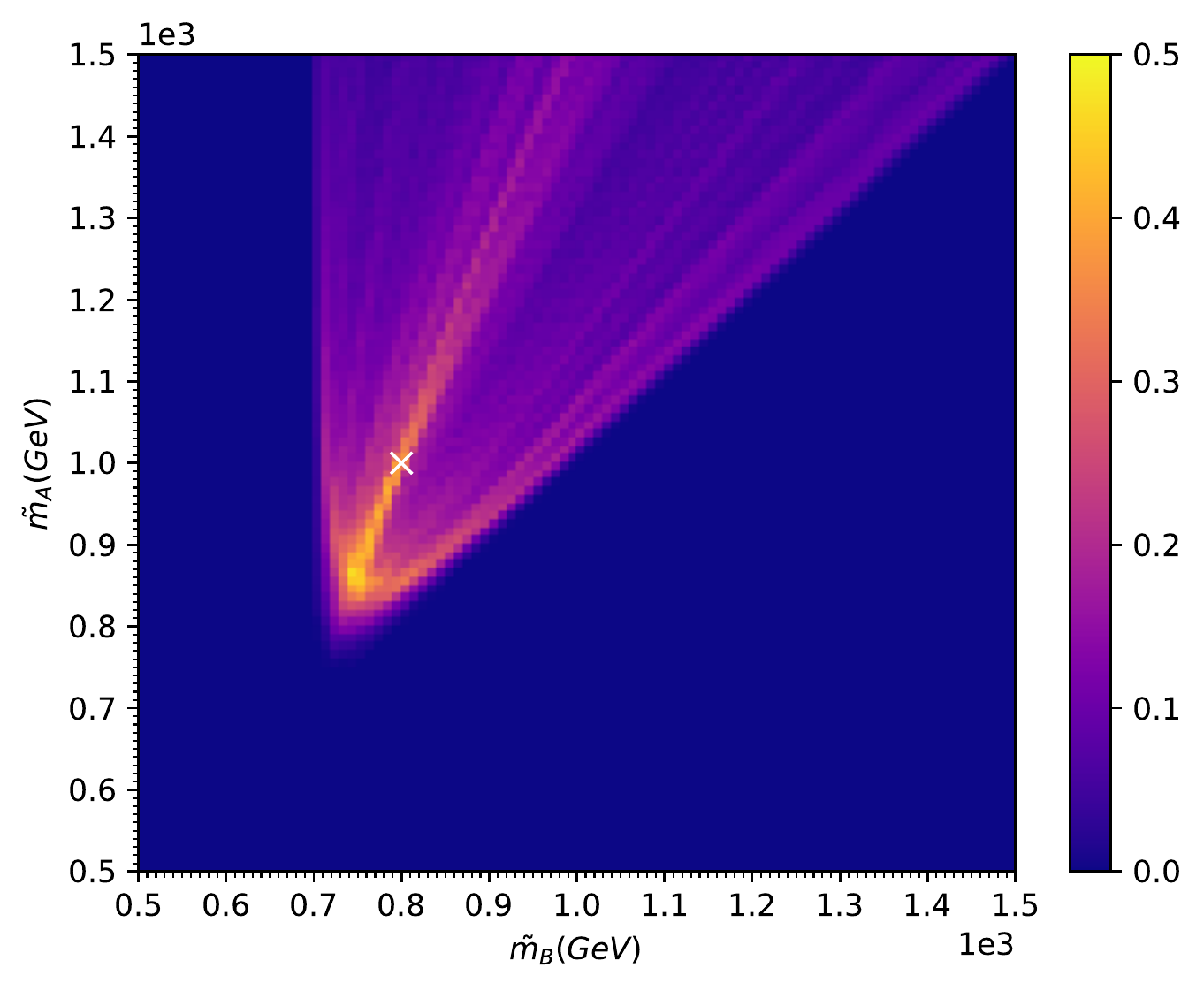}
 \caption{\label{fig:newmoneyplots} The same as Fig.~\ref{fig:moneyplot1}, for Standard Model dilepton $t\bar{t}$ background events (left panel)
 and preselected signal events (right panel), both for $\tilde m_C=700$ GeV. 
 The signal events entering the plot on the right have been preselected to obey all four kinematic endpoints (\ref{endpoints})
 predicted for the Standard Model dilepton $t\bar{t}$ background. The white $\times$ symbol marks the location $(m_B,m_A)$ of the true masses 
 in the signal topology. }
\end{figure}
The main advantage stems from the fact that we are able to extract useful information from {\em all} events and not just those which
populate the vicinity of a kinematic endpoint. In order to illustrate this idea, we can look at the performance of the focus point method on 
a subset of events which would not assist in the standard extraction of the kinematic endpoints of the signal through a fit. 
The mass spectrum for our study point ($\mA=1000\GeV$, $\mB=800\GeV$ and $\mC=700\GeV$) happens to be such that 
all four kinematic endpoints (\ref{endpoints}) for the signal are higher than the respective endpoints for the 
$t\bar{t}$ background. Therefore, by restricting ourselves to signal events which satisfy all four {\em background} endpoints,
we can select signal events far away from the signal endpoints; such events are also hard to distinguish kinematically from the background. 

In Fig.~\ref{fig:newmoneyplots} we contrast the heat maps (created as in Fig.~\ref{fig:moneyplot1}) 
for Standard Model dilepton $t\bar{t}$ background events (left panel) and so preselected signal events (right panel), both for $\tilde m_C=700$ GeV. 
The white $\times$ symbol marks the location $(m_B,m_A)$ of the true masses  in the signal topology.
The signal events entering the plot on the right have been preselected to obey all four kinematic endpoints (\ref{endpoints})
predicted for the Standard Model dilepton $t\bar{t}$ background. Therefore, as far as kinematic endpoints are concerned, 
these signal events are very hard to distinguish from the SM background, and in particular, cannot be used for discovery \cite{Cho:2014yma}.
Yet the heat map for these ``low utility" signal events in the plot on the right still shows a distinct feature through the correct masses, 
which is not present for the SM case shown in the left panel.\footnote{The absolute maximum in the right plot of Fig.~\ref{fig:newmoneyplots}
is shifted away from the white $\times$ symbol due to the bias introduced by the applied preselection. 
Once the whole signal sample is considered, the peak will shift back to the correct position, as shown in Fig.~\ref{fig:moneyplot1},
and will include a substantial contribution from the low utility signal events from Fig.~\ref{fig:newmoneyplots}.}

The focus point method is based purely on event kinematics, and is therefore quite model-independent.
Note that in addition to measuring the masses of particles, this technique can also be used as a new physics search technique
by designing an analysis which would look for a bump feature in the density of the kinematic boundaries.

\acknowledgments
DK was supported in part by the Korean Research Foundation (KRF) through the CERN-Korea Fellowship program, and is presently supported by the Department of Energy under Grant No. DE-FG02-13ER41976/DE-SC0009913.  The work of KM and PS is supported in part by the United States Department of Energy under Grant No. DE-SC0010296.

\appendix

\section{The Flat Direction of Low Sensitivity in Mass Space}
\label{app:flatdirection}

The numerical results in Sections~\ref{subsec:mc} and \ref{sec:fpmc} suggested the presence of a region (``flat direction") in mass space
which has relatively low sensitivity. In other words, the analysis is least sensitive to the degree of freedom (in mass parameter space) parametrizing the flat direction.
In this Appendix we identify and parametrize the flat direction.

One of the key ideas in deriving the flat direction is that the kinematic constraints \eqref{eqn:met} and \eqref{eqn:on-shell} are a) Lorentz invariant and b) invariant under separate longitudinal boosts of the two branches in 
Fig.~\ref{fig:feynmandiag}. Therefore, without loss of generality we can restrict our attention to events in which both $A_1$ and $A_2$ 
are produced at rest in the longitudinal direction, and have equal and opposite transverse momenta. The other key idea is that 
the flat direction arises because there is an additional hidden constraint which is approximately satisfied by events in our sample,
but is unaccounted for in our analysis. This constraint arises due to the fact that 
at LHC energies, the transverse momenta of the parent particles  are small relative to their masses. 
This is a good approximation at hadron colliders like the LHC, provided that the $\A{i}$ particles are not too light. 
In this case, one can show that there exists a flat direction near the true mass point which satisfies
\begin{subequations}
\bea
\frac{d\tmA^2}{d\tmC^2} &=& \frac{\tmA^2+\tmB^2}{\tmC^2+\tmB^2}, \\ [2mm]
\frac{d\tmB^2}{d\tmC^2} &=& \frac{2\tmB^2}{\tmC^2+\tmB^2}.
\eea
\end{subequations}
These equations can be solved to give
\begin{subequations}
\bea
\frac{\tmA^2-\tmB^2}{\tmB} &=& \frac{\mA^2-\mB^2}{\mB} = \textrm{constant}, \\ [2mm]
\frac{\tmB^2-\tmC^2}{\tmB} &=& \frac{\mB^2-\mC^2}{\mB} = \textrm{constant}. 
\eea
\label{eq:flat}
\end{subequations}

Table \ref{tab:flat} lists the mass spectrum predicted by (\ref{eq:flat}) for the three representative values for $\tmC$ used on the plots.

\begin{table*}[h]
\begin{center}
\setlength{\tabcolsep}{1.1mm}
\renewcommand{\arraystretch}{1.4}
\begin{tabular}{|c|c|c|}
\hline   
$\tmC$ (GeV) & $\tmB$ (GeV) & $\tmA$ (GeV)
\\
\hline \hline
400 & 505  & 694         \\ \hline
700 & 800 & 1000 \\ \hline
1000 & 1098 & 1304
\\
\hline
\end{tabular}
\end{center}
\caption{ Mass spectrum along the flat direction (\ref{eq:flat}) for the three $\tmC$ values used in the paper. } 
\label{tab:flat}
\end{table*}

\end{document}